\documentclass[
  aps,
  rmp,
  preprint,
  longbibliography,
  nofootinbib,
  amsmath,
  amsfonts,
  amssymb,
  floatfix
]{revtex4-2}

\usepackage{fullpage,amsmath,xparse,amssymb,mathtools,graphicx,braket,xcolor,verbatim}

\usepackage[colorlinks=true,linkcolor=blue,citecolor=blue]{hyperref}

\usepackage[colorlinks=true,linkcolor=blue,citecolor=blue]{hyperref}

\newcommand{\cev}[1]{\reflectbox{\ensuremath{\vec{\reflectbox{\ensuremath{#1}}}}}}

\usepackage{multirow}
\usepackage{cancel}

\begin{document}

\title{Ultralight Dark Matter: Progress and Current Constraints}

\author{Andrew Eberhardt}
\thanks{Kavli IPMU Fellow.}
\email{\\ andrew.eberhardt@ipmu.jp}
\author{Elisa G. M. Ferreira}
\email{elisa.ferreira@ipmu.jp}
\affiliation{Kavli Institute for the Physics and Mathematics of the Universe (WPI), UTIAS, The University of Tokyo, Chiba 277-8583, Japan}

\begin{abstract}

Ultralight dark matter refers to the lightest potential dark matter candidates. We will focus on the mass range that has been studied using gravitational signatures on astrophysical and cosmological observations, corresponding to a mass $10^{-24} \, \mathrm{eV} \lesssim m \lesssim 10^{-18} \, \mathrm{eV}$. We will discuss the motivations for this mass range. The most studied model in this range corresponds to a minimally coupled, single, classical, spin-0 field comprising all dark matter, originally proposed as ``fuzzy" dark matter. However, the work exploring extensions of this model (for example, higher spin, self-coupled, multiple field, and mixed models) will be one of the focuses of this review. The phenomenology associated with ultralight dark matter is rich and includes linear effects on the linear power spectrum, solitonic core structures forming at the center of halos, nonlinear effects resulting in heating of stellar distributions, and relativistic effects relating to pulsar signals and black hole superradiance, to name a few. This set of effects has been studied using an equally extensive set of numerical tools. We will summarize the most common ones and discuss their applications and limitations. Ultralight dark matter also has a wide variety of observational constraints, including those from halo mass functions, the Lyman-alpha forest, halo density profiles, stellar dynamics, and black hole spins. We will review them, focusing on the observations made, the method of study, and the major systematics. We will end with a discussion of the status of the field and future work needed. 

\end{abstract}

\maketitle
\tableofcontents

\section{Introduction}

The current cosmological model $\Lambda$CDM has been very successful at describing the observed large-scale structure \cite{Planck2020}. One of the main components of this model is dark matter which is thought to make up approximately 5/6ths of all matter in the universe \cite{Planck2020}. The evidence for the existence of dark matter is quite strong and includes galaxy rotation curves \cite{Rubin1970, Freeman1970}, redshift \cite{Zwicky1933, Smith1936}, the cosmic microwave background \cite{Bennett2013, Planck2020}, weak lensing \cite{Hoekstra_2004, Clowe2006}, and other probes. A large body of effort has been dedicated to creating a plethora of
models and experiments to identify the correct dark matter model \cite{Feng2010}. The models themselves have a large level of variety and complexity but it is often useful to parameterize them according to the dark matter particle mass, which spans a range from $\sim 10^{-22} \, \mathrm{eV}$ to $\sim 10 \, \mathrm{M}_\odot$ \cite{Feng2010}. Despite a global effort, there has been no confirmed non-gravitational detection of dark matter.

The term ``ultralight" dark matter can refer to a wide class of models with masses ranging from $10^{-22} \, \mathrm{eV} \lesssim m \lesssim \, \mathrm{eV}$. Specific models may include the QCD axion \cite{Peccei1977, Kim1979}, dark photon \cite{Fabbrichesi2021}, or light particles from compactified extra dimensions in string theory \cite{Arvanitaki2010, Svrcek2006, Halverson2019,Bachlechner2019}, among others. The couplings of these models to other particles in the dark sector or to the standard model are also varied. However, a unifying characteristic of all of them is that they are sufficiently light that the
occupation number per de Broglie wavelength in galactic systems is large
compared to unity. This is generally true for dark matter candidates with $m \lesssim \, \mathrm{eV}$. This, combined with the Pauli exclusion principle, then implies that all ultralight dark matter candidates must be bosonic if they are the dominant dark matter component\cite{Tremaine1979}. However, multi-component ultralight Fermionic has also been considered \cite{Davoudiasl2020}. When occupation numbers are large compared to unity, we are in the ``wave dark matter" regime, which has a field rather than a particle description on scales smaller than or similar to the de Broglie (or in some cases Compton) wavelength and timescale of the field.

In this review article, we will focus on the work done on the lower mass bound of ultralight dark matter, around $\lesssim 10^{-19} \, \mathrm{eV}$. This sub-region of parameter space is often referred to as ``fuzzy" dark matter \cite{Hu2000}. In this mass regime, the dark matter manifests wave-like phenomena on astrophysical scales \cite{Hu2000}. Often, this model is treated as a minimally coupled classical field constituting all of the dark matter. This straightforward model is attractive because much of the phenomenology depends only on the field mass (although the choice of initial conditions can also be important for some observations). And despite its simplicity, this model contains a rich phenomenology producing constraints from a wide range of astrophysical observables.

Historically, this model was originally motivated by small-scale structure discrepancies between observations and dark matter only simulations \cite{Weinberg2015, Bullock2017}. These small-scale structure problems are usually summarized as the core-cusp \cite{Navarro1995, Persic1995, Gentile2004}, missing satellites \cite{Klypin1999, Moore1994}, and too-big-to-fail problems \cite{Boylan-Kolchin2011}. It was proposed that a dark matter with mass around $10^{-22} \, \mathrm{eV}$ would alleviate some of these discrepancies \cite{Hu2000}. And while this is historically interesting, these motivations are currently less relevant for the study of ultralight dark matter. The reasons being 1) the small-scale structure problems are now thought to admit baryonic or observational solutions \cite{Weinberg2015, Bullock2017} and 2) a $10^{-22} \, \mathrm{eV}$ mass field comprising all the dark matter is now significantly disfavored by a wide variety of probes \cite{Nadler2021,Nadler_2025,Garland2024,Bar2022,zimmermann2025,Rogers:2020ltq,Powell2023,teodori2025}. Instead observations of small-scale structure now provide some of the leading constraints on the lower mass bound of ultralight dark matter models and much of the modern work in the fuzzy dark matter field is concerned with improving this lower bound.

Ultralight dark matter phenomenology affects a wide array of astrophysical observations. Gradient energy in the non-relativistic field equations generates an effective pressure, the so-called ``quantum'' pressure (in analogy to condensed matter physics), which resists the formation of structure below certain scales \cite{Hu2000}\footnote{This term is often heuristically associated with the uncertainty principle in the condensed matter and Bose--Einstein condensate literature. In the cosmological ultralight dark matter context, however, it arises directly from gradient energy in the classical non-relativistic field equations.}. This impacts the linear power spectrum of density fluctuations and changes the halo mass function \cite{Nadler2021, Garland2024}. This can also impact the scales at which we would expect to observe fluctuations in the Lyman-alpha forest \cite{Rogers:2020ltq, Irsic2017, Armengaud2017, Kobayashi2017}. The ``quantum pressure'' also results in solitonic cores forming at the center of galaxies, which can impact stellar rotation curves \cite{Bar2018, Bar2019, Bar2022}. On the scale of de Broglie wavelength, density granules in ultralight dark matter halos can heat stars and disrupt compact objects \cite{dalal2022, Marsh2019, Chowdhury2023}. Fluctuations in the field can produce metric perturbations which may be observable with lensing or pulsar timing \cite{Porayko2018, Powell2023, Kim2024, Eberhardt2024}. All of these effects and constraints exist in the minimally coupled classical spin 0 (vanilla) model by making very simple arguments about the existence of a single ultralight field that constitutes all the dark matter. These phenomena are not independent but are connected through the characteristic scale set by the particle mass. The suppression of small-scale power in the linear matter power spectrum propagates into a cutoff in the halo mass function and determines the scales probed by the Lyman-$\alpha$ forest. On nonlinear scales, the same physics leads to solitonic cores whose size is related to the Jeans scale, while interference patterns within halos occur on the local de Broglie wavelength.

Building on this basic model, a large body of more recent work has studied the implications of model extensions for ultralight dark matter phenomenology and constraints. These extensions include studying the impact of multiple fields \cite{Tellez-Tovar2021, Street2022, Guo2021, Luu2020, Huang2023, Glennon2023, Vogt2022, Chen2023, Gosenca2023,Mirasola2024,Luu2023,luu2024, Jain2023c}, self-interacting models \cite{Glennon2022,Glennon2023b,Mirasola2024, Dave2024}, higher spin fields \cite{Amin2022, Amin2023, LopezSanchez2025, Jain2022, Jain2023, Jain2023b, Jain2023c, Amaral2024, Adshead2021, Schiappacasse2025}, and quantum corrections \cite{Eberhardt2023, Eberhardt2021, Eberhardt2022Q, Eberhardt2022, Eberhardt_testing, sikivie2017, Hertzberg2016, Allali_2020,Allali2021,Allali2021b, Lentz2019,Lentz2018,Kopp2021, Bernal2024}. The interaction of the increased model parameter space complexity with constraints and phenomena has yielded a large number of interesting results, which we will cover in this review. 

The most recent major review of ultralight dark matter is 4 years old at the time of the publication of these reviews \cite{Ferreira2021,Hui2021}. The main purpose of this review will be to present the necessary background to understand the current state of the field and its apparent trajectory. In addition to summarizing the relevant physics behind the most important phenomena (which remains relatively unchanged), we will devote specific effort to understanding the current constraint space. For each constraint, we will discuss the observational data used, on which numerical and analytic results it relies, which systematic errors are most relevant, and how it applies to extensions of the vanilla ultralight dark matter model. Each section is intended to be presented as pedagogically as possible, with relevant references provided for more complete analyses when desired. 

The review is organized as follows. In Section~\ref{sec:ULDM}, we define ultralight dark matter. Section \ref{sec:motivations} contains a discussion of the particle physics and observational motivations for ultralight dark matter. Section \ref{sec:uldmModels} summarizes the ultralight dark matter models and extensions. Section \ref{sec:pheno} discusses the physics relating to ultralight dark matter phenomenology. Section \ref{sec:numMethods} describes some of the most popular numerical methods used to study these phenomena. Section \ref{sec:constraints} discusses the constraints at the lowest mass end of ultralight dark matter. Finally, Section \ref{sec:conclusion} contains conclusions and a summary of the status of the field. 

\section{Ultra-light dark matter} \label{sec:ULDM}

\begin{figure*}[!ht]
	\includegraphics[width = .90\textwidth]{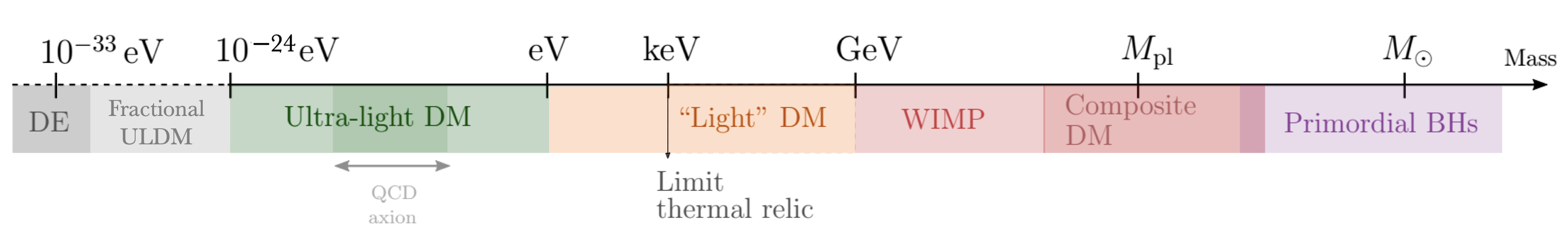}
	\caption{Illustration of the wide range of dark matter models proposed in the literature, spanning many orders of magnitude in mass. Adapted from~\cite{Ferreira2021}.}
	\label{fig:mass_scale}
\end{figure*}

Ultralight dark matter is the name given to the lightest possible class of dark matter candidates. This is described by a massive boson (of mass $m$), produced non-thermally in the early universe\footnote{Because this component is very light, it could not have been in thermal equilibrium with the universe plasma at late times, otherwise it would be very hot today, which is not allowed by observations, as we discuss in Section~\ref{subsec: DM_small_scales}.}. This bosonic particle can describe dark matter today if they have masses $10^{-32}\, \mathrm{eV} < m <  \mathrm{eV}$\footnote{This is the definition of ULDM that we use in this review. Other literature might use a different definition, so the reader should be aware of possible variations in terminology.} In the lower end, ULDM can be all the dark matter if $m > 10^{-24}\, \mathrm{eV} $, being only a subdominant part of DM for masses lighter than this (we will review these constraints in Section~\ref{sec:constrain_extremely_light}). For $m < 10^{-32}\, \mathrm{eV} $, this massive bosonic field behaves like dark energy. These limits are illustrated in Fig.~\ref{fig:mass_scale}.

But why study such a small mass particle? This mass range is interesting since not only does it have a strong motivation in particle physics, as we will see in the next section, but also since a particle with such a small mass presents a large de Broglie wavelength. The de Broglie wavelength of a particle is given by
\begin{align}
\lambda_{\mathrm{dB}} &\equiv \frac{2\pi \hbar}{m\, v} \\
& \sim 0.48 \mathrm{kpc} \left( \frac{10^{-22}\, \mathrm{eV}}{m} \right) \left( \frac{250\, \mathrm{km/s}}{v} \right)\,,
\end{align}
where $v$ is the velocity (in the non-relativistic limit) of the particle, or, in the case of dark matter, the velocity dispersion of the galactic halo, and $\hbar = h/2\pi$ is the reduced Planck's constant. Therefore, in the case of ULDM, the de Broglie wavelength can be very large, e.g., $\sim \mathrm{kpc}$ when $m \sim 10^{-23}\, \mathrm{eV}$ for a Milky Way-like halo (see Appendix A for other masses). Therefore for $m \lesssim \mathcal{O}(1-10) \, \mathrm{eV}$ (depending on the velocity dispersion of the galactic halo), in galaxies, the interparticle distance of the dark matter particles is smaller than the de Broglie wavelength of each particle, and the waves will superimpose creating a coherent superposition of waves describing a macroscopic wave with a high occupation number of these bosons in galaxies. With this high occupation in galaxies, we can describe ULDM as a coherent (classical) wave (fixed frequency, given by the mass, and constant but random phase). We will discuss this in detail in Section~\ref{sec:uldmModels}.

This large de Broglie wavelength leads to distinctive and testable signatures on galactic and sub-galactic scales. The main goal of this review is to describe the behavior of ULDM, its phenomenology, and observational constraints.

\begin{figure*}[!ht]
	\includegraphics[width = .67\textwidth]{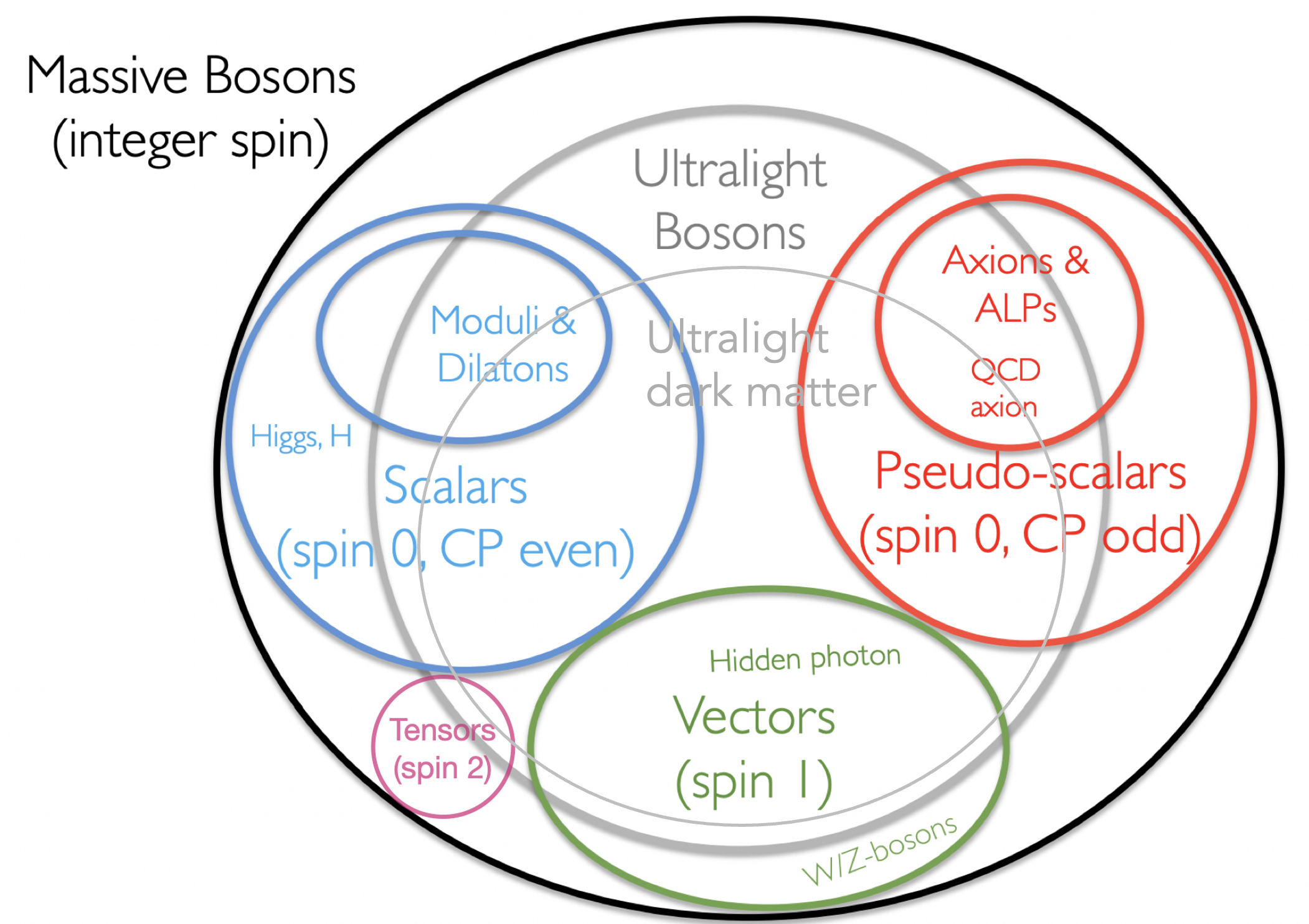}
	\caption{Massive bosons that can be the particles that explain the ULDM. Figure adapted from~\cite{Chadha-Day:2021szb}.}
	\label{fig:microphysics}
\end{figure*}

ULDM is the general name given to a wide class of models that can be described by different particle candidates. ULDM is also known in the literature as wave dark matter, ultra-light axions, scalar field dark matter, among other names.
There are many possible candidates for the particle that describes the ULDM, as we can see in Fig.~\ref{fig:microphysics}. Any ultralight massive boson that can be produced in enough abundance in the universe can be a candidate for the microphysical description of this component: from the QCD axion, to ALPs and other massive scalars, to even vectors or higher spin bosons. We will discuss in Section~\ref{sec:motivations} how these ultralight particles are predicted and even expected in extensions of the standard model of particle physics, leading to a good motivation for this dark matter candidate. 

Since this review focuses on gravitational signatures, we adopt a general phenomenological framework and do not commit to specific microphysics unless needed.  We are going to describe the behaviour of a massive scalar field (or higher spin field) under gravity. Within this phenomenological description, we have different phenomenological models for ULDM~\cite{Ferreira2021}. The simplest model, the vanilla ULDM, is the fuzzy dark matter (FDM)\footnote{Some authors use the term fuzzy dark matter for massive scalars with masses $10^{-22}\, \mathrm{eV} < m <  10^{-20}\, \mathrm{eV}$~\cite{Hui2021}. This name was given since the wave behaviour of this component makes the galaxy look fuzzy, which is more prominent for larger de Broglie wavelengths coming from such a light particle.} model, a weakly coupled massive scalar field under gravity. We can also have extensions like adding self-interaction to this model, the self-interacting fuzzy dark matter, or describe it with higher spin fields.  The mass range and other parameters are defined for each of these models.
We dedicate the entire Section~\ref{sec:constraints} to using observations to constrain the possible mass and fraction of ULDM.

There is also another class of these models, the superfluid dark matter model~\cite{Berezhiani:2015bqa, Berezhiani:2015pia} (see also~\cite{Ferreira2021}), which we do not discuss in this review. These models are also composed of ultra-light bosonic fields where wave behaviour plays a central role in the dynamics. However, dark matter forms a superfluid phase whose phonon excitations mediate an additional force, and they are therefore described within a different phenomenological framework.

In cases where the production mechanism or particle interactions are important—e.g., for isocurvature bounds or couplings to Standard Model fields—it is necessary to specify the particle content. These constraints can be stringent and are addressed in particle-focused reviews~\cite{Marsh:2015xka, Chadha-Day:2021szb}. We will only touch on these cases where relevant.

\section{Motivations} \label{sec:motivations}

Historically, the leading candidate for dark matter has been the weakly interacting massive particle (WIMP)—a heavy, beyond-the-Standard-Model particle introduced to account for the observed DM abundance. However, in recent years, a variety of alternative candidates have gained traction, including ultralight dark matter, which has emerged as a particularly intriguing possibility.

ULDM is compelling not only from the perspective of particle physics, where it naturally arises in theories involving light scalar fields, but also due to its distinctive astrophysical implications. In what follows, we review the key motivations driving interest in the ULDM framework.

\subsection{Particle physics}

One of the strongest motivations for ULDM arises from high-energy physics. Historically, interest in light bosonic fields began with the axion, originally proposed to solve the strong CP problem in quantum chromodynamics (QCD), entirely independently of the dark matter puzzle. However, it was soon recognized that the axion could also account for the observed dark matter abundance, linking two seemingly unrelated problems in particle physics and cosmology.

More broadly, it was soon realized that axion-like particles (ALPs) and other ultra-light bosonic fields (besides the SM ones) are quite a generic prediction in many beyond-the-Standard-Model (BSM) scenarios, including string theory.  These light bosons span a wide mass range, and many fall into the ultra-light regime relevant for dark matter. As a result, ULDM is now motivated not just by the specific case of the QCD axion, but by the broader expectation that light, weakly-coupled bosons are a generic prediction of fundamental theory.

\subsubsection{QCD axions}

Here we briefly review the QCD axion. The axion\footnote{In this review, we reserve the term axion for the QCD axion. However, the term is often used more broadly in the literature to refer to ALPs or other ULDM candidates.} is the prototypical example of ultralight dark matter, and remains one of the most theoretically well-motivated candidates.

The axion was proposed as a solution to the strong CP problem in QCD. CP symmetry refers to the combination of charge conjugation (C) and parity (P). It is known that CP is violated in the weak interactions, but the QCD action also allows for a CP-violating term. The strong CP problem arises from the fact that the QCD Lagrangian allows a CP-violating term proportional to a parameter $\theta$. While there is no theoretical reason for $\theta$ to be small, experimental bounds, in particular from measurements of the neutron electric dipole moment, require it to be unnaturally suppressed, with $|\theta| \lesssim 10^{-10}$~\cite{Abel:2020pzs}. This fine-tuning problem is known as the strong CP problem.

A solution was proposed by Peccei and Quinn, who introduced a new global $\mathrm{U}(1)_{\mathrm{PQ}} $ symmetry, now known as Peccei--Quinn (PQ) symmetry. This symmetry is spontaneously broken at some high energy scale \( f_a \), leading to the appearance of a \textit{pseudo-Goldstone boson}: the axion.

The key idea is to promote the static CP-violating angle $\theta$ to a dynamical field. In this framework, the axion field $\phi(x)$ couples to the QCD topological charge density via the term $\mathcal{L} \supset \frac{\phi(x)}{f_a} \frac{g_s^2}{32\pi^2} G_{\mu\nu}^a \tilde{G}^{\mu\nu,a} \,$,
where $ G_{\mu\nu}^a $ is the gluon field strength and $ \tilde{G}^{\mu\nu,a}$ its dual. Due to nonperturbative QCD effects (instanton corrections), the axion acquires a potential that is minimized at a field value that cancels the effective CP-violating term: $\theta_{\text{eff}} = \theta + \langle \phi \rangle/f_a \rightarrow 0$.
Therefore, the axion dynamically relaxes to a vacuum expectation value that ensures CP conservation in QCD, solving the strong CP problem without fine-tuning.

However, this mechanism not only explains the absence of observable CP violation in the strong interaction, but also predicts the existence of a new light particle, the axion, whose properties are determined by the Peccei--Quinn scale $f_a$. In particular, the axion's mass and its interactions with Standard Model particles are inversely related to $f_a$: the higher the symmetry-breaking scale, the lighter and more weakly coupled the axion becomes~\cite{Weinberg:1977ma,Wilczek:1977pj}.

From a cosmological perspective, the axion can also play the role of dark matter. There are two main production mechanisms, depending on when the PQ symmetry breaks relative to inflation:
\begin{itemize}
    \item Pre-inflationary symmetry breaking: If the PQ symmetry is broken before or during inflation, the axion field is homogenized over our observable Universe. It can be initially displaced from the minimum of its potential, and as the Universe expands and cools, the Hubble parameter eventually drops below the axion mass. At this point, the field begins to oscillate coherently around the minimum. These oscillations behave like cold, pressureless matter and contribute to the dark matter density, the misalignment mechanism~\cite{Preskill:1982cy,Abbott:1982af,Dine:1982ah}. For a broad range of initial field values and PQ scales, this process can yield a relic abundance consistent with the observed dark matter:
    \begin{align}
        \Omega_\mathrm{uldm} \sim 0.1 \left( \frac{f_a}{10^{17} \, \mathrm{GeV}} \right)^2 \left( \frac{m}{10^{-22} \mathrm{eV}} \right)^{1/2} \, .
    \label{eq:abundance_axions}
    \end{align}
    
    \item Post-inflationary symmetry breaking: If the PQ symmetry breaks after inflation, different regions of the Universe acquire different initial values for the axion field. In this case, topological defects such as cosmic strings and domain walls form. As these defects evolve and decay, they emit axions, contributing to the dark matter density. This mechanism is more complex and involves additional uncertainties in the contribution from strings and domain walls, but it can still account for the observed dark matter density for certain ranges of $f_a$~\cite{Davis:1989nj,Battye:1994au,Hiramatsu:2012gg,Kawasaki:2014sqa}.
\end{itemize}

Both scenarios highlight the axion as a predictive and well-motivated dark matter candidate, with its cosmological history deeply connected to the early Universe and the nature of symmetry breaking.

We point to the beautiful and clear Figure 4 from~\cite{Chadha-Day:2021szb} for a schematic view of the production mechanisms for axion DM.

\subsubsection{ALPs and other massive bosons}

 A wide range of ultralight bosonic fields, pseudo-scalars, scalars, and vectors, are motivated from high-energy theory and can behave as dark matter on cosmological scales.

 Axion-like particles are pseudo-Nambu–Goldstone bosons associated with spontaneously broken global symmetries. Like the QCD axion, they are light pseudo-scalars, but unlike it, their mass and decay constant are not related by QCD dynamics. In general, for ALPs, the mass $m$ and decay constant $f_a$ are independent parameters. This makes them flexible dark matter candidates with a wide parameter space.

ALPs are only one example of a broader class of ultralight bosonic fields that naturally appear in theories beyond the Standard Model. In particular, string theory compactifications predict not only pseudo-scalars like ALPs, but also scalars such as moduli and dilatons. These fields parameterize the shape and size of the extra dimensions in string theory and are generically expected to be light. If stable and produced with the right abundance, they can also act as ultralight dark matter~\cite{Svrcek2006}.

This motivates what is known as the axiverse scenario~\cite{Arvanitaki2010, Sheridan2025}. In this framework, string theory compactifications lead to the appearance of a large number of light bosonic fields~\cite{Green1987, Svrcek2006, Arvanitaki2010, Dine2007, Halverson2019, Bachlechner2019, Sheridan2025}. These particles can include axions, ALPs, and moduli, spanning many orders of magnitude in mass. The axiverse naturally predicts a landscape of ultralight fields, some of which may behave as dark matter.

Furthermore, string theory and other extensions of the Standard Model can also predict ultralight vector bosons, such as hidden photons. Some attempts were also made to produce ultralight tensor (spin-2) dark matter.

Although equation (\ref{eq:abundance_axions}) assumes the misalignment mechanism, it is usually used as a first approximation for the abundance of ALPs and other ultralight bosons since it gives the correct scaling behavior. However, the precise abundance depends on the microphysics of this massive boson. ALPs arising in string theory or other BSM models, additional dynamics (e.g., coupling to other sectors, mass generation mechanisms) can modify the effective relic abundance, see, for example \cite{Cyncynates_2022}.

\subsection{Dark matter on small scales}
\label{subsec: DM_small_scales}

The evidence for the existence of dark matter is overwhelming, coming from different observations on different scales. From rotation curves of galaxies, to clusters, gravitational lensing, to the study of large-scale structures. Although we still do not know the nature of dark matter, all of these measurements reveal different properties of dark matter, and the evolution and composition of our universe. In particular, given the very precise measurements of the large-scale structure and the cosmic microwave background (CMB), we were able to construct the standard cosmological model. The standard model of cosmology, $\Lambda$CDM, is known to be extremely predictive of large-scale structure given a few parameters \cite{Planck2020}. In this, dark matter is described by the cold dark matter (CDM) model, a coarse-grained description where DM is given by a perfect fluid which is cold, pressureless, dark (it does not or weakly interacts with other forces besides gravity), and with no or small self-interaction. This model is known to explain the large-scale observations very accurately, which can be seen by the exquisite fit to the matter power spectrum up to scales of $k \lesssim 10 - 20 \, \mathrm{Mpc}^{-1}$. The large-scale measurements provide the strongest bounds on the dark matter properties.  However, we do not have the same level of precise observations for the smaller scales, which allows deviations from the CDM model and extra properties for the DM candidate to have. For example, the fact that the matter power spectrum is highly unconstrained on scales $k \gtrsim 10 - 20 \, \mathrm{Mpc}^{-1}$ allows DM to be e.g. warm ($m > \mathrm{keV}$), or have a small self-interaction, or a small electromagnetic charge, among other properties. This, together with the still evolving measurements of the small galactic properties of DM, leads to a quest to search for the properties of DM on small scales, which can lead us to unravel the nature of DM.

This can also be summed up to a number of discrepancies between observations and simulations that exist at small scales (see \cite{Bullock2017} for a review). Collectively, these were formerly known as the small-scale structure problems and include the too-big-to-fail, missing satellites, profile diversity, and core-cusp problems. 
There is ongoing discussion in the literature about the extent to which these discrepancies indicate limitations of the CDM model, with several studies suggesting that baryonic effects and improved comparisons with observations may alleviate or even solve them (see e.g.~\cite{Bullock2017,Sales:2022ich}). 

\subsubsection{Core-cusp problem}

This problem refers to the fact that simulated halos have ``cuspy" inner profiles, whereas observed halos are generally ``cored" \cite{Navarro1995,Walker2011}. Early dark matter only, n-body simulations demonstrated that the inner profile of halos goes as $\rho(r) \sim r^{-1}$ \cite{Dubinski1991}. Later studies indicate that dark matter only simulations produced halos fit by the Navarro-Frenk-White (NFW) profile \cite{Navarro1995}, given
\begin{align} \label{eqn:NFW_profile}
    \rho_{\mathrm{NFW}} = \frac{\rho_s}{(r/r_s)(1 + r/r_s)^2} \,,
\end{align}
where $\rho_s$ and $r_s$ are the scale density and radius, respectively.

The solution to this problem can be thought of in several ways: (a) within the CDM model, baryonic effects are responsible for creating these cores; (b) improved comparisons between simulations and observations alleviate the discrepancy~\cite{Zentner:2022xux,Roper:2022umd,Sands:2024vvd}; (c) alternative dark matter models produce cores as part of their dynamics. 

Regarding (a), many groups have been improving their simulations to include baryons and their effects in order to answer whether baryons alone can explain the observed cores. Including baryons in these simulations is a very daunting task, and currently, this is done by modelling their effects. Different simulations use different models, with different levels of baryonic feedback included, leading to different results (see ~\cite{Vogelsberger:2019ynw} and references within for a review). Baryonic processes can generate cores in some simulations, although whether baryons alone fully resolve the cusp-core problem remains debated (see e.g.~\cite{DelPopolo:2021bom,Sales:2022ich,Straight:2025udg,Jahn:2021gpf}). This leads to the second possibility that we need to change the DM model in order to produce cores from the DM dynamics, with a combination of both effects, from DM and baryons, being the final description of the cored profile.

It was in this context, and of the other small-scale problems, that ULDM was proposed as a possible solution~\cite{Hu2000}. 
As we are going to see in detail in the following sections, due to the wave nature of this DM candidate on small scales, ULDM halos have solitonic cores in their centers supported against further collapse by ``quantum" pressure. We will see in Section \ref{subsec:Pheno:solitons} a description of these cores and in Section~\ref{sec:constraints} how we can use observations of these cores (or their absence) to put bounds on the mass of the ULDM.

\section{Ultra light dark matter models} \label{sec:uldmModels}

\subsection{Sch\"odinger-Poisson equations}
Much of the work studying ultralight dark matter relies on the Schr\"odinger-Poisson equations. These are the equations of motion governing a single, non-relativistic, classical, spin-0 field. There also exists a large body of work investigating multi-field, relativistic, quantum, or higher spin fields which will be discussed in later sections. 

Typically, we start the derivation of these equations from a classical Klein-Gordon field \cite{Gordon1926,Klein1926}, whose Lagrangian is given as
\begin{align}
    \mathcal{L} = \frac{1}{2} \partial_\mu \phi(x) \, \partial^\mu \phi(x) - \frac{1}{2} m^2 \phi^2(x) - \frac{\lambda}{4} \phi^4(x) \, , 
\end{align}
and equation of motion is given as
\begin{equation} \label{eqn:KG}
    \frac{1}{\sqrt{-g}} \partial_\mu (\sqrt{-g} \, g^{\mu \nu} \partial_\nu) \phi(x) + m^2 \phi(x) + \lambda \phi^3(x) = 0 \, .
\end{equation}
Where $\phi$ is the real Klein-Gordon field, $g$ the metric, $m$ is the field mass, and $\lambda$ a constant representing the interacting strength. It will be useful to write the Klein-Gordon field as a sum of a complex field, $\psi(x)$ and its complex conjugate as
\begin{align} \label{eqn:KG2Schr}
    \phi(x) = \frac{1}{\sqrt{2m}} \left( e^{-imt} \psi(x) + \mathrm{c.c.}\right) \, .
\end{align}
The normalization above is typical, but arbitrary. The exponential component, $e^{-imt}$, oscillates at on the Compton timescale, i.e., $\tau_c = 2 \pi \hbar / m c^2$. However, in the non-relativistic regime, the complex field, $\psi(x)$, oscillates on the generally much slower de Broglie timescale, i.e. $\tau_\mathrm{db} = 2\pi \hbar / m v^2$, where here $v$ is some characteristic velocity in the system. By writing the field in this way, we can separate the fast and slow components. We will usually want to focus on $\psi$, the Schr\"odinger field, which is usually more important for most astrophysical and cosmological considerations (though there are at least a few notable exceptions).

The derivative term in equation \eqref{eqn:KG} is contracted using the weak gravity and non-relativistic FRW metric, which is given using the Newtonian gauge as
\begin{equation} \label{eqn:weakFieldMetric}
    ds^2 = (1 + 2 \Phi(x)) \, dt^2 + a^2(t) \, \delta_{ij} \,  dx^i dx^j .
\end{equation}
Here $a(t)$ is the scale factor and $\Phi(x) = V(x) / c^2$ is the dimensionless Newtonian potential which solves Poisson's equation
\begin{align} \label{eqn:Poisson}
    \nabla^2 V(x) = 4 \pi G m |\psi(x)|^2 \, .
\end{align}
Here we have chosen the normalization of $\psi$ such that it is the nonrelativistic density, i.e., $|\psi(x)|^2 = \rho(x)$. Using equation \eqref{eqn:weakFieldMetric} then we can rewrite equation \eqref{eqn:KG} as
\begin{align}
    \left[ \frac{3 H \partial_t + \partial_t^2}{(1+ 2 \Phi)} + \frac{\nabla^2}{a^2(t)} \right] \phi(x) + m^2 \phi + \lambda \phi^3 = 0 \, .
\end{align}
Where $H \equiv \dot a / a$ is the Hubble constant. To study the nonrelativistic portion of the field we now plug in equation \eqref{eqn:KG2Schr}. This gives
\begin{align}
    0 = &-i \partial_t \psi(x) + \frac{\nabla^2 \psi(x) }{2m \, a^2(t)} - m V(x) \psi(x) + \frac{\lambda}{4 m^2} \psi^3 \label{eqn:SP_exp} \\ 
    &+ \frac{3 H}{2(1+ 2 \Phi)} \left( -i\psi(x) + \frac{\partial_t \psi}{m} \right) + \frac{\partial^2_t \psi(x)}{2m} \, .
\end{align}
The first line is the Schr\"odinger Poisson equation in an expanding background with an interaction term, and the second line contains terms which are typically assumed to be small. The Hubble friction term can be removed by rescaling the field, and for the non-expanding equations of motion, we simply set the scale factor to unity, i.e., $a(t) = 1$.

In the non-expanding, non-relativistic, classical case, we generally write the Schr\"odinger-Poisson equations as
\begin{align} \label{eqn:SP_eqn_w_a}
    \partial_t \psi(x,t) &= \frac{-i}{\hbar} \left(\frac{-\hbar^2 \nabla^2}{2 m } + m \, V(x,t) + \frac{4 \pi \hbar^2 a_s}{m} |\psi(x,t)|^2 \right) \psi(x,t) \, , \\
    \nabla^2 V(x,t) &= 4 \pi G |\psi(x,t)|^2 \, .
\end{align}
where here we have written the self-interaction term in terms of the scattering length,$a_s$, i.e., $\lambda = 16 \pi \, m \, \hbar^2 a_s$. Without the self-interaction term, we write
\begin{align} \label{eqn:SP_eqn}
    \partial_t \psi(x,t) &= \frac{-i}{\hbar} \left(\frac{-\hbar^2 \nabla^2}{2 m } + m \, V(x,t) \right) \psi(x,t) \, , \\
    \nabla^2 V(x,t) &= 4 \pi G m |\psi(x,t)|^2 \, ,
\end{align}
where $m$ is the field mass, $p = -i \hbar \partial_x$ is the momentum operator, $a_s$ is the scattering length, and $V(x)$ is the gravitational potential, given by the Poisson equation. Here we have parametrized the interaction term using $a_s$, we note that this is not the only common parametrization. Much of the work studying ultralight dark matter will also set the interaction term to $0$ as well. 

\subsection{Extensions: self-interactions, multiple fields, particle spins}

\subsubsection{Multiple fields}

The simplest ultralight dark matter model contains a single complex classical field which describes all the dark matter. However, there also exists a large body of work looking at the phenomenological implications of relaxing the single field assumption \cite{Tellez-Tovar2021, Street2022, Guo2021, Luu2020, Davoudiasl2020,Huang2023,Glennon2023, Vogt2022, Chen2023, Gosenca2023, Mirasola2024, Luu2023, luu2024}. Generally, this is motivated by the ``axiverse" scenario \cite{Arvanitaki2010, Sheridan2025} in which the existence of many ($\sim \mathcal{O}(100)$) of ultralight fields may exist. In this scenario, we write the Schr\"odinger-Poisson equations as
\begin{align} \label{eqn:multiFieldSP}
    \partial_t \psi_j (x,t) &= \frac{-i}{\hbar} \left( \frac{-\hbar^2 \nabla^2}{2 m_j } + m_j \, V(x,t) \right) \psi_j(x,t) \, , \\
    \nabla^2 V(x,t) &= 4 \pi G \sum_j m_j |\psi_j(x,t)|^2 \, .
\end{align}

If the multiple fields have uncorrelated density fluctuations, then one of the implications of this extension is that the amplitude of density fluctuations is decreased \cite{Gosenca2023} with the reduction simply going as $\sqrt{N_\mathrm{fields}}$. Isolated halo simulations in \cite{Gosenca2023} determined that density fluctuations did become correlated over time. This extension would then be most important for constraints which rely on density fluctuations, for example, the heating of stellar dispersions, see Section \ref{subsec:granuleDensityPatterns}. The implications of multiple fields for these constraints was considered in \cite{Gosenca2023}.

The impact of multiple fields on halo cores has also been studied in some depth \cite{luu2024, Huang2023, pozo2024, Mirasola2024}. The shape and ability of nested cores to form can be complicated by the oscillations of one of the cores \cite{luu2024} or self-interactions between multiple fields \cite{Mirasola2024}. Interestingly, it was pointed out that the dwarf galaxy halo cores can be fit as two populations if by a two-field model in which a single field dominates in each halo \cite{pozo2024}. 

Cosmological simulations of multifield dark matter have also been performed \cite{Huang2023,Luu2023,luu2024}. Interestingly, it was demonstrated that the halo cores can have a diversity of different final constituent field fractions \cite{luu2024} (i.e. different halos have different fractions of each field). A full characterization of this phenomenon has yet to be performed. But if multifield ultralight dark matter does provide a wide diversity of halos, it would likely alleviate tensions associated with observations of individual halos (for example \cite{Marsh2019, dalal2022, teodori2025, Bar2022, zimmermann2025}, many of which make up some of the strongest current constraints).

\subsubsection{Self-interactions}
Generally, when working with ultralight dark matter, we only consider gravitational couplings. However, axiverse and QCD axion models, which may motivate the existence of ultralight fields, generally include a quartic self-coupling or couplings to other ultralight fields. While much of the work studying ultralight dark matter ignores the self-interaction term there also exists a body work which has looked at the effect the self-interaction has on a range of phenomena, including the effect on grounds state solutions \cite{Mirasola2024, Dave2023}, halo core structure \cite{Painter2024, Desjacques2018,mocz2023, Dawoodbhoy2021, Lee1995, Indjin2025, Indjin2024}, soliton condensation \cite{Chen2021}, tidal disruption \cite{Glennon2022, Dave2024}, the Jeans scales \cite{Shapiro2021}, dynamical friction \cite{Glennon2023b} and soliton interactions \cite{Amin2019, Glennon2021}. The one field equations of motion is given as equation \eqref{eqn:SP_eqn}. The equations of motion generalized in the multi-field case to \cite{Mirasola2024}
\begin{align} \label{eqn:multiFieldGPP}
    \partial_t \psi_j (x,t) &= \frac{-i}{\hbar} \left( \frac{-\hbar^2 \nabla^2}{2 m_j } + m_j \, V(x,t) \right. \\
    &\left. + \frac{\hbar^3}{2 m_j^2} \lambda_{jj} |\psi_j(x,t)|^2 + \frac{\hbar^3}{4 m_j} \sum_k \frac{\lambda_{jk}}{m_k}|\psi_k(x,t)|^2 \right) \psi_j(x,t) \nonumber \, , \\
    \nabla^2 V(x,t) &= 4 \pi G \sum_j |\psi_j(x,t)|^2 \, .
\end{align}
Where the square matrix $\lambda_{jk}$ describes the coupling between the $j$th and $k$th field. The above equation is the muilti-field Gross-Pitaevskii-Poisson equations. 

For the QCD axion, which is often studied using the one-field version of equation \eqref{eqn:multiFieldGPP}, the interaction term is not a free parameter and depends on the mass. However, in general, the self-coupling and inter-field couplings for ultralight dark matter and studied without this constraint. 

The main implication of self-interactions is generally to provide another source of potential energy to ultralight dark matter solutions. Attractive self-interactions may make solitonic cores denser \cite{Painter2024, Desjacques2018,mocz2023, Dawoodbhoy2021} or reduce tidal stripping for example \cite{Glennon2022} and vice versa for repulsive self-interaction, see Section \ref{subsec:soliton:selfinteract}. Self-interactions may also alter the granular density structure. This can occur by additional interactions altering the velocity dispersion and thus de Broglie scale of granules \cite{may2025} or by altering the typical density contrast \cite{Capanelli2025}, see Section \ref{subsec:granules:selfInteract}.  This may provide additional compactness to ultralight dark matter solutions which may help avoid constraints which rely on ultralight dark matter reducing clustering below certain scales, such as Lyman-alpha constraints, see for example discussion in \cite{Painter2024, may2025, Arvanitaki_2015}, see Section \ref{subsec:structure:selfInteractions}. 

\subsubsection{High spin fields}

Typically, the ultralight dark matter field is assumed to be a spin-0 field. However, extensions to higher spin models have also been studied \cite{Amin2022, Amin2023, LopezSanchez2025, Jain2022, Jain2023, Jain2023b, Jain2023c, Amaral2024, Adshead2021, Schiappacasse2025}. The equations of motion for a spin-$s$ field are worked out in \cite{Jain2023} to be
\begin{align} \label{eqn:SP_spin}
    \partial_t \psi_j(x,t) &= -i \left( \frac{-\nabla^2}{2 m} + m V(x,t) + \frac{\lambda }{2m} |\psi_j|^2 \right. \\
    &+ \left. \frac{\alpha}{m^2} \sum_i \mathbf{S}\cdot \hat{\mathbf{S}}_{ij} + \frac{\xi}{2s+1} \sum_{ik} \hat{A}_{ji} \psi^\dagger_i \psi_k(x,t) \hat{A}_{ki} - i \sum_{ikl} g_{lk} [\hat{S}_l]_{ji} \nabla_k \right) \psi_j(x,t) \, . \nonumber \\ 
    \nabla^2 V &= 4 \pi G \sum_j |\psi_j|^2 \, .
\end{align}
Here we have set $\hbar \equiv 1$. $\psi_j$ represents the $j$th spin component of the field. The first line of equation \eqref{eqn:SP_spin} is the standard Schr\"odinger-Poisson equation with a self-interaction term. The first term in the second line describes the spin-self-coupling. $\mathbf{S} = \sum_{ij} \psi^\dagger_j \hat{\mathbf{S}}_{ji} \psi_i$ is the spin density and $\hat{S}_x = \hat{x} \cdot \hat{\mathbf{S}}$. $\alpha$ is describes the spin self-coupling. The second term in second line describes the two-body spin singlet interaction. The spin singlet matrix, $\hat A_{ij} = (-1)^{s-i} \delta_{i,-j}$, and has the following properties
\begin{align}
    &\hat A^{-1} = \hat A = \hat A^T \, , \\
    &\hat A \hat S_i \hat A = - \hat S^\dagger_i \, , \\
    &\psi^T \hat A \hat S_i \psi = 0 \, .
\end{align}
The final term in equation \eqref{eqn:SP_spin} accounts for additional allowed couplings. The $g_{ij} \propto \epsilon_{ij3}$ are real constants and are proportional to the Levi Civita symbol. 

If studied without the spin-coupled or self-interaction terms, as done in \cite{Amin2022}, then higher spin dark matter is equivalent to multifield dark matter with $2s + 1$ fields of mass $m$. Like multifield dark matter, the observational implication of this would be a reduction in the amplitude of overdensities \cite{Amin2022,LopezSanchez2025}, see Section \ref{subsec:granules:selfInteract}. Also similar to multifield dark matter the kinetic relaxation time of solitons is extended by a the number of degrees of freedom \cite{Jain2023c}, see Section \ref{subsec:pheno:multiAndSpin}. Unlike multifield models however, higher spin fields may also exhibit large scale polarization \cite{Jain2022, Amin2022} which can be present even when large scale polarization is absent in the initial conditions.

\subsection{Quantum and classical physics} \label{sec:ULDM:quantum}
Ultralight dark matter is generally treated as a classical field. That is to say that at every point in space we associate a complex number whose amplitude carries information about the density, and phase about the velocity. In a truly quantum description, we would instead promote the field to a quantum field, which instead places an operator at every point in space. The classical field approximation is useful for many analyses as it makes them much more tractable. However, the details and implications of this approximation have been the focus of a large body of work \cite{Eberhardt2021,Eberhardt2022,Eberhardt2022Q,Eberhardt_testing,Eberhardt2023,Kopp2021,Sikivie2009,sikivie2017, Hertzberg2016,Allali_2020,Allali2021,Allali2021b,Lentz2018,Lentz2019, Guth2015, Erken2012,Noumi2014,chakrabarty2021, Dvali2017,Dvali2018}. And unlike the non-relativistic, or minimally coupled, approximations discussed in this review, there remains much debate in the field relating to this approximation. 

Often this approximation is simplified to an argument based solely on the occupation number or on whether or not the field is described by a Bose-Einstein Condensate \cite{sikivie2017, Guth2015, Erken2012, Sikivie2009,Noumi2014}, with the latter seeming to provide arguments both in favor of and opposed to \cite{sikivie2017}, the classical field approximation \cite{Guth2015}. At this point, there exist many demonstrations that a large occupation alone is insufficient to guarantee the accuracy of the classical field approximation \cite{Guth2015, Allali_2020,Allali2021,Allali2021b}. 
A detailed derivation of the classical field theory from the quantum description can be found in Appendix \ref{app:ULDM:quantum}.

\section{Phenomenology} \label{sec:pheno}

The phenomenology of ultralight dark matter is very rich and covers a large range of scales and effects. In this section, we will summarize some of the most important ones. We will start with a discussion of the correspondence with cold dark matter on large scales and show the limits in which this model agrees with the predictions of $\Lambda$CDM. Following this we will discuss the linear ultralight dark matter effects. The impact of ``quantum pressure" on the transfer function can impact the halo mass function and structure growth. This supports the formation of cores in galactic density profiles which are persistent structures, called solitons, that random walk and oscillate. We also discuss the nonlinear dynamics of the field. This results in oscillating granular densities that can interact with stellar distributions, alter the path of light, or introduce quantum corrections. Finally, we discuss relativistic effects, covering both the Compton scale relativistic pressure and interactions with compact objects like black holes. Throughout this section, we will also try to cover some of the recent work exploring how model extensions can affect each phenomenon. 

\subsection{Cold dark matter correspondence} \label{subsec:cdmCorrespondence}

We know that the standard model of cosmology, $\Lambda$CDM is very successful on large scales. Any viable dark matter candidate must recover the predictions of $\Lambda$CDM in some limits. Historically, ultralight dark matter was introduced as a model that would agree with $\Lambda$CDM on large scales but which introduced a tunable scale (the de Broglie wavelength) that could alter predictions on the small scales. It was hoped that this model could explain the small-scale discrepancies between simulations and observations that existed at that time. In this subsection, we will discuss the different ways in which we can understand the CDM limit (i.e., the Vlasov-Poisson equations) of ultralight dark matter. 

We note here that the Schr\"odinger-Poisson equations were once studied as a numerical tool to approximate the Vlasov-Poisson system. The details of this endeavor are not particularly relevant to the study of ultralight dark matter as a model, but are interesting nonetheless. For more details on this subject see \cite{Eberhardt2020, Garny2018,Kopp2017,Davies1997,Widrow1993,Uhlemann2014, mocz2018}.

\begin{figure*}[!ht]
	\includegraphics[width = .97\textwidth]{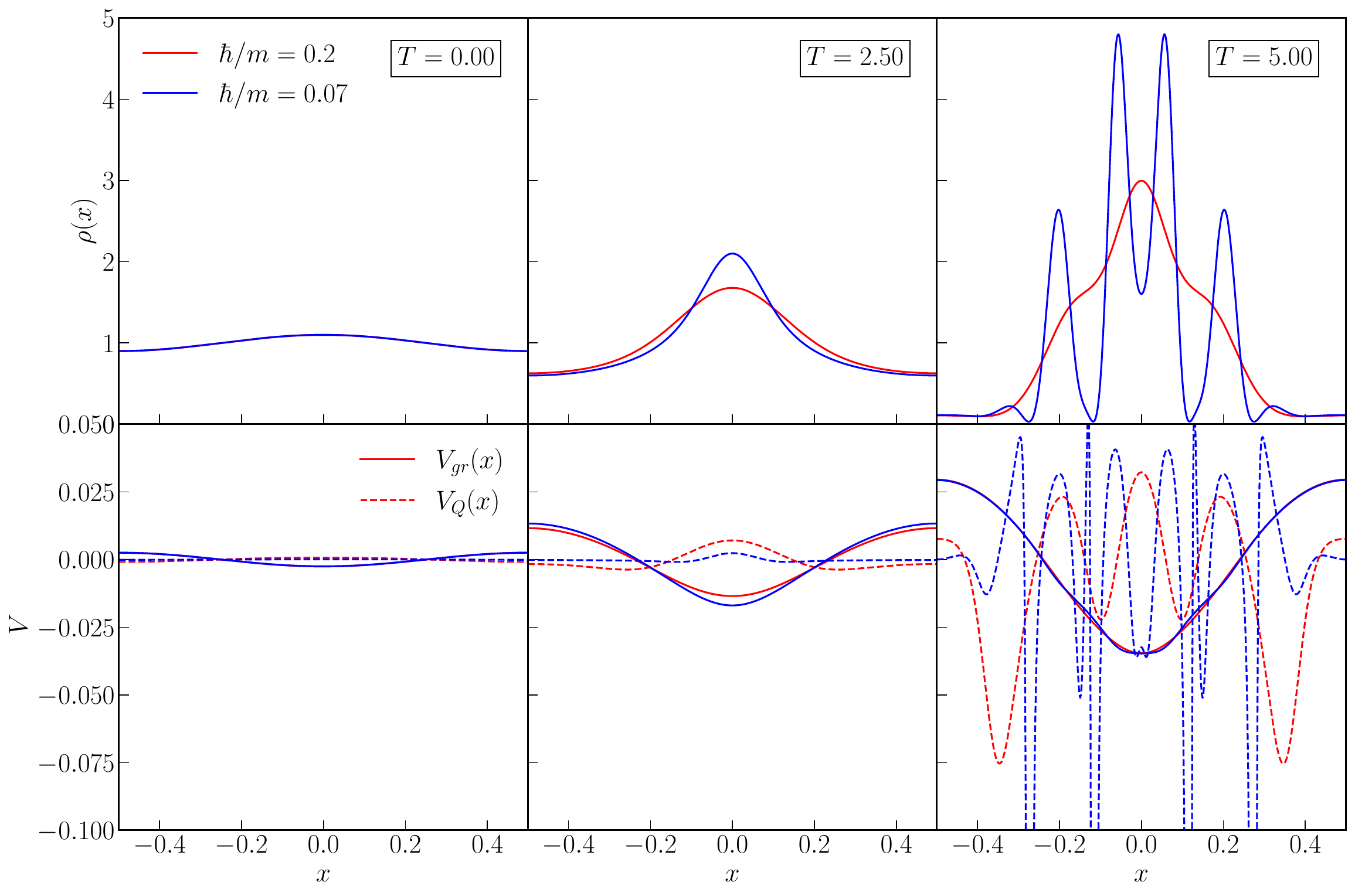}
	\caption{ Two simulations of the gravitational collapse of an initial overdensity in a single spatial dimension. Two values of $\hbar / m$ are plotted. Each column shows a different time, $T$. Quantities are presented in simulation units. \textbf{Top row:} shows the spatial density. \textbf{Bottom row:} compares the gravitational and ``quantum" potential terms, $V_{gr}$ and $V_Q$, respectively. Larger values of $\hbar / m$ correspond to larger ``quantum" pressure effects which provide a resistance against collapse. }
	\label{fig:quantumPressure}
\end{figure*}

One of the most straightforward ways to understand how ultralight and cold dark matter dynamics differ is the ``quantum pressure". The ``quantum pressure" describes the effect of the Heisenberg uncertainty relation on the dynamics of classical fields. We can also think of this term as arising from having a phase space that is multi-valued in velocity as a function of position. However, the form of the term in the equations of motion is simplest to see in the Hamilton-Jacobi (or Madelung) representation of the Schr\"odinger-Poisson equations \cite{Madelung1927}. In this formulation, we start by splitting the complex field, $\psi$, into an amplitude and complex phase as 
\begin{align} \label{eqn:psiComplex}
    \psi(x,t) = A(x,t) e^{iS(x,t) / \hbar} \, . 
\end{align}
Then the Hamilton-Jacobi equation is written by plugging equation \eqref{eqn:psiComplex} into equation \eqref{eqn:SP_eqn}. This gives us
\begin{align}
    \partial_t A(x,t) &= \frac{-1}{2m} \left( A(x,t) \, \nabla^2 S(x,t) + 2 \nabla A(x,t) \cdot \nabla S(x,t) \right) \, , \\
    -A(x,t) \, \partial_t S(x,t) &= \frac{-\hbar^2}{2 m} \nabla^2 A(x,t) + \frac{1}{2m} A(x,t) (\nabla S(x,t))^2 + V(x,t) \, A(x,t) \, . 
\end{align}
If we now identify the spatial density with the square amplitude, i.e.  $\rho(x,t) = |A(x,t)|^2$, the velocity as $v(x,t) = \frac{1}{m} \nabla S(x,t)$, then we can rewrite the above equations as 
\begin{align} \label{eqn:Madelung}
    \partial_t \rho + \nabla \cdot (\rho \, v) = 0 \, ,\\
    \partial_t v + v \cdot \nabla v + \nabla V - \frac{\hbar^2}{2 m^2} \nabla \frac{\nabla^2 \sqrt{\rho}}{\sqrt{\rho}} = 0 \, . \nonumber 
\end{align}
Which are the fluid equations for a collisionless fluid with an additional term, which we identify as the ``quantum pressure". We sometimes will write the quantum pressure as an additional potential term as
\begin{align}
    V_Q = \frac{-\hbar^2}{2m^2} \frac{\nabla^2 \sqrt{\rho}}{\sqrt{\rho}} \, .
\end{align}
We note that this potential is not well defined as $A/\sqrt{\bar \rho} \rightarrow 0$. This may appear problematic, but it is perhaps unsurprising as the phase of the field is also not well defined in this same limit. We can see that its inclusion in equation \eqref{eqn:Madelung} is the only term that distinguishes those equations from the standard collisionless fluid equations. It is then perhaps intuitive that this term should be proportional to $(\hbar/m)^2$ as we expect this to weight our terms which introduce wave-like phenomena to otherwise corpuscular-like evolution. We recall that in the limit that $\hbar/m$ is small compared to the dynamical scales in the system, we must recover the dynamics of cold dark matter in order for ultralight dark matter to agree with large-scale observations. 

The quantum potential term is compared with the gravitational potential term for two simulations of the gravitational collapse of an initial spatial over-density in Figure \ref{fig:quantumPressure}. We can see that the quantum pressure delays the collapse of the overdensity and resists the growth of structure below certain scales. 

It is often pointed out that the naming of the ``quantum pressure" is improper, and so we will also note the reason here. First, the term arises completely within the classical field theory. And while it reflects the existence of an uncertainty relation, we note that any two variables related by the Fourier transformation would have a similar uncertainty relation. Second, the term is not the result of a pressure but instead a stress tensor, $\Sigma_{ij} = \frac{\rho}{2} \partial_i \partial_j \ln \rho$, i.e.,
\begin{align}
    \nabla \frac{\nabla^2 \sqrt{\rho}}{\sqrt{\rho}} = \rho^{-1} \partial_i \frac{\rho}{2} \partial_i \partial_j \ln \rho \, .
\end{align}
Work exploring using the Madelung formulation as an ultralight dark matter simulation method includes \cite{Li_2019, Uhlemann2014, Garny2018, mocz2015, Veltmaat2016, Haas2002, Kopp2017}. Though this method admits numerical difficulty when the field goes to $0$, which happens any time the velocity is multi-valued in phase space, such as shell crossing. When this occurs, the last term in the equation above is not well defined. 

The CDM-ULDM correspondence has also been studied in terms of phase space descriptions. Two additional methods of parameterizing the difference between CDM and ULDM are explored in Appendix \ref{app:cdm_correspond}.

\subsection{Structure growth} \label{subsec:transferFunction}

\subsubsection{Suppression of small scales}

\begin{figure*}[!ht]
	\includegraphics[width = .70\textwidth]{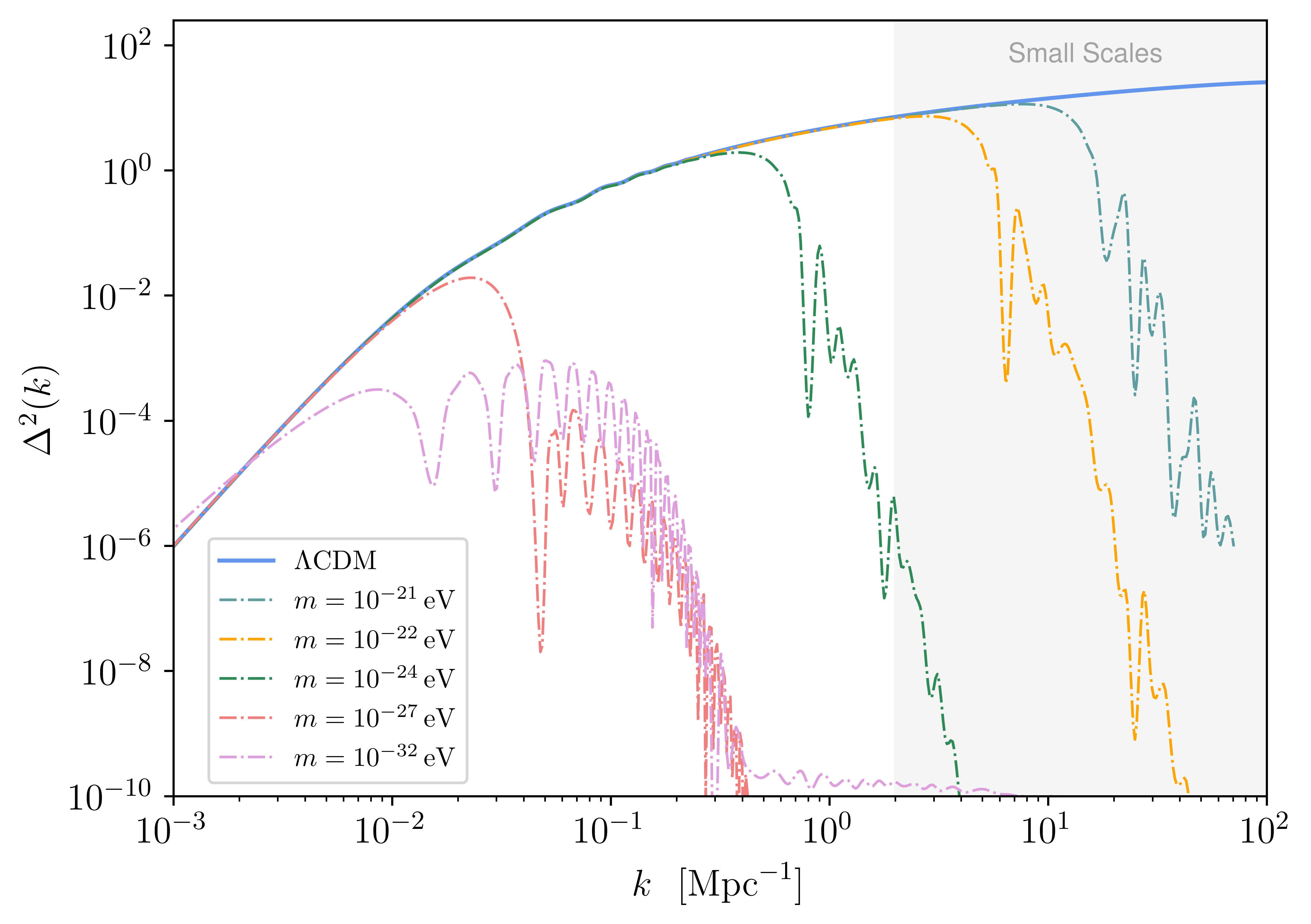}
	\caption{Dimensionless matter power spectrum $\Delta^2(k)$ for FDM models with different masses, compared with the $\Lambda$CDM prediction, where $\Delta^2(k) \equiv \frac{k^3}{2\pi^2} P(k)$ and $P(k)$ is the matter power spectrum. Adapted from Ref.~\cite{Ferreira2021}.}
	\label{fig:transferFunc}
\end{figure*}
Like warm and dark matter with standard model interactions, ultralight dark matter suppresses the linear power spectrum compared to cold dark matter; comparison is plotted in Figure \ref{fig:transferFunc}. However, the specific shape and cause of this suppression are unique to the ultralight dark matter case. The quantum pressure resists the collapse of small scales in the universe. This results in a different linear power spectrum of density fluctuations. In the cold dark matter case, Jean's scale analysis would tell us the rate at which modes grow, i.e., the mode $k$ grows according to $e^{\gamma t}$ where $\gamma^2 = 4 \pi G \delta \rho_k$, where $\delta \rho_k$ is the overdensity associated with that spatial mode. $1/\gamma$ is the normal dynamical time associated with cold gravitational collapse. The effect of field fluctuations on the Jeans scale was worked out in the original ultralight dark matter paper \cite{Hu2000}. We modify the normal dynamical time by the energy of a free field ($E = k^2/2m$), giving $\gamma^2 = 4 \pi G \delta \rho_k + (k^2/2m)^2$. This results in the modified Jean's scale, $r_J$, \cite{Hu2000, Khlopov1985}
\begin{align}
    r_J &= \pi^{3/4} (G \delta \rho)^{-1/4} m^{-1/2} \\ 
    &\approx 50 \left( \frac{m}{10^{-22} \, \mathrm{eV}} \right)^{-1/2} \left( \frac{\rho}{\rho_b} \right)^{-1/4} \left( \Omega_m h^2 \right)^{-1/4} \,,
\end{align}
where $\rho_b = 2.8 \times 10^{2} \Omega_m h^2 \mathrm{M_\odot kpc^3}$ is the background density, $\Omega_m$ the matter density, and $h = H/100 \, \mathrm{km/s/Mpc^{-1}}$ the Hubble constant. On scales below $r_J$ the uncertainty relation, i.e., the ``quantum pressure," prevents gravitational collapse. 

This finite and scale-dependent Jean's scale impacts the linear density power spectrum causing a suppression in the transfer function and power spectrum. This suppression is related to the Jeans length, which is determined by the mass of the ultralight dark matter. The ULDM mass determines where this cutoff scale is located in the power spectrum, see Figure \ref{fig:transferFunc}.
This can be calculated analytically using the modified Boltzmann codes \href{https://github.com/dgrin1/axionCAMB}{axionCAMB}~\cite{Hlozek2015}, or its successor \href{https://github.com/Ra-yne/AxiECAMB}{axionECAMB}~\cite{Liu:2024yne}, or \href{https://github.com/PoulinV/AxiCLASS}{AxiCLASS}~\cite{Poulin:2018dzj,Smith:2019ihp}. One can also parametrize this FDM suppression by the analytical approximation~\cite{Hu2000, Passaglia_2022}
\begin{align}
    P_\mathrm{ULDM}(k) = T_\mathrm{ULDM}^2(k) P_\mathrm{CDM}(k) \,, \qquad T_\mathrm{ULDM}^2(k) \approx \frac{\sin(x^{5/2})}{x^{5/2}(1 + B x^{7/2})} \,,
\end{align}
where $x = A \, k/k_J$, $k_J = 9 m_{22}^{1/2}$, $A = 2.22 m_{22}^{1/25 - 1/1000 \ln m_{22}}$, $B = 0.16 m_{22}^{-1/20}$. 
It is common notation that $m = m_{22} \times 10^{-22} \, \mathrm{eV}$.

This suppression of the power spectrum, together with the evolution of this ultra-light field, was studied in detail in~\cite{Cookmeyer2020, Hlozek2015,Hlozek2017, Rogers:2020ltq,Rogers2023}, where they used observations to constrain the mass and fraction of this ultra-light component (we will describe these in Sec.~\ref{sec:constrain_extremely_light}). 

A consequence of this suppression is a reduction in the amount of structure at small scales.  This has been studied in most depth by looking at the abundance of small-scale satellites and the suppression of small-scale galaxies in halo mass functions in simulations \cite{Nadler_2025,Nadler2021}. 

Additionally, the impact of ULDM cosmologies on halo properties has been studied in \cite{Dome2023} with filaments, walls, and voids studied in \cite{Dome2023b}. They found that ULDM halos had lower concentrations, were more elongated around the virial radius, and had stronger intrinsic alignment correlations. 

\subsubsection{Halo mass function}

The suppression of small scales results in a reduction of halos below a certain halo mass. It is for this reason that ultralight dark matter is often discussed in the context of the missing satellites problem. It was suggested as an explanation for the apparent lack of satellites \cite{Du_2016}, but has since been found in tension with the existing satellite abundance \cite{Nadler_2025, Nadler2021}. This cutoff depends both on the fraction of dark matter in an ultralight field and on the mass of the ultralight field, with lower mass resulting in halo mass cutoffs at larger halo mass \cite{Du_2016,Kulkarni_2021, May2022}. The deviation from the CDM halo mass function was estimated using the analytical fitting formula as a function of halo mass, $M_h$, and redshift $z$, given as 

\begin{align}
    \left. \frac{dn}{dM_h} \right|_{\psi} (M_h, z) = \left. \frac{dn}{dM_h} \right|_{CDM} (M_h, z) \left[ 1 + \left( \frac{M_h}{M_0} \right)^{-1.1} \right]^{-2.2} \, . 
\end{align}

Which was originally fit to N-body simulations with an altered power spectrum \cite{Schive2016}. The characteristic halo scale $M_0 = 1.6 \times 10^{10} m_{22}^{-4/3} \, \mathrm{M_\odot}$. Later, fully Schr\"odinger-Poisson simulations corroborated this result \cite{May2022}. This has been studied both numerically \cite{May2022, Schive2016, Stucker_2021} and analytically \cite{Du_2016,Kulkarni_2021}. The latter is typically done using the Press-Schechter formalism \cite{PressSchechter1974}. Though the details of this calculation can produce results that differ quite substantially from one another \cite{Kulkarni_2021} and calculations have not been corroborated by full nonlinear simulations \cite{May2022, elgamal2023}.

\subsubsection{Structure growth in multifield scenarios}

Cosmological simulations of structure growth in multifield settings have also been studied \cite{luu2024,Lague2024, Tellez-Tovar2021}. This includes both the effect on the density power spectrum \cite{Lague2024} and halo mass function \cite{luu2024}. An interesting and nontrivial result of this work is that the halos that formed in these simulations could be dominated by one or the other species of dark matter particle as opposed to simply maintaining the fraction of each species put in at the initial conditions. The generality of this result has not been studied, but it may complicate the ability to apply observations of individual galaxies or dwarf galaxies to mixed dark matter models. For example, \cite{pozo2024} demonstrated that the current dwarf galaxy halo diversity could be fit by two different mass species of ultralight dark matter.

\subsubsection{Structure growth with self interactions} \label{subsec:structure:selfInteractions}

A lot of the literature surrounding self-interacting ultralight dark matter is concerned with the affect the interactions have on late time structures. However, self-interactions have also been studied in terms of linear structure growth \cite{may2025, Arvanitaki_2015}. The linear power spectrum for overdensities $\delta_k = \delta \rho_k / \bar \rho$ is given \cite{Arvanitaki_2015}

\begin{align}
     \ddot \delta_k + 2 H \,  \dot \delta_k \approx \left( 4\pi G \bar \rho - \frac{k^2}{4 m^2 a^4} - \frac{3 \lambda \bar \rho k^2}{8 m^4 a^2} \right) \delta_k \, .
\end{align}

For comoving wavenumber $k$, where $\bar \rho$ is the average matter density, $a$ the scale factor, $\lambda$ the quartic self coupling, and $m$ the field mass. 

\subsubsection{Dependence on production mechanism}

The misalignment mechanism is one of the most studied production mechanisms for ultralight dark matter candidates. The details of this production method has been demonstrated to have important potential implications for structure. For example, the scenario in which in the initial misalignment angle of the field is large has been shown to produce power spectrum enhancements via a parametric resonance. This effect would actually result in more dense and numerous small halos than in $\Lambda$ CDM \cite{Arvanitaki_2020}. 

Additionally, whether or not the misalignment occurs after or before inflation can also effect structure. When misalignment occurs before inflation the misalignment angle in a single patch is relevant to the production of dark matter. However, if misalignment occurs after inflation then the power spectrum can have both a free-steaming cutoff and white-noise portion at sufficiently small scales \cite{amin2024lowerbounddarkmatter}.

\subsection{Solitons} \label{subsec:Pheno:solitons}

\subsubsection{Summary}

\begin{figure*}[!ht]
	\includegraphics[width = .94\textwidth]{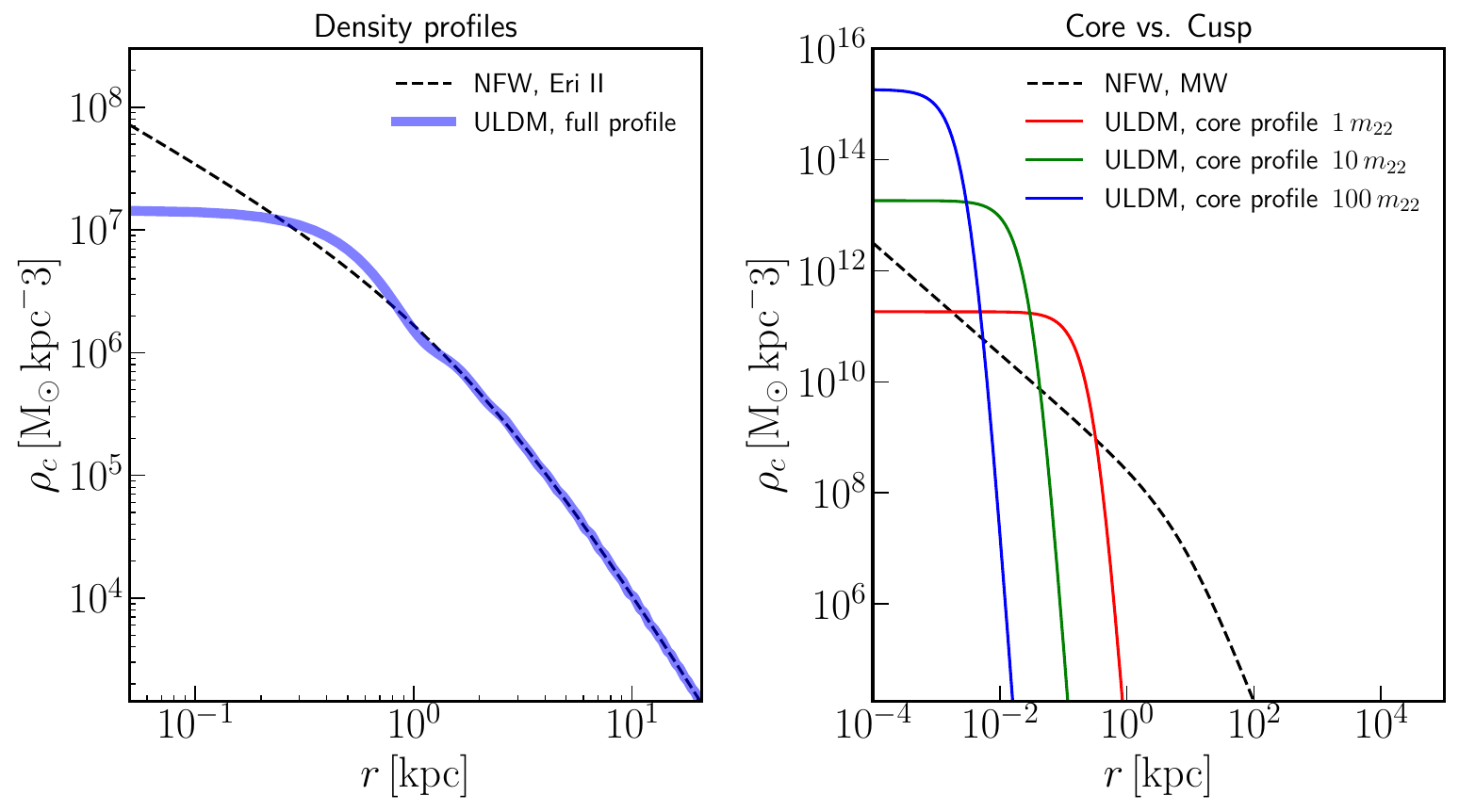}
	\caption{ \textbf{Left.} A cored density profile associated with ultralight dark matter, blue, compared with the NFW profile of a dwarf galaxy (black). We can see the core density profile, which then matches the NFW at large scales. Comparisons of the predicted and observed density profiles of dwarf galaxies are often used to constrain ultralight dark matter \cite{Bar2018, Bar2019, Bar2022, Safarzadeh2020, zimmermann2025}. \textbf{Right.} Comparison of NFW profile cusp (dotted black line) and predicted ultralight dark matter cores for different ultralight dark matter masses for Milky Way parameters. Milky Way parameters taken from \cite{Lin2019}. Core parameters come from cosmological simulations \cite{Schive2014_CoreHalo} and shape from equation \eqref{eqn:solitonicCore}. We can see that lower masses correspond to larger and less dense central cores. The presence of cores in ultralight dark matter was among the original motivations for the model \cite{Hu2000}. Note that the core densities are approximately flat within the core radius. Solitons are discussed in Section \ref{subsec:Pheno:solitons}.}
	\label{fig:core_vs_cusp}
\end{figure*}

``Solitons" are the persistent cores of galactic dark matter halos that form in ultralight dark matter scenarios. They are a consequence of the finite Jean's length of FDM on non-linear scales. Solitons have been found at the core of all ultralight dark matter halos \cite{Schive2014, Mocz2017, Manita2024}. In these systems, the core is mostly comprised of the ground state of the gravitational potential of the dark matter halo. The soliton has a breathing and random walking behavior given interaction with the higher energy eigenmodes. The shape of the core can be found by solving the eigenvalue problem of time and angular averaged gravitational potential, $V(r)$, of the dark matter halo, i.e.
\begin{align}
    &H_l \phi_{il} = E_{il} \phi_{il} \, , \\
    &H_l = \frac{- \hbar^2 \nabla_r^2}{2m} + m V(r) + \frac{\hbar^2}{2m} \frac{(l + l)l}{r^2} \, , \\
    &\nabla^2 V(x) = 4 \pi G |\psi(x)|^2 \,,
\end{align}
where $i$ and $l$ are the energy and angular momentum quantum numbers, respectively. $\phi$ is the corresponding eigenmode and $E$ the eigenenergy. The core profile of the core is given by $\phi_{00}$. Simulations of the core have found the following density profile \cite{Schive2014_CoreHalo} 
\begin{align}
    \rho_c(x) = \frac{1.9 a^{-1} (m / 10^{-23} \, \mathrm{eV})^{-2} (x_c/\, \mathrm{kpc})^{-4}}{[1 + 9.1 \times 10^{-2} (x/x_c)^2]^8} \, \mathrm{M_\odot \, pc^{-3} } \,,
    \label{eq:soliton_density}
\end{align}
where $x_c$ is the core radius in co-moving coordinates. For the simulation in this work~\cite{Schive2014_CoreHalo}, a fitting function relating the physical radius $r_c = a x_c$ and core mass $M_c$ was found: 
\begin{align} \label{eqn:solitonicCore}
    r_c \approx 1.6 m_{22}^{-1} a^{1/2} \left( \frac{\zeta(z)}{\zeta(0)} \right)^{-1/6} \left( \frac{M_h}{10^9 \, M_\odot} \right)^{-1/3} \, \mathrm{kpc} \, , \\
    M_c \approx \frac{1}{4} a^{-1/2} \left( \frac{\zeta(z)}{\zeta(0)} \right)^{1/6} \left( \frac{M_h}{M_\mathrm{min,0}} \right)^{1/3} M_\mathrm{min,0} \,,
\end{align}
where $M_h$ is the halo mass, $M_\mathrm{min,0} \sim 4.4 \times 10^7 m_{22}^{-3/2} M_\odot$, and $\zeta$ is a slowly varying parameter.
The problem is generally treated numerically by studying the properties of cores that form in simulated halos \cite{Zagorac2023, Schive2014_CoreHalo, Schive2014}. The specific soliton scaling relations above are actually subject to some debate with cosmological simulations \cite{Schive2014, Schive2014_CoreHalo} generally finding different results from halos constructed by soliton collisions \cite{Zagorac2023, Mocz2017,Chen2017}, usually depending on how the simulations were made. This discrepancy was investigated in Ref. \cite{Blum_2025} which shows the relation in Ref. \cite{Schive2014} is more commonly achieved and the scaling \cite{Mocz2017} is an upper bound. It was shown in Ref. \cite{Chan:2021bja} using the largest Schrodinger-Poisson simulations to date, which produce many galaxies with different formation histories, that the relation between $M_h \times r_c$ has a large dispersion for different galaxies. This is also compared to simulations using soliton collisions. Given that, obtaining an exact $M_h \times r_c$ fitting function from simulations might not be a good description of this relation that has a distribution depending on the galaxy, its history, and its environment. This is confirmed in Ref. \cite{Manita2024} where they claim that the scatter in the core-halo relation cannot be solely explained by the concentration mass relation of the halo. 

The core profile, however, is confirmed by many simulations (including the ones cited above) to agree with (\ref{eq:soliton_density}). Therefore, the profile for FDM has a universal shape given by
\begin{align}
   \rho (r) = \begin{cases}
        \rho_c (r) & \text{for } r < r_t\,, \\
        \rho_{\mathrm{NFW}} (r) & \text{for } r > r_t \,,
    \end{cases}
    \label{eq:full_FDM_profile}
\end{align}
where $r_t$ is the transition radius. The halo profile of ULDM then is expected to have a core in the center and be given by an NFW profile in the outskirts of the galaxy, as expected from observations, as we can see in Figure~\ref{fig:core_vs_cusp}. This profile has to be continuous in $r_t$. It was also shown in Ref. \cite{Chan:2021bja} that this profile has to be smooth (the first derivative is also continuous) to describe realistic profiles. In this work, it was also shown that although this profile shape is universal, the transition is longer or shorter depending on the different galaxies and their formation histories. The different properties of the galaxies and of FDM were shown to influence $r_t$ in Ref. \cite{Manita2024}.

It has also been demonstrated that the soliton profile is sensitive to the presence of baryons \cite{Veltmaat2019}. Comparison of core vs cusp profiles for a Milky Way-size galaxy is given in Figure \ref{fig:core_vs_cusp}. 

The rate at which mass condenses in its ground state has also been the focus of a large body of work \cite{Levkov2018}. The general consensus is that solitons do form quickly enough for them to reasonably exist in any dark matter halo. The condensation time, $\tau_s$, was estimated (in the kinetic regime \cite{Levkov2018}) to be
\begin{align}
    \tau_s \sim \frac{\sqrt{2}}{12 \pi^3} \frac{m^3 \sigma^6}{ G^2 \rho^2 \Lambda} \, .
\end{align}
Where $\Lambda$ is the Coulomb logarithm. For typical dwarf galaxy parameters, this timescale is \cite{Levkov2018}
\begin{align}
    \tau_s \sim \left( \frac{m}{10^{-22} \, \mathrm{eV}} \right)^3 \left( \frac{\sigma}{30 \, \mathrm{km/s}} \right)^6 \left( \frac{0.1 M_\odot}{\rho} \right)^2 \, \mathrm{Myr} \, . 
\end{align}

It has been argued that the presence of solitonic cores would produce a core in galaxies observable through stellar velocity curves \cite{Bar2018, Bar2019, Bar2022, Hayashi2021}. Similarly, it was argued that the oscillation of solitonic cores would destabilize certain stellar orbits \cite{Marsh2019}. The presence of solitonic cores in all halos also has implications for the diversity of galaxy rotation curves. Observations of dwarf galaxies indicate a wide range of inner density profile amplitudes and slopes, including systems that appear consistent with cuspy profiles. In contrast, vanilla ultralight dark matter generically predicts solitonic cores at the center of all halos, with properties tied to the halo mass through the core–halo relation. This can lead to tension with the observed diversity of rotation curves, and has been used to place constraints on the ultralight dark matter mass (see e.g.~\cite{Chan:2021bja,Hayashi2021,Wardana:2026qag}).

There is a debate in the literature about whether a Bose-Einstein condensate is formed in these solitonic cores. This will be discussed in Sec.~\ref{sec:ULDM:quantum} and is also discussed in Ref. \cite{Ferreira2021} and references therein.

The ability of ultralight dark matter to produce galactic cores was one of its original motivations. Some works have argued for a potential ``catch-22'', namely that the masses required to reproduce the observed cores would also suppress the formation of the corresponding host halos. However, current constraints on the ultralight dark matter mass allow for a broad range of values in which cores can form without significantly suppressing halo formation. In particular, analyses of dwarf galaxy kinematics show that viable mass ranges remain compatible with both the presence of cores and the abundance of their host halos \cite{Hayashi2021,Wardana:2026qag}. Therefore, while soliton-based arguments provide important constraints, the existence of a generic ``catch-22'' is not supported once current bounds are taken into account.

\subsubsection{Tidal disruption and tunneling}

Given the amount of information we gain about ultralight dark matter from satellite galaxies, the details of tidal disruption are extremely important. It has been demonstrated that tidal disruption can be quite different in ultralight dark matter scenarios \cite{Du2018}. It was demonstrated that the tidal stripping can runaway. Physically, this is due to the fact that as mass is removed from the halo, the characteristic size of the soliton increases, leading to increased tidal stripping.

\subsubsection{Multifield and mixed dark matter and higher spin solitons} \label{subsec:pheno:multiAndSpin}

Solitons have also been studied in the multiple ultralight field case \cite{Street2022, luu2024, Mirasola2024, Huang2023, Dome2024, Schwabe2020,Guo2021}. These have been studied numerically in fixed resolution simulations \cite{Luu2023}, expanding universe simulations \cite{Huang2023}, and with mixed analytic and 1D simulations in Ref. \cite{Mirasola2024}. The simulations were limited to two-field systems. We will describe the fields in terms of their masses, $m$, and mass fractions $\beta$. 

We may have naively expected that in the two-field case, the resulting solitons are simply a superposition of each soliton we would have expected in the one-field case. However, all the numerical results seem to indicate that this is not the case. A scheme for finding the ground state of the two-field system was presented in Ref. \cite{Luu2023}. However, it was demonstrated that it is not necessarily the case that the system will find this ground state \cite{Huang2023, Luu2023}. Simulation results comparing different $\beta$ parameters for fields of mass $m_1 = 0.8 m_{22}$ and $m_2 = 1.6 m_{22}$ is summarized in Ref. \cite{Luu2023}, they find 
\begin{itemize}
    \item $\beta_1 \gg \beta_2:$ When the light field dominates, the profile of the light field is consistent with both the two and one field ground solutions. However, the oscillations in the light field prevent the formation of the heavy field soliton, and as a result, the expected heavy field soliton fails to form. 
    \item $\beta_1 \sim \beta_2:$ When the abundance of each field is comparable, each field's soliton is well described by the ground state solution of the two-field system (a solution which is distinct from a superposition of the one field solutions). 
    \item $\beta_1 \ll \beta_2:$ When the heavy field dominates, the soliton of both fields is well described by the two-field ground state. Likewise, the soliton for the two-field solution is also well described by the one-field solution. 
\end{itemize}
The solitons in multifield systems are therefore more complicated than the addition of multiple one-field solutions. The oscillations of a light soliton may provide a core potential with time variation such that the formation of the heavy field soliton is confounded. Likewise, even when both solitons form, the altered potential due to the existence of each soliton will mean that the two-field ground state solution is necessary to describe the shape of each soliton, unless one field is sufficiently dominant that the potential approaches the one-field result. Several works have demonstrated that even the extension to two fields alone can provide a variety of halo cores \cite{Guo2021,Luu2023,Mirasola2024}.

\subsubsection{Self-interaction} \label{subsec:soliton:selfinteract}
The interactions between solitons with inter-field interactions can also lead to interesting modifications to the ground state (soliton) profile \cite{Mirasola2024,Guo2021, Dave2023}. For example, the profile of the ground state separates into ``phases" depending on the relative amount of energy in ``quantum pressure" energy, gravitational potential energy, and self-interaction potential energy. If the self-interaction was weak, the ground states simply superimposed. However, as the (repulsive) self-interaction strength is increased, the core of one of the nested solitons can become underdense, creating a ``hollow" core. As the self-interaction strength is increased further, the two solitons can separate, forming two spatially distinct solitons. 

In full halo simulations, self-interactions can alter the size of the central soliton, with repulsive interactions causing it to expand \cite{Indjin2025} and vice versa. This can also alter the susceptibility to tidal stripping. With attractive self-interactions tending to reduce tidal stripping \cite{Glennon2021}.

\subsection{Granules and interference in density patterns} \label{subsec:granuleDensityPatterns}

\begin{figure*}[!ht]
	\includegraphics[width = .94\textwidth]{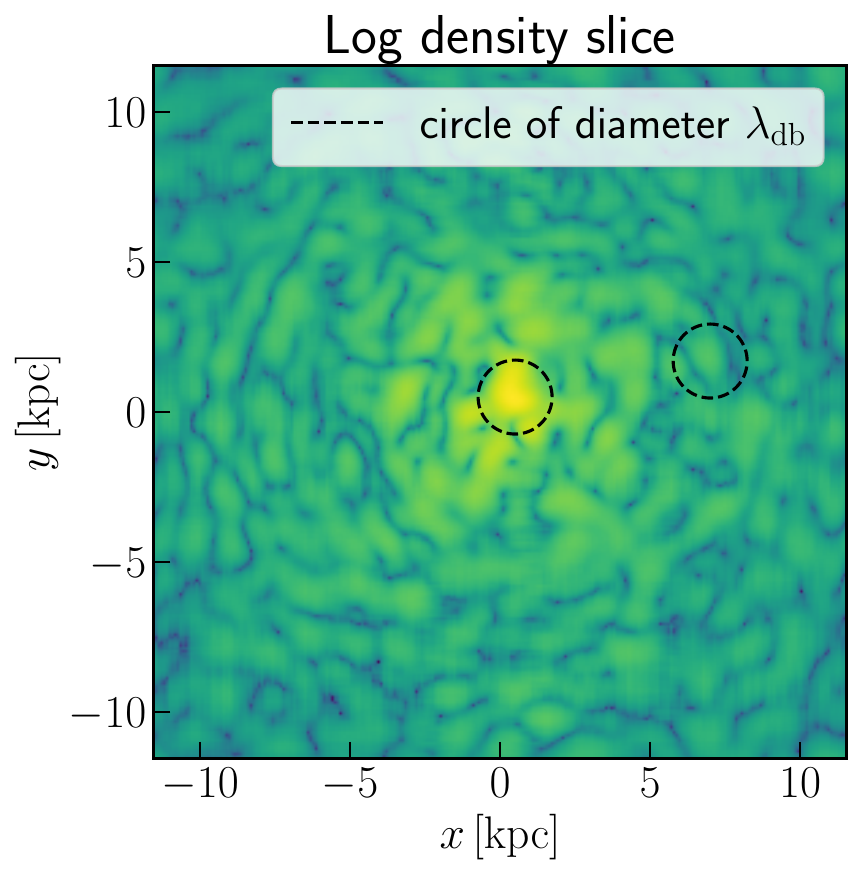}
	\caption{ A density slice through an ultralight dark matter halo. The log density is plotted. The solitonic core is visible in the center of the halo and density granules throughout the skirt. A circle with a diameter equal to the de Broglie wavelength is shown in dotted black for comparison. We can see that this size roughly describes both the central soliton and the density granules. The halo parameters are for Eri II \cite{Li2017,Crnojevic2016}. }
	\label{fig:labeled_density}
\end{figure*}
\subsubsection{Summary}

One of the most prominent features of ultralight dark matter is the granular density pattern resulting from the interference of different velocity modes. This has been demonstrated to be present in any ultralight dark matter systems with a velocity dispersion \cite{Schive2014, Veltmaat2016,Mocz2017,Du2018,Li_2019, Edwards2018, Nori2018,Veltmaat2019,Mocz2019, Liu2022}, e.g., halos. It has also been studied in the context of filaments \cite{Zimmermann2024_filaments}. This effect is often described as ``quantum" in the literature because the granules are on the de Broglie scale, \cite{Hu2000}. However, it is entirely contained within the classical field theory, and in fact, large quantum corrections to the classical theory have been shown to reduce the amplitude of granular fluctuations in the density \cite{Eberhardt2023}. 

These fluctuations come from the interference between different ultralight dark matter modes. If we decompose the field into a superposition of plane waves, then we will find that it is the velocity differences that characterize the interference. It is instructive to look at the simplest example, a field which is a superposition of two free plane-waves in a single spatial dimension, as follows
\begin{align}
    \psi(x) = \sqrt{\frac{\rho_0}{2} } \left(e^{-i m\,v_1\,x/\hbar -im \, v_1^2 \, t /2\hbar} + e^{-im\,v_2\,x/\hbar -i \, m \, v_2^2 t /2\hbar} \right) \, .
\end{align}
For two velocities $v_1$ and $v_2$, field of mass $m$, average density $\rho_0$. We can rewrite the field then as
\begin{align}
    \psi(x) = \sqrt{\frac{\rho_0}{2} } \, e^{-im \, \bar v \, x /\hbar - im \bar v^2 t / \hbar } \left( e^{-im\, \Delta v \,x/ \hbar - i m (\Delta v^2 - 2 \bar v \Delta v) t / 2 \hbar} + e^{im\,\Delta v\,x/2 \hbar - i m (\Delta v^2 + 2 \bar v \Delta v) t / 2 \hbar} \right) \, .
\end{align}
Where $\bar v = \frac{v_1 + v_2}{2}$ and $\Delta v = \frac{v_1 - v_2}{2}$ are the average of and difference between the velocities, respectively. Recall that $\rho(x) = |\psi(x)|^2$, the density for this field is then 
\begin{align}
    \rho(x) = \rho_0 \left(1 + \cos(2 m \Delta v x / \hbar) \, \cos(m \Delta v^2 t / \hbar) \right) \, .
\end{align}
From this, we see a few things. First of all, the mean density is $\rho_0$ as expected. Secondly, the average velocity is not relevant to the density, this is good since we would like our system to be Galilean invariant. Next, we can see the density oscillate between twice its average value and $0$ with a wavelength described by the velocity difference $\lambda = \pi \hbar / m \Delta v$ and that these interference patterns oscillated on a timescale equal to the approximate crossing time of a granule $ \tau = \lambda_\mathrm{db} / \Delta v = 2 \pi \hbar / m \Delta v^2$. 

For systems with velocity dispersions, $\sigma$, we usually write the de Broglie length and time as
\begin{align}
    &\lambda_\mathrm{db} = 2 \pi \hbar / m \sigma \sim 0.6 \left( \frac{10^{-22} \, \mathrm{eV}}{m} \right) \left( \frac{200 \, \mathrm{km/s}}{\sigma} \right) \, \mathrm{kpc} \, , \\
    &\tau_\mathrm{db} = 2 \pi \hbar / m \sigma^2 \sim 3 \left( \frac{10^{-22} \, \mathrm{eV}}{m} \right) \left( \frac{200 \, \mathrm{km/s}}{\sigma} \right)^{2} \, \mathrm{Myr} \, .
\end{align}
We note that factors of $2$ and $\pi$ differ in the literature, but that these are arbitrary, as long as the results are consistent when applied to observables. 

\subsubsection{Modeling granules}

Ultralight dark matter density granules alone are associated with a wide range of observational constraints and phenomena. This has motivated a large body of study, particularly in recent years \cite{Church2019, Marsh2019, Powell2023, dalal2022, teodori2025, Eberhardt2024, Kim2024,Kim2024_astro,dror2024,Chowdhury2023, Eberhardt2023, Gosenca2023, Amin2022, eberhardt2025}. These are studied using a variety of methods, which we summarize here.

\textbf{Gaussian random field and power spectra.}
For a superposition of plane waves with a Gaussian distribution, i.e. 
\begin{align}
    \psi(x) = \int dv^3 \, e^{-(\vec v/\sqrt2\sigma)^2} \, e^{-i m  \vec v \vec x / \hbar} \, ,
\end{align}
the over densities, $\delta \rho = \rho -\braket{\rho}$, are well described by a Gaussian random field, where $\braket{\rho}$ is the average value of the density. In $k$ space these fluctuations are given by the Maxwell-Boltzmann energy distribution \cite{Chan2020}, i.e. 
\begin{align}
    \rho(k) = |\psi(k)|^2 \propto\exp[-(\hbar k /\sqrt{2} m \sigma)^2] \, .
\end{align}
This has been shown to describe the fluctuations around the mean density in collapsed objects such as halos as well as long as one is careful to use the local velocity dispersion and not look at length scales on which the halo potential changes are large (if we restrict ourselves to length scales for which the change in the gravitational potential $\delta V$ is small compared to the potential $V$, i.e. $\delta V / V \ll 1$) \cite{Chan2020}. The Gaussian random field approximation has been applied to the study of strong \cite{Chan2020, Powell2023} and stochastic \cite{eberhardt2025} lensing.

The power spectrum describing these fluctuations goes as 
\begin{align}
    P^\rho (k) \propto e^{-\hbar^2 k^2 / 2\sqrt{2} \sigma^2 m^2} \, . \label{eqn:rhoPS}
\end{align}
The effect of these modes has been studied in the context of dynamical heating of stellar distributions \cite{Church2019, Marsh2019} as well as pulsar Doppler shifts \cite{Kim2024} and stellar astrometry \cite{dror2024, Kim2024_astro,Furlanetto2024}. 

\textbf{Eigenvalue approximations.}
We showed above how fluctuating interference patterns arise from superpositions of plane waves. In the halo picture, we tend to think of the granular fluctuations in the ``skirt" of the halo, i.e., outside the central soliton, as arising from interference between the higher energy modes in the eigenvalue decomposition \cite{Dalal2021, Li2021,Zagorac_2022}. 

It has been argued that simulations using the eigenvalue approximation scheme appropriately model the interaction between stars and ultralight dark matter for collapsed halos \cite{Dalal2021}. This approximation has been used to study the dynamical heating of stellar distributions \cite{dalal2022}.

\textbf{Quasi particle approximation. }
A useful approximation for the influence of the granules is to treat them as quasi-particles with radius and effective mass given, respectively.
\begin{align}
    r_\mathrm{eff} \sim \lambda_\mathrm{db} \, , \\
    m_\mathrm{eff} \sim \lambda_\mathrm{db}^3 \rho \, .
\end{align}
Where $\rho$ is the local density. The ultralight dark matter can then be treated as a gravitationally interacting gas of these particles. The quasi-particle approximation has been applied successfully analytically in a number of works \cite{Eberhardt2024, dalal2022, Kim2024_astro, dror2024, Kim2024, eberhardt2025}. However, it is important to keep in mind the limitations of this approximation. The fluctuations include spatial modes at scales other than the de Broglie wavelength associated with the stellar dispersion, $\sigma$, see equation \eqref{eqn:rhoPS}, and are not persistent but condense and evaporate on the de Broglie time scale. The limitations of this approximation were studied in Ref. \cite{zupancic2023}.

\textbf{Full simulations.} The most accurate method for simulating granular effects is full-scale simulations. This has been done in a number of works \cite{Gosenca2023, teodori2025, Chowdhury2023, zupancic2023, eberhardt2025, Eberhardt2024}. Generally, these works find agreement with approximations in the appropriate regimes. 

\textbf{High spin and mixed dark matter models}

The amplitude of granular over-densities has been studied in the context of multi-field ultralight dark matter \cite{Gosenca2023} and higher spin ultralight dark matter \cite{Amin2022}. In both cases, it was determined that if the degrees of freedom remained uncorrelated then the reduction in the amplitude of density fluctuations should go as $\delta\rho \rightarrow\delta\rho / \sqrt{N_\mathrm{df}}$ where $N_\mathrm{df}$ is the number of degrees of freedom. In the multifield case the is the number of fields is and in the $s$ spin case is $2s + 1$.

\subsubsection{Stellar deflections} \label{subsec:heating}

\begin{figure*}[!ht]
	\includegraphics[width = .94\textwidth]{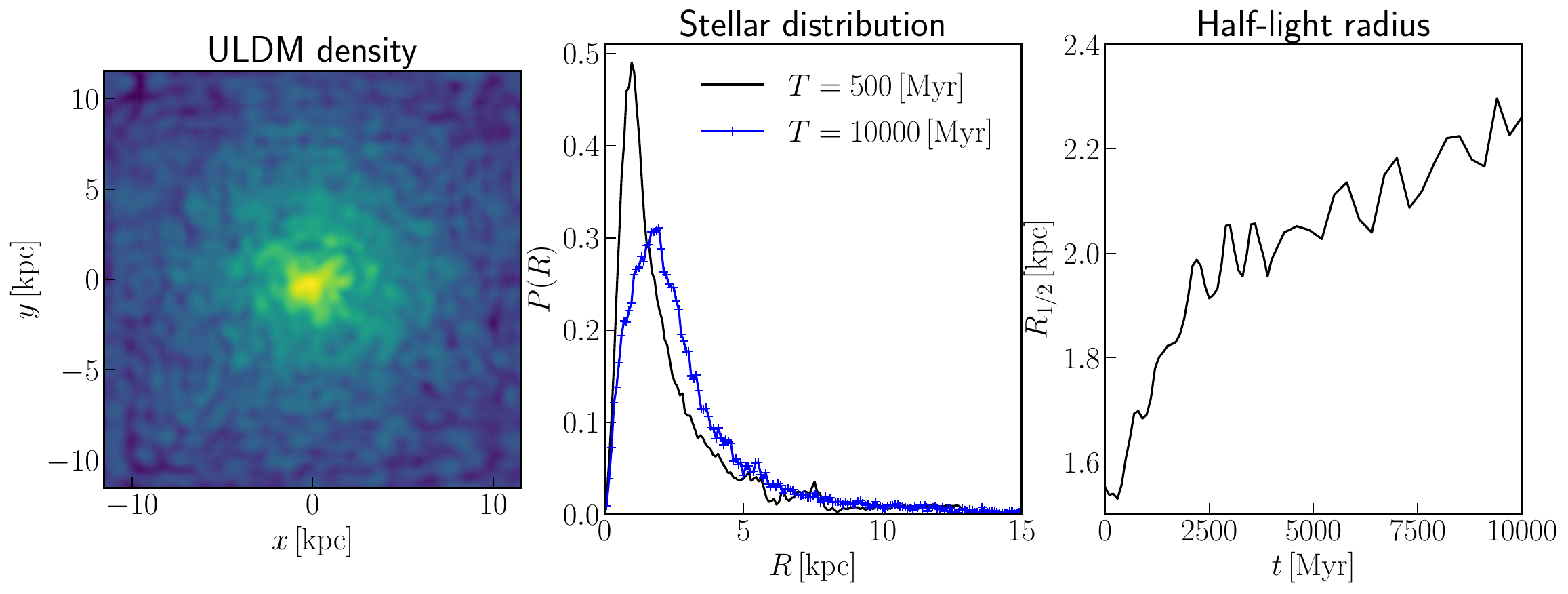}
	\caption{ \textbf{Left:} the log density of an ultralight dark matter halo. \textbf{Center:} distribution of stellar particle radii embedded in the halo at two different snapshots. \textbf{Right:} Half-light radius of stellar particles overtime. We can see the half-light radius of the particles increases overtime. In this simulation $m_{22} = 5$ and $f_\mathrm{fdm} = 0.5$. The increase in the half-light radius of the stars overtime is consistent with dynamical heating from the ultralight dark matter granules. Figure adapted from \cite{Eberhardt2025b}.}
	\label{fig:heating}
\end{figure*}
The large de Broglie wavelength of these quasi-particles means they could be many times more massive than stars. It has been shown analytically \cite{Marsh2019} and numerically \cite{dalal2022, Chowdhury2023, Eberhardt2025b} that the gravitational interaction between the dark matter granules and stellar mass particles results in a transfer of energy to the stellar particles, heating their distribution or randomly walking their velocities. This effect is straightforward to understand using the quasi-particle approximation \cite{dalal2022}. The change in velocity of a particle passing another particle is given to linear order
\begin{align} \label{eqn:kicks}
    \delta v = \frac{MG}{vb} \, .
\end{align}
Where $M$ is the mass of the perturbing particle and $b$ is the impact parameter of the encounter. These kicks are assumed to be random with mean $\braket{\delta v} = 0$ and variance $\braket{\delta v^2} = \delta v^2$. A distribution of stars undergoing $N_\mathrm{enc}$ will then have the variance of its distribution, $\sigma^2$, increase linearly
\begin{align}
    \Delta \sigma^2 = N_\mathrm{enc} \delta v^2 \, .
\end{align}
as the variances of a sum of random variables add linearly. Systems whose velocity dispersion is described by a Maxwell-Boltzmann distribution, like stars in a halo, have their kinetic energy related to the variance of their distribution. Therefore, the variance of the stellar distribution increases is related to an increase in the energy of the stellar distribution. For stars in a bound system, this will result in an increase in the half-light radius of the system. This effect is plotted for an example simulation in Figure \ref{fig:heating}. We can see that the half-light radius of the stellar distribution increases over time as expected.  

If we assume that the ultralight dark matter quasi-particles are causing the stellar kicks, then we can rewrite equation \eqref{eqn:kicks} in terms of the dark matter parameters
\begin{align}
    \delta v \sim \frac{\rho_\mathrm{ULDM} \, \hbar^2 G}{m^2 \sigma^3} \, , \\
    N_\mathrm{enc} \sim \frac{\sigma^2 m T}{\hbar} \, .  
\end{align}
meaning the velocity dispersion should increase at a rate
\begin{align}
    \partial_t \Delta \sigma^2 \sim \frac{G^2 \rho_\mathrm{ULDM}^2 \hbar^3}{m^3 \sigma^4} \, . 
\end{align}
A more detailed analysis starting from the power spectrum is done in Ref. \cite{Marsh2019}. Here we have worked out the increase for free particles interacting with a granular density field described by a single length scale, $\lambda_\mathrm{db} = \hbar / m \sigma$. In general, the actual change in the velocity dispersion of the stars will depend on the specific potential they find themselves in. If, for example, the stars are in a potential dominated by the dark matter, the energy transferred to the stars will go both into the kinetic and potential components of the stellar distribution and will not be as simple as we have described here. 

\subsubsection{Gravitational metric effects}

A number of granular effects on the gravitational metric perturbations of ultralight dark matter have been studied as well in the context of pulsar timing \cite{Porayko2018, Eberhardt2024}, fast radio burst timing \cite{Gao2025}, and gravitational lensing. 
Shapiro delays induced by the changing proper distance between the Earth and pulsar \cite{Eberhardt2024}, and redshift delays induced by the changing gravitational redshift between the Earth and pulsar positions \cite{Eberhardt2024}. The motivation behind studying these effects is generally that such probes would potentially be sensitive to ultralight dark matter masses where the de Broglie oscillations of the field were similar to the experimental runtime, i.e., $m \sim 10^{-17} \, \mathrm{eV}$. The mass scales have been difficult to probe with other methods. However, present estimates show that these effects are far too small to be observed by current probes. The amplitude and mass scaling of each method is well described by the quasi-particle approximation. 

\textbf{Graviational redshift} The time delay created by the random potential difference between Earth and the pulsars created by the granular density field was studied in Ref. \cite{Eberhardt2024}. This potential difference creates a gravitational redshift which fluctuates on the de Broglie time. The time delay was estimated to be 
\begin{align}
    \delta t_\mathrm{rms}^\mathrm{z} &\sim \frac{ \braket{\Phi}_{\mathrm{rms}} \tau_\mathrm{db}}{8 \pi c^2} \left( \frac{T}{\tau} \right)^{3/2} \sim 4 \times  10^{-4} \left( \frac{10^{-17} \, \mathrm{eV}}{m} \right)^{3/2} \left( \frac{200 \, \mathrm{km/s}}{\sigma} \right)^4 \left( \frac{\rho}{10^7 \, \mathrm{M_\odot / kpc^3}} \right) \, \left( \frac{T}{30 \, \mathrm{yrs}} \right)^{3/2} \mathrm{ns} \, .
\end{align}
Where $\braket{\Phi}_{\mathrm{rms}} / c^2 \approx  \frac{G m_{\text{eff}}}{\lambda_\mathrm{db} \, c^2}$ is the typical size of the potential fluctuation.

\textbf{Shapiro delay} The time delay created by the fluctuating proper distance between Earth and the pulsar created by the granular density field was studied in Ref. \cite{Eberhardt2024}. The time delay was estimated to be 
\begin{align}
        \delta t_\mathrm{rms}^\mathrm{sh} \sim \frac{G \, m_\mathrm{eff}}{c^3} \left(\frac{D}{\lambda_\mathrm{db}}\right)^{2/3} &\sim 5 \times 10^{-2} \left( \frac{10^{-17} \, \mathrm{eV}}{m} \right)^{7/3} \left( \frac{200 \, \mathrm{km/s}}{\sigma} \right)^{7/3} \left( \frac{\rho}{10^7 \, \mathrm{M_\odot / kpc^3}} \right) \left( \frac{D}{\mathrm{kpc}} \right)^{2/3} \, \mathrm{ns} \, .
\end{align}
$D$ is the pulsar-earth separation.

\textbf{Lensing} The impact of density granules on lensing observations has been studied for strong lensing \cite{Powell2023} of radio-imaged jets, weak lensing along the critical curve of A370 \cite{Broadhurst_2025}, and stochastic lensing of stars \cite{eberhardt2025}. The works generally use the Gaussian random field approximation to make arguments about the statistics of the density perturbation along the line of sight. In the case of strong lensing, the density granules can confound the ability of the lens's image to be focused on the scales observed in data \cite{Powell2023}. Similarly, for weak lensing, in Ref. \cite{Broadhurst_2025}, it was argued that ultralight dark matter density granules would produce the corrugated critical curves observed in the ``Dragon Arc" system. For stochastic lensing, it was argued that the fluctuating density along the line of sight would make a small oscillating signal in the brightness of objects observed through ultralight dark matter halos \cite{eberhardt2025}. The oscillation in the brightness is given 
\textbf{\begin{align} \label{eqn:lensing_granules}
    \delta \kappa_\mathrm{rms}^\mathrm{gr} \sim 3 \times 10^{-12} \left( \frac{10^{-17} \, \mathrm{eV}}{m} \right)^{1/2} \left( \frac{200 \, \mathrm{km/s}}{\sigma} \right)^{1/2} \left( \frac{\rho}{10^7 \, \mathrm{M_\odot / kpc^3}} \right) \left( \frac{D}{\mathrm{kpc}} \right)^{3/2} \, .
\end{align}}

\subsubsection{Multi-field/multi-component, self-interactions, and higher-spin models} \label{subsec:granules:selfInteract}

Granules have been the focus of a number of studies of extensions of the vanilla ultralight dark matter model. This includes looking at the effect of multiple fields or multiple component dark matter \cite{Gosenca2023, Eberhardt2025b}, self-interacting dark matter \cite{Capanelli2025}, and higher-spin models \cite{Amin2022}.

For multi-component dark matter, the heating rate of each species is thought to superimpose with the rate of injected energy for the $i$th species proportional to $\rho_i^2 m_i^{-3}$. This implies that when the masses are logarithmically distributed, as is expected to be the case in the axiverse \cite{Arvanitaki2010, Sheridan2025}, a single species will dominate the heating. This scaling was tested numerically in Ref. \cite{Eberhardt2025b}.

In the case of multiple fields with similar masses and density fractions or for higher-spin ultralight dark matter the amplitude of the overdensity is reduced by the root of the number of degrees of freedom, i.e. as $1/\sqrt{n_\mathrm{fields} }$ in the case of $n_\mathrm{fields}$ fields \cite{Gosenca2023}, or $1/\sqrt{2s + 1}$ in the case of spin-$s$ dark matter \cite{Amin2022}. This relies on the fact that the overdensities in each degree of freedom do not become correlated which has been found to be true numerically \cite{Gosenca2023}. 

Self-interactions may produce multiple effects. It has been shown in place-wave box simulations that the presence of an attractive self-interaction can increase the amplitude of overdensities (and vice versa) \cite{Capanelli2025}. However, it was also argued that an attractive self-interaction would increase the velocity dispersion in a halo for a given total enclosed mass; this would then decrease the de Broglie wavelength of the granules in that halo and therefore reduce the size and impact of the granules in the system \cite{may2025}. The change in granule size was studied numerically for the repulsive case in Ref. \cite{Indjin2025} which concluded that the change in the granule size was minimal.

\subsection{Relativistic potential} \label{sec:relativistic_potential} 

\subsubsection{Summary}

Much of the work on ultralight dark matter considers the non-relativistic phenomenology. However, we expect that there should also be fluctuations on the Compton scales, i.e. 
\begin{align}
    \tau_c = \frac{2 \pi \hbar}{m c^2} \, , && \lambda_c = \frac{2\pi}{mc} \, .
\end{align}
where $\tau_c$ is the Compton time and $\lambda_c$ the wavelength.

The form of the relativistic potential was worked out in \cite{Khmelnitsky2014}. In order to derive the full relativistic effects, we will first write the field as a real scalar as 
\begin{align} \label{eqn:compton_ansatz}
    \phi(x,t) = A(x)\cos(mt + \alpha(x)) \, .
\end{align}
Which describes a field oscillating at the Compton frequency with arbitrary spatial dependence. In the following, we will set $\hbar = c = 1$. The energy-momentum tensor of a free scalar is
\begin{align}
    T_{\mu \nu} = \partial_\mu \phi \,  \partial_\nu \phi- \frac{1}{2} g_{\mu\nu}((\partial \phi)^2 - m^2 \phi^2) \, .
\end{align}
Where $g_{\mu\nu}$ is the metric. Plugging in equation \eqref{eqn:compton_ansatz} we find in the energy density, $T_{00}$ a time independent component
\begin{align}
    \rho_{DM} \equiv T_{00} = \frac{1}{2} m^2 A^2 \, ,
\end{align}
and an oscillating portion arising from the spatial derivatives of $\phi$, i.e. 
\begin{align}
    \rho^\mathrm{osc}_{DM} \sim \frac{k^2}{m^2} \rho_{DM} = v^2 \rho_{DM} \, .
\end{align}
As expected, the oscillating portion of the energy density is suppressed by $v^2/c^2$.

The spatial component of the tensor $T_{ij}$ is given 
\begin{align}
    T_{ij} = -\frac{1}{2} m^2 A^2 \cos (2mt + 2 \alpha) \,  \delta_{ij} \equiv p(x,t) \, \delta_{ij}\, .
\end{align}
In the weak gravity limit, we then write the Newtonian metric
\begin{align} 
    dx^2 = (1 + 2\Phi(x,t)) dt^2 - (1 - 2 \Psi(x,t))\, \delta_{ij} \, dx^i \, dx_j \, .
\end{align}
The potentials can be written to leading order in the oscillations as 
\begin{align} \label{eqn:psi_osc}
    \Psi(x,t) \approx \Psi_0(x) + \Psi_c(x) \, \cos(2mt + 2 \alpha(x)) + \Psi_s (x) \, \sin(2m t + 2 \alpha(x)) \, .
\end{align}
The $00$ component of the linearized Einstein equations is given 
\begin{align}
    \nabla^2 \Psi = 4 \pi G T_{00} = 4 \pi G \rho_{DM} ( 1+ \mathcal{O}(v^2)) \, .
\end{align}
The oscillating portions of equation \eqref{eqn:psi_osc} can be found using the trace of the spatial component of the Einstein equations
\begin{align}
    -6 \partial_t^2 \Psi + 2 \nabla^2(\Psi + \Phi) = 24 \pi G \,  p(x,t) \, .
\end{align}
Which gives $\Psi_s = 0$ to leading order and 
\begin{align}
    \Psi_c(x) = \frac{1}{2} \pi G A(x)^2 = \frac{\pi G \, \rho_{DM}(x)}{m^2} \, .
\end{align}
We know from Poisson's equation that $\Psi_0 \propto k^2 \rho_{DM}$, and here we see that $\Psi_c = (k/m)^2 \Psi_0 = v^2 \Psi_0$ is related to the time-independent portion of the potential by the expected factor of $v^2/c^2$. Relativistic corrections were also studied in Ref. \cite{Namjoo2018}.

\subsubsection{Effect on pulsar timing} 
The relativistic potential provides $0$ pressure over a full period and is suppressed at the non-relativistic velocities we expect to describe galactic dark matter. It is therefore unlikely to provide much impact on structure formation. However, systems with extreme sensitivity to the spacetime metric, such as pulsars, have been suggested as probes of this potential. 

In principle, this oscillating potential creates both a redshift delay, due to the changing observed potential difference between Earth and the pulsar, and also a Shapiro delay, due to the fluctuating proper distance between Earth and the pulsar. However, because of the leading order, the Shapiro delay is $0$ over a period; this effect is generally neglected, and only the redshift, $z(t)$, is considered. The redshift can be written as
\begin{align}
    z(t) \equiv \frac{\Omega(t) - \Omega_0}{\Omega_0} = \Psi_c \left( \cos(\omega t + 2 \alpha(x_e)) - \cos(\omega(t-D) + 2\alpha(x_p) \right) \, .
\end{align}
Where $x_e$, and $x_p$ are the positions of the Earth and pulsar, respectively, which are separated by a distance $D$. $\Omega(t)$ is the observed frequency of the pulsar and $\Omega_0$ the unperturbed frequency. This can be related to an observed time delay in the pulsar signal by integration 
\begin{align}
    \Delta t(t) = -\int_0^t z(t') dt' \, .
\end{align}
The resulting time delay residual goes as
\begin{align}
    \Delta t(t) = \frac{2 \Psi_c}{\omega} \sin \left( m D + \alpha(x_e) + \alpha(x_p) \right) \cos \left( 2 m t + \alpha(x_e) + \alpha(x_p) - m D \right) \, .
\end{align}
This produces a root-mean-square signal with amplitude and frequency given
\begin{align}
    \delta t_\mathrm{rms} = \sqrt{2} \frac{\Psi_c}{2 m} \, , && f = 4 \pi m \, .
\end{align}
This produces a characteristic strain, $h_c$, as
\begin{align}
    h_c = 2 \sqrt{3} \Psi_c = 2 \times 10^{-15} \left( \frac{\rho_\mathrm{DM}}{0.3 \, \mathrm{GeV/cm^3}} \right) \left( \frac{10^{-23} \, \mathrm{eV}}{m} \right)^2 \, ,
\end{align}
with frequency
\begin{align}
    f = 5 \times 10^{-9} \left( \frac{m}{10^{-23} \, \mathrm{eV}} \right) \, \mathrm{Hz} \, .
\end{align}

\subsection{ULDM clouds}

Since ULDM consists of ultralight bosonic particles, various mechanisms can lead to the accumulation of these fields around massive astrophysical objects. We describe two such mechanisms below.

\subsubsection{Superradiance}

Another probe that can lead to constraints on the mass of light bosons arises from a process known as black hole superradiance. This phenomenon occurs around rotating (Kerr) black holes and allows ultralight bosonic fields to extract energy and angular momentum from the black hole if certain conditions are met.

The effect is strongest when the Compton wavelength of the boson, $\lambda_C = h/mc$ is comparable to the size of the black hole's ergoregion. The ergoregion is the spacetime region just outside the event horizon where no object can remain stationary due to the dragging of inertial frames by the black hole’s rotation. In this region, bosonic waves can undergo superradiant amplification, analogous to the classical Penrose process, if their frequency satisfies the superradiance condition: $\omega < m_{\ell} \, \Omega_{BH}$, where $ \omega $ is the wave frequency, $ m_{\ell} $ is the azimuthal quantum number of the mode, and $ \Omega_{BH} $ is the angular velocity of the black hole horizon.

For a massive boson, the amplified modes can become trapped around the black hole, forming long-lived gravitational bound states, often referred to as a \textit{gravitational atom}. These modes grow exponentially by repeatedly extracting spin from the black hole, generating an axion (or ALP) cloud.

This leads to an instability that causes the black hole to spin down on astrophysically short timescales. The absence of spin-down in observed astrophysical black holes provides a way to constrain ultralight bosons, in a mass range that complements other search methods, as we will show in Section~\ref{sec:constraints}. This is a gravitational process and quite general. It applies to all ULDM candidates, with the details, such as growth rates and observational signatures, depending on the spin and interactions of each specific model.

\subsubsection{Axion halos}

Axion halos refer to gravitationally bound configurations of ultralight bosonic fields that form in the external gravitational potential of massive objects, such as stars or compact objects. Unlike superradiance, these halos do not rely on black hole spin or wave amplification. Instead, ambient ultralight fields become gravitationally bound, forming quasi-stationary configurations that oscillate with a frequency set by the particle’s mass~\cite{Budker:2023sex}.

While often discussed in the context of axions or ALPs, this phenomenon applies generally to any ultralight bosonic field. The formation of these halos requires modifying the Schrödinger–Poisson equations typically used for ULDM that only consider the ULDM self-gravity, and include the external gravitational potential of the host object. These configurations can lead to local enhancements in the field density and potentially observable time-dependent signals near astrophysical objects~\cite{Banerjee:2019xuy}.

\subsection{Axion birefringence}

We are now going to discuss an effect that requires specifying the particle that composes ULDM. The results discussed in this section do not apply to general ULDM models, but only to those in which the dark matter is an axion or ALP. In this case, because birefringence is tightly connected to specific observational searches, we briefly summarize current observational results alongside the phenomenology.

Axion birefringence refers to the rotation of the polarization plane of light as it travels through space in the presence of a background axion or axion-like particle (ALP) field. This effect arises from the parity-violating axion–photon coupling, $\mathcal{L} \supset \frac{1}{4} g_{a\gamma\gamma} a F_{\mu\nu} \tilde{F}^{\mu\nu}$, where $a$ is the axion field, where $F_{\mu\nu}$ is the electromagnetic field strength, and $\tilde{F}^{\mu\nu}$ is its dual. This coupling modifies the propagation of electromagnetic waves causing the left- and right-circularly polarized photons to travel at different speeds, resulting in a net rotation of linear polarization. The total rotation angle is given by
\begin{equation}
\beta= \frac{1}{2} g_{a\gamma\gamma} \left[\phi(t_{\text{obs}}) - \phi(t_{\text{em}})\right],
\end{equation}
where $\phi(t)$ is the background axion field and $t_{\text{em}}$, $t_{\text{obs}}$ are the times of emission and observation. This phenomenon is referred to as cosmic birefringence and provides a direct observational probe of axion-like fields and parity violation in the universe.

Recent analyses of cross-correlations between the even-parity CMB E-modes and odd-parity B-modes have provided evidence for isotropic cosmic birefringence, reporting a rotation angle of $\beta \sim 0.35^\circ$ at $3.6\sigma$ confidence level \cite{Minami:2020odp,Eskilt:2022cff,Cosmoglobe:2023pgf}. Ground-based experiments, including \textsc{BICEP}, \textsc{ACTPol}, and \textsc{SPTpol}, have placed stringent constraints on anisotropic cosmic birefringence \cite{BICEP2:2017lpa,BK-LoS:2023,Namikawa:2020ffr,SPT:2020cxx,Bortolami:2022whx,Zagatti:2024,Namikawa:2024:BB, Greco:2022xwj}. Additionally, measurements of the Crab Nebula by \textsc{POLARBEAR} suggest a nonzero birefringence signal \cite{Adachi:2024lga}.

The observed birefringence is compatible with axion masses $m \lesssim 10^{-28} \, \mathrm{eV}$, where the field remains effectively frozen during the history of the Universe. Recently, this mass range was expanded, $10^{-32}\, \mathrm{eV} \lesssim m \lesssim 10^{-26}\, \mathrm{eV}$, showing that heavier axions can also explain the signal~\cite{Zhang:2024dmi}. 

For heavier axion masses, the field begins to oscillate on timescales shorter than the light propagation time. In this regime, birefringence no longer produces a net static rotation but instead leads to an oscillating polarization angle. This effect induces a time-dependent birefringence signal that can be searched for in the polarization of light from pulsars, fast radio bursts, active galactic nuclei, and the cosmic microwave background.

Axion birefringence is a unique signature of axion or ALP fields that couple to photons, and is especially powerful for probing ultralight axions on cosmological scales. 

\subsection{Quantum effects}

\begin{figure*}[!ht]
	\includegraphics[width = .97\textwidth]{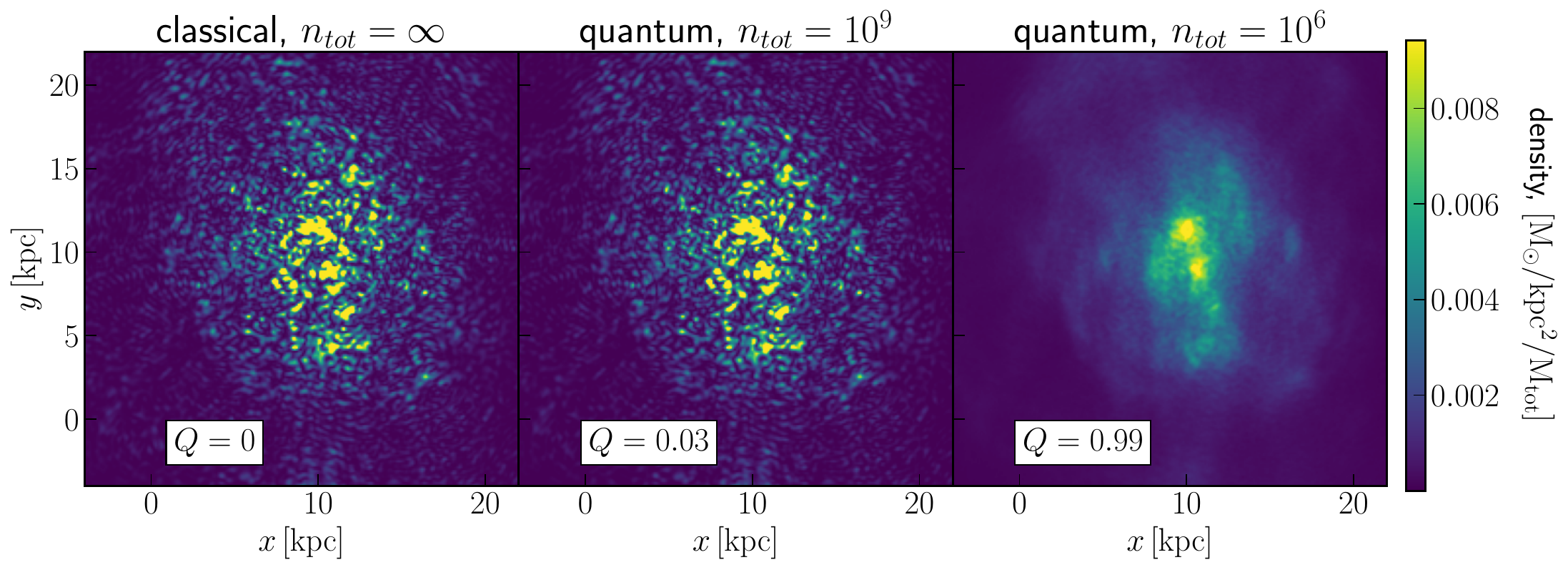}
    \caption{Figure taken from Ref.~\cite{Eberhardt2023}. Here we plot the condensed object resulting from the collapse of a momentum-space Gaussian density in two spatial dimensions for different values of $n_{\rm tot}$. The density shown here is the result of 1 Gyr of evolution. Each plot also shows the simulated value of the $Q$ parameter at this time. The left panel shows the classical field evolution, the condensed object in this case exhibiting the expected granular structure. The right panel shows the evolution for a simulation using the truncated Wigner approximation with $n_{\rm tot}=10^6$ particles. Here $n_{\rm tot}$ denotes the total particle number in the simulation and sets the size of the initial fluctuations sampled from the Wigner distribution. In the TWA, each stochastic realization is evolved using the classical field equations, and the density shown is obtained from the ensemble average. The comparison is therefore between the classical field evolution and the TWA approximation to the quantum expectation value, rather than an exact quantum evolution. Here, as in the one-dimensional case, we see that the quantum corrections have removed most of the granular structure. This is due to the fact that the quantum corrections resulting from the spread of the quantum state around the mean-field value have become of similar order to the amplitude of the field. In the middle panel, we show the same simulation with $n_{\rm tot}=10^9$ particles. The quantum corrections in this case are much smaller, and the resulting density is almost identical to the classical case.}
	\label{fig:2d_collapseQuantum}
\end{figure*}
Ultralight dark matter is usually represented as a classical field. The way in which this approximation is taken and motivated is discussed in Section \ref{sec:ULDM:quantum}. In this section, we will discuss the effect that quantum corrections to the classical field theory may have on observables. Work in this subfield often relies on different metrics of classicality, which measure the existence of quantum corrections in different ways; a detailed discussion of some of the more prominent metrics is discussed in Appendix \ref{app:matricsClassicality}. 

One way in which we may proceed is to identify some specific correction term to the classical field equations and look at how long it takes for this term to become large in comparison to the classical term. Efforts of this type are then typically concerned with identifying a ``quantum breaktime" after which the classical field equations no longer make accurate predictions. 

Typically, we expect that the initial quantum state should be well described classically. Generally, we will take this to mean that the initial quantum state is something like a coherent state, which at high occupation is tightly centered on the classical field value in phase space. If however, we start with an initial quantum state which is poorly described by the classical, such as a number eigenstate, then the classical field theory will admit corrections on the dynamical time of the system regardless of occupation number. This was demonstrated in simulations of number eigenstates \cite{sikivie2017, Hertzberg2016}. 

We will typically then assume that the initial conditions are well described classically, usually that it is in a coherent state. In the presence of nonlinearity in the Hamiltonian, this will generically cause the quantum phase space to spread around the classical value, introducing corrections to the classical field equations. However, as the occupation number increases, the quantum breaktime grows longer and the quantum evolution approaches the classical field theory. This has been demonstrated in simulations of coherent states and field number states \cite{Eberhardt2021, Eberhardt2022, Eberhardt2022Q, Eberhardt2023}. 

The rate at which number changing process \cite{Dvali2018} and the fields quartic self-interaction \cite{Dvali2018,Eberhardt2022Q} introduce quantum corrections has been demonstrated to be slow, $\propto t^{1/2}$, and therefore, unlikely to introduce large corrections to the classical field equations in the lifetime of the universe. 

It is well known that systems that are classically chaotic will have quantum corrections that grow exponentially over time. This is related to the way in which a quantum state, to first order, can be approximated as an ensemble of states with small perturbations in the initial conditions. For systems which are collapsing non linearly due to gravity, it was demonstrated that the quantum corrections grew exponentially \cite{Eberhardt2023} on the scale of the system dynamical time, i.e. $\propto e^{t/t_d}$. In a closed quantum system then there would exist certain dynamical systems for which the quantum corrections would be large in the lifetime of the universe. Likewise, it has been shown that the quantum state can quickly become ``squeezed" even if the phase space remains Gaussian \cite{Kopp2021}. 

All of these analyses neglect the impact of quantum decoherence, which is anticipated to project the quantum state back onto some basis of pointer states. It has been estimated analytically \cite{Allali_2020, Allali2021, Allali2021b} and numerically \cite{Eberhardt2023} that the decoherence rate is fast compared to any of the breaktimes discussed here. These pointer states are generally assumed to be classical, though there currently is no study identifying the pointer states of ultralight dark matter. 

While the quantum breaktime and growth of corrections have been very thoroughly studied, it is not immediately from these analyses what observables quantum corrections, if large, would actually impact. Numerical and analytic work has argued that quantum corrections could result in the development of long-range correlation \cite{Lentz2018, Lentz2019}. Additionally, its simulations of coherent states demonstrated that large quantum corrections can reduce the amplitude of density fluctuations \cite{Eberhardt_testing, Eberhardt2023}. The effect of quantum corrections on density granules in a simulated halo is plotted in Figure \ref{fig:2d_collapseQuantum}. It was argued that in these circumstances the haloscopes may be sensitive to the quantum state of the dark matter \cite{Marsh2022}. 

\section{Numerical methods} \label{sec:numMethods}

\begin{table*}
\begin{center}
\begin{tabular}{| c | c | c | c | c | c | c }

\hline
\rule{0pt}{12pt} Tool & Repo & reference\\
 
\hline
\hline

\rule{0pt}{12pt} Schr\"odinger-Poisson solver & \href{https://github.com/auckland-cosmo/PyUltraLight}{PyUltraLight} & \cite{Edwards2018} \\[2ex]
\hline

\rule{0pt}{12pt} Differentiable Schr\"odinger-Poisson solver & \href{https://github.com/JaxionProject/jaxion}{jaxion} & \cite{Mocz_Jaxion_2025} \\[2ex]
\hline

\rule{0pt}{12pt} Schr\"odinger-Poisson/CDM (adaptive) solver & \href{https://github.com/axionyx/axionyx_1.0}{axionyx} & \cite{Schwabe2020} \\[2ex]
\hline

\rule{0pt}{12pt} Schr\"odinger-Poisson (adaptive) solver & \href{https://github.com/gamer-project/gamer}{gamer} & \cite{kunkel2024} \\[2ex]
\hline

\rule{0pt}{12pt} ULDM cosmological initial conditions & \href{https://github.com/dgrin1/axionCAMB}{axionCAMB} & \cite{Hlozek2015} \\[2ex]
\hline

\rule{0pt}{12pt} Sch\"odinger-Poisson, (adaptive) full physics baryons simulations & \href{https://gitlab.mpcdf.mpg.de/vrs/arepo}{AREPO} & \cite{Weinberger_2020} \\[2ex]
\hline

\hline
\end{tabular}
\caption{\label{tab:num} Numerical tools and related repositories commonly used in ultralight dark matter work. }
\end{center}
\end{table*}

This section describes some of the most common numerical tools used to model ultralight dark matter. Popular repositories associated with these tools are listed in Table \ref{tab:num}. In this section, we cover the most common method of simulation, the pseudo-spectral method. We also discuss ways to generate mock ultralight dark matter halos. Finally, we discuss approximation schemes used to circumvent the numerical limitations of the pseudo-spectral method. 

\subsection{Pseudo spectral methods} \label{sec:num:SP}

\begin{table*}
\begin{center}
\begin{tabular}{| c | c | }

\hline
Cause & Requirements \\
 
\hline \hline 

\rule{0pt}{15pt} Kick operator aliasing (gravity)  & $\delta t < \frac{ \hbar}{m V_\mathrm{max}}$    \\[2ex]
\hline

\rule{0pt}{15pt} Kick operator aliasing (self-interaction)  & $\delta t < \frac{ m^2 }{\lambda \hbar^2 \rho_\mathrm{max}}$    \\[2ex]
\hline

\rule{0pt}{15pt} Drift operator aliasing & $\delta t < \frac{m L^2}{{\hbar}N^2}$ \\[2ex]
\hline

\rule{0pt}{15pt} Momentum aliasing & $N > \frac{m v_\mathrm{ph} L}{ \hbar \pi}$  \\[2ex]
\hline

\end{tabular}
\caption{\label{tab:contr_solver} A list of numerical resolution requirements and related causes for a fixed resolution pseudo-spectral Schr\"odinger-Poisson solver. The first three rows are conditions related to temporal aliasing of the update operators associated with the momentum (kick) and position (drift) updates. The timestep needs to be chosen to be sufficiently small such that the phase updated by each operator is small. Adding additional interactions or external potentials will each require their own additional kick operator condition. Here we have listed two of the most frequent (associated with gravitational and self-interacting potentials), but in principle it is straightforward to add more. The last condition is spectral aliasing where the maximum and minimum representable momenta identify. The spatial resolution needs to be chosen such that the wavelength associated with the highest momentum mode is resolvable. We have written these conditions in terms of the simulation parameter time step, $\delta t$, and grid resolution, $N$. $V_\mathrm{max}$ is the maximum value of the potential, $m$ the mas of the field, $L$ the box size, and $v_\mathrm{ph}$ the maximum physical velocity in the system. }
\end{center}
\end{table*}

The pseudo-spectral method is the workhorse solver for many ultralight dark matter simulations \cite{Schive2014,Schwabe2016, Mocz2017,Du2018,Li_2019, Edwards2018, Mocz2019, Schwabe2020,teodori2025, May2021}. It has been used widely in the field to simulate isolated halos \cite{Gosenca2023, Schwabe2020,teodori2025}, the formation of cores \cite{Mocz2017,Levkov2018,Schwabe2016}, and cosmological initial conditions \cite{Schive2014, Mocz2019, May2021}. A comparison of this method with the fluid ``Madelung" method is contained in \cite{Li_2019, mocz2015, Veltmaat2016}. Extensions of this method to relativistic systems is contained in \cite{Felder2008, Giblin2010, Amin2012, Helfer2017, Widdicombe2020, Buschmann2020, Guo2021, Ling2024} but will not be discussed here.

Example public repositories include \href{https://github.com/auckland-cosmo/PyUltraLight}{PyUltraLight} (a lightweight psuedo spectral integrator) \cite{Edwards2018}, \href{https://github.com/JaxionProject/jaxion}{jaxion} (a differentiable psudo spectral integrator) \cite{Mocz_Jaxion_2025}, \href{https://github.com/axionyx/axionyx_1.0}{axionyx} (an adaptive mesh integrator) \cite{Schwabe2020}, and \href{https://github.com/gamer-project/gamer}{gamer} (a hybrid fluid and wave-like solver) \cite{kunkel2024}. 

\subsubsection{Initial conditions}

The complex field is initialized on a grid. We will define grid positions, in a simulation box of size $L$, as 
\begin{align}
    x_i = -L/2 + (i + 1/2)dx \, .
\end{align}
Where $dx = L/N$ is the grid resolution and $N$ is the number of grid cells. We have chosen to center this box at $x=0$, but this choice is arbitrary. The above is for a system with a single spatial dimension but the generalization to multiple dimensions is straightforward. The initialization of the field values varies widely, but a general procedure involves superimposing different parts of the field phase space, i.e.
\begin{align}
    \psi(x_i) = \sum_S A(x_i, v_i) \, e^{iS(x_i)/\hbar} \, .
\end{align}
This construction is often useful as the interpretation of the amplitudes as phase space densities, i.e. $|A(x_i, v_i)|^2 = f(x_i, v_i)$, and derivative of the phase as relating to the phase space velocity, i.e. $\nabla S(x_i) / m = v(x_i)$. We note also that, because this method relies heavily on Fourier transforms as we will see the following subsections, it is necessary that $\psi(-L/2) = \psi(L/2)$ the field identifies on the boundaries of the box, other specious small scale modes will be generated in the system. These generally manifest as ripples from the box edges.

\textbf{Cosmological initial conditions}
For cosmological simulations, the initial matter power spectrum needs to be consistent with the modified transfer function created by ultralight dark matter ``quantum" pressure. Codes like \href{https://github.com/dgrin1/axionCAMB}{axionCAMB} \cite{Hlozek2015} produce these initial conditions by solving the evolution of the equation for the field, i.e. 
\begin{align}
    \partial_t^2 \phi +2 \mathcal{H} \partial_t \phi + m^2 a^2 \phi = 0 \, ,
\end{align}
and integrating them to the desired time and then drawing a realization from the predicted power spectrum. 

Codes like \textit{N-GenIC} \cite{Springel2015} generate particle initial conditions using the Zel'dovich approximation. These can be converted to field initial conditions using the relationships for the amplitude and phase of the given by
\begin{align}
    |\psi(x)| = \sqrt{\rho(x)} \, , && \nabla \mathrm{Arg}(\psi(x)) = \frac{m}{\hbar} v(x) \, ,
\end{align}
where $v(x)$ is the velocity field defined at position $x$.

\subsubsection{Solver}

We start from the Schr\"odinger-Poisson equations
\begin{align}
    \partial_t \psi(x,t) &= -\frac{i}{ \hbar}\left( \frac{\hat p^2}{2m } + m V(x,t) \right) \psi(x,t) \, , \\
    \nabla^2 V(x,t) &= 4 \pi G \rho(x,t) \, .
\end{align}
Where $\hat p = -i\hbar \nabla$ is the standard field momentum operator. The solution to this is
\begin{align}
    \psi(x,t + T) = e^{-\frac{i}{\hbar} \int_t^{t+T}  dt' \left( \frac{\hat p^2}{2m } + m V(x,t) \right) } \psi(x,t) \, .
\end{align}
For time intervals $\delta t$ which are small compared to the dynamical times in the system (we will quantify this limit in a later section), we can approximate the evolution as 
\begin{align}
    \psi(x,t + \delta t) \approx e^{-\frac{i}{\hbar} \delta t \left( \frac{\hat p^2}{2m } + m V(x,t) \right) } \psi(x,t) \, .
\end{align}
In this same limit, we can split the exponential into a kinetic and a potential component
\begin{align}
    \psi(x,t + \delta t) = e^{-i \delta t \frac{\hat p^2}{2m} / \hbar} e^{-i \delta t \, m V(x,t) / \hbar} \left( 1 + \mathcal{O}(\delta t^2 [\nabla^2, V(x)]) \right) \psi(x,t) \, .
\end{align}
Since each component of the exponential is diagonal in a different basis (the kinetic in the momentum basis and the potential in the position basis), we can act on them separately on the field in the appropriate basis. Realizing that the kinetic portion updates the phase space position, and the potential portion the phase space velocities (see Section \ref{subsec:hamEqns}), we can write the update of the field in terms of the symplectic leap-frog update, i.e.
\begin{align} \label{eqn:pseudoSpecUpdate}
    \psi_1(x) &= e^{-i \delta t \, m V(x,t,\psi(t)) / \hbar / 2} \psi(x,t) \, \, \, \mathrm{\textbf{[half-step kick update]}}  , \\
    \psi_2(p) &= e^{-i \delta t \frac{\hat p^2}{2m} / \hbar} \mathcal{F}\left[ \psi_1(x) \right](p) \, \, \, \mathrm{\textbf{[full-step drift update]}} , \\
    \psi(x, t + \delta t) &= e^{-i \delta t \, m V(x,t,\psi_2) / \hbar / 2} \mathcal{F}^{-1}\left[ \psi_2(p) \right](x) \, \, \, \mathrm{\textbf{[half-step kick update]}} .
\end{align}
Here we have used the kick-drift-kick scheme. $\mathcal{F}$ is the Fourier transform. The potential on the first line is calculated using $\psi(x,t)$ and on line 3 using $\mathcal{F}^{-1}[\psi_2(p)](x)$. The potential is calculated using the spectral method, i.e. 
\begin{align}
    V(x,t,\psi) = \mathcal{F}^{-1}\left[ 4 \pi G \,  \frac{\mathcal{F} [|\psi(x',t)|^2] (\vec k)}{k^2} \right](x) \, . \label{eqn:Poisson_k_space}
\end{align}

\textbf{Boundary conditions.} Because of the use of Fourier transforms, the above scheme implicitly assumes that our simulation box has periodic boundary conditions. We can avoid this with the usual technique, i.e., we define a new field on a new grid, $x'_j$, twice the size of the previous, when calculating the potential, i.e., equation \eqref{eqn:Poisson_k_space}, $\psi'$, as follows
\begin{align}
    \psi'(x_j') = 
\begin{cases}
\psi(x_j), \, 0 \le j < N \, \\
0 , \, \mathrm{else}
\end{cases}
\end{align}
Where $x_j' \in [-L/2, L/2 + L]$, the index runs from $j\in [0,2N)$. The non-periodic potential is then 
\begin{align}
    V(x_i,t,\psi) = \mathcal{F}^{-1}\left[ 4 \pi G \,  \frac{\mathcal{F} [|\psi'(x_j',t)|^2] (\vec k)}{k^2} \right](x_i) \, , \, 0 \le i < N . 
\end{align}
An important note is that the edges of the box still identify in the sense that mass leaving the box through the right edge will re-enter the box on the left edge. This can be solved either by having $\psi/\bar \psi \rightarrow 0$ at the boundary of the box, or by implementing an absorbing boundary condition. 

\textbf{Adaptive mesh schemes.}
There has been work done developing adaptive resolution code to try and circumvent some of the intrinsic numerical difficulties of ultralight dark matter, for example \href{https://github.com/axionyx/axionyx_1.0}{axionyx} \cite{Schwabe2020} and \href{https://github.com/gamer-project/gamer}{gamer} \cite{Schive2018}. 

The adaptive scheme in \cite{Schwabe2020} used higher resolution  (``refined") grids in high-density regions nested in a low-resolution root grid with up to 6 levels of refinement in the halo simulations tested. The standard pseudo-spectral method is used on the lowest resolution grid. On the refined grids, a fourth-order Runge-Kutta integrator and a finite difference scheme are used. The method has proved particularly useful for simulating isolated individual halos, see for example \cite{Schwabe2020, Gosenca2023}. This is due to the fact that isolated halos have a naturally telescoping hierarchy of scales where higher resolution is necessary inside the halo and lower outside the central regions. This method has proved less beneficial for cosmological simulations where high resolution is necessary in the centers of dense regions but also in low-density regions where the velocity of the dark matter may still be very high meaning that there is not an obvious separation of low resolution and high resolution regimes.

\subsubsection{Extensions} 

Here we show how one can extend the solver above to include self-interactions, multiple fields, and higher spins. 

\textbf{Multiple, self-interacting fields.}
For multiple and interacting fields, we can write the Gross-Pitaevskii-Poisson as equation \eqref{eqn:multiFieldGPP}. In this case, we write our update as 
\begin{align} \label{eqn:pseudoSpecUpdate_general}
    \psi^1_j (x) &= e^{-i \delta t \, m_j V(x,t,\psi(t)) / \hbar / 2 -i\delta t \hbar^2 \sum_k \lambda_{jk} \, |\psi_k(x,t)|^2 / 8 m_j^2 } \psi_j(x,t) \, \, \, \mathrm{\textbf{[half-step kick update]}}  , \\
    \psi^2_j(p) &= e^{-i \delta t \frac{\hat p^2}{2m_j} / \hbar} \mathcal{F}\left[ \psi^1_j(x) \right](p) \, \, \, \mathrm{\textbf{[full-step drift update]}} , \\
    \psi^2_j (x) &= e^{-i \delta t \, m_j V(x,t,\psi(t)) / \hbar / 2 -i\delta t \hbar^2 \sum_k \lambda_{jk} \, |\psi_k(x,t)|^2 / 8 m_j^2 } \psi_j(x,t) \, \, \, \mathrm{\textbf{[half-step kick update]}}  ,
\end{align}
for each $\psi_j$. Where the gravitational potential is now given 
\begin{align}
    V(x,t,\psi) = \mathcal{F}^{-1}\left[ 4 \pi G \,  \frac{\mathcal{F} [\sum_k |\psi_k(x',t)|^2] (\vec k)}{k^2} \right](x) \, . \label{eqn:Poisson_k_space}
\end{align}

\textbf{Higher spins.} 
For a spin $s$ field, with self-coupling $\lambda$, we can write the equations of motion as in \eqref{eqn:SP_spin} \cite{Jain2023}. In that work, they write the kick and drift Hamiltonians respectively as
\begin{align}
    &[H_\mathrm{drift}]_{nn'} = \delta_{nn'} \left( \frac{-1}{2m} \nabla^2 \right) - i g_{ij} [\hat S_i]_{nn'} \nabla_j \, , \\
    &[H_\mathrm{kick}]_{nn'} = \left( m V - \frac{\lambda}{m^2} \rho \right) \delta_{nn'} -  \frac{\alpha}{m^2} \sum_i \mathbf{S}\cdot \hat{\mathbf{S}}_{ij} \frac{\xi}{2s+1} + \sum_{ik} \hat{A}_{ji} \psi^\dagger_i \psi_k(x,t) \hat{A}_{ki} \, .
\end{align}
Recall that a kick/drift update is given $e^{-i \delta t H_\mathrm{kick/drift} / \hbar}$. Note that the authors have here set $\hbar = 1$. $\delta_{ij}$ is the Kronecker delta. Constraints on the timestep are further discussed in \cite{Jain2023}.

\subsubsection{Numerical properties}

This method is generally considered one of the most accurate when simulating the non-relativistic classical field and can easily be extended to include solvers that interact with corpuscular matter or expanded to include other ultralight dark matter extensions, including self-interactions and multiple fields. The integrator is unitary to machine precision, and the leap-frog integrator is symplectic and accurate to $\mathcal{O}(\delta t^2)$. However, the method comes with a few numerical drawbacks. We summarize the resolution requirements for this solver in Table \ref{tab:contr_solver}.

In order to better understand how these properties impact actual simulations let us roughly consider a test problem of simulating a galaxy with characteristic mass $M$, length $L$, comprised of ultralight dark matter of mass $m$. Let us say that the typical velocity in this system is $v \sim  \sqrt{GM/L}$. The momentum aliasing condition described above requires that we resolve the length scale corresponding to the de Broglie wavelength associated with this velocity, i.e. $dx \sim \hbar /mv$. This implies then that the resolution of our simulation should be $N = L / dx$. Because we are in 3D the computational requires are going to scale as $N^3 \log N$ (where the logarithm comes from Fourier transforms when computing the momentum and potential). From this we notice a few different trends even before we start integrating. First, we are punished for simulating larger galaxies which increases both the characteristic length and velocity scales in the system requiring a larger resolution to simultaneously resolve the largest and smallest scales in the problem. Second, we are punished for increasing the field mass $m$ which decreases the associated de Broglie wavelength $dx$ at fixed characteristic velocity. So if we consider a single galaxy, the resolution $N \propto m$ still scales linearly with increasing field mass. 

The temporal conditions give similar scalings. The gravity kick operator will require a smaller timestep as gravitational potentials deepen. However, often we are in the limit where the kinetic energy limits our timestep. In this case, the timestep goes as $\delta t \propto m / v^2 \propto 1/m$. This means that increasing the size of the galaxy or increasing the field mass will both result in a smaller timestep. This combination of larger resolution requirements and a smaller timestep is why pushing to simulate larger galaxies and higher masses has been numerically challenging. 

\textbf{Spectral aliasing:} 
If we first look at the naive scaling of the update in equation \eqref{eqn:pseudoSpecUpdate}, we see the scaling goes as a Fourier transformation, which in $D$ spatial dimensions means the update goes as $\mathcal{O}(N^3 \log N)$. However, there is additional numerical complexity introduced by the time-step. The time-step in equation \eqref{eqn:pseudoSpecUpdate} involves updates of phases in position and momentum space. Because any phase $\theta \equiv \theta + 2 \pi$ identifies with itself mod $2 \pi$ we need to choose a timestep such that the phase changes are small compared to $2 \pi$. We can separate this as two conditions on the timestep from kinetic and potential components of the update, respectively.
\begin{align}
    \delta t \, \tilde \hbar k_\mathrm{max}^2 / 2 &\ll 2 \pi \, ,\\
    \delta t \, V_\mathrm{max} / \tilde \hbar &\ll 2 \pi \, .
\end{align}
Where $\tilde \hbar \equiv \hbar / m$ is a common definition used in ultralight dark matter simulations. $k_\mathrm{max} = \pi / dx = \pi N /L$ is the maximum wavenumber resolvable in the system. This potential aliasing of the timestep is generally referred to as ``spectral aliasing". Usually, the kinetic condition (the first line) is the more restrictive. To understand the restriction this places on simulations of ultralight dark matter, let us consider how the minimal resolvable scale needs to scale with the dark matter mass $m$. Assuming the physical parameters of interest remain the same, e.g., the velocity dispersion, maximum spatial scale, etc., but that we wish to run multiple simulations at different masses to study how some phenomena depend on field mass we have a deBroglie wavelength that goes as 
\begin{align}
    \lambda_\mathrm{db} = 2 \pi \tilde \hbar / \sigma \, \propto m^{-1}.
\end{align}
A straightforward resolution requirement is then that $dx \lesssim \lambda$. This means that if we are running a simulation at the resolution limit, to increase the mass we need to increase the number of grid cells, $N$, proportionally. This invokes a memory cost then that scales as $\mathcal{O}(m^3)$ and a runtime cost that scales as $\mathcal{O}(m^5 \log m)$. Therefore, in a typical parameter search, an increase in the mass by a factor of $2$ increases the memory requirements by a factor of $8$ and the runtime by a factor of $\sim 30$. For most applications, in which constraints exist in a logarithmic mass parameter space this is simply infeasible and generally other methods are preferred. 

\textbf{Kinetic aliasing:} 
Because much of this method involves representations in Fourier space, we should consider the Fourier space representation of the field
\begin{align}
    \psi(k_i) = \sum_j \psi(x_j) \, e^{ik_i x_j} \, .
\end{align}
Where $k_j \in [-\pi (N-1)/L, -\pi (N-2)/L, \dots, \pi N/L]$ is the $j$th wavenumber. And $x_j = -L/2 + j\, dx$ as before. If we consider two values of $k$ and $k'$ where $k \ne k'$ we would say that the two modes identify when represented on our grid if plane waves with these wavenumbers are equal everywhere on the grid. This will be true when 
\begin{equation} \label{test}
    e^{ikx_j} - e^{ik'x_j} = 0
\end{equation}
is satisfied for all $j \in [0,N)$. We first make the following change of variables 
\begin{align}
    k_{CM} &= (k + k')/2 \\
    \Delta k &= (k - k')/2 \, .
\end{align}
we can then rewrite equation \eqref{test} as 
\begin{align}
    e^{ik_{CM}x}(e^{i \Delta k x} - e^{-i \Delta k x}) &= 0 \\ 
    2i e^{ik_{CM}x} \sin(\Delta k x) &= 0
\end{align}
which implies that 

\begin{equation}
    \sin(\Delta k x_j) = 0,
\end{equation}

the easiest way for this to happen is for the argument at every grid point $x_j$ to be equal to some multiple of $\pi$, which we write $m j \pi$ for $m \in \mathbb{Z} \ge 0$. We can now write our equivalence condition as 

\begin{align}
    &\Delta k x_j = m j \pi \\
    &\Delta k L/N = m \pi \\
    &\Delta k = m \pi N / L \\
    &k-k' = m \frac{2\pi N}{L} \\
    &k = k' + m \frac{2\pi N}{L}
\end{align}

recall that our maximum momentum is given $k_{max} = \frac{\pi N}{L}$ and we find that 

\begin{equation}
    k \equiv k + 2 m k_\mathrm{max} .
\end{equation}

$2 m k_{max}$ is just the momentum width of our phase space, which implies that the maximum negative and positive momenta, $\pm k_{max}$, are identified, which implies that our phase space is periodic in momentum space.

\subsection{Mock halo generation}

In this section, we discuss different numerical methods for generating mock halos. Many of the phenomenological efforts discussed in this review study the properties of mock halos generated using one of the following methods.
\subsubsection{Eigenvalue decomposition} \label{sec:eigen_halos}
\begin{figure*}[!ht]
	\includegraphics[width = .97\textwidth]{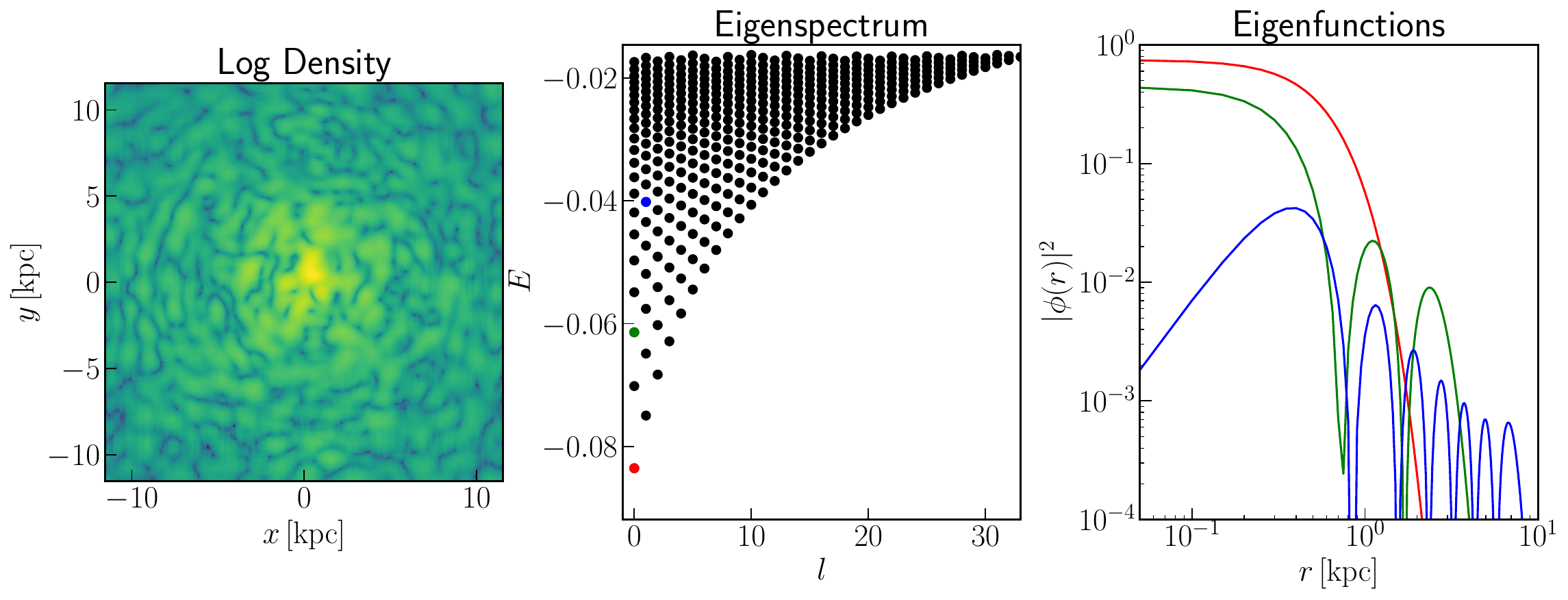}
	\caption{Eigenspectrum of halo generated using eigenvalue decomposition method. \textbf{Left:} log density slice of an ultralight dark matter generated using eigenvalue decomposition. \textbf{Center:} spectrum of included eigenmodes, energy on y-axis, and angular momentum quantum on x-axis. \textbf{Right:} A few example eigenfunctions. We note the ground state eigenfunction (red) is consistent with the cored density profile, equation \eqref{eqn:solitonicCore}. The number of $0$ crossings in the eigenmode is equal to its energy quantum number, and modes with non-zero angular momentum go to $0$ at $r=0$, consistent with our expectations from similar eigenvalue problems in quantum mechanics (e.g., the spectrum of the hydrogen atom). }
	\label{fig:spectrum}
\end{figure*}

The eigenvalue decomposition method \cite{Zagorac_2022, Yavetz2022, Yang2025} relies on the fact that the halo profile is well described by semi-analytic models. Therefore, if we construct a halo as a superposition of the eigenvalues of the predicted halo profile, the resulting halo should be approximately stable on the largest scales but will correctly model the nonlinear dynamics of the de Broglie scale effects, e.g., granules or soliton oscillations. 

We start by identifying a ``target" radial density profile, $\rho_t(r)$, which may be a cored NFW or other interesting profile. The specific target profile is arbitrary; however, using a target profile that is too cuspy or has discontinuities in density or derivatives may require a large number of eigenmodes to reproduce, and so it is advisable to pick a profile without these problematic features. The target density profiles give us a Hamiltonian via the radial component of the Schr\"odinger-Poisson equations
\begin{align}
    H_l = -\frac{\hbar^2}{2m} \nabla_r^2 + m V_t(r) + \frac{\hbar^2}{2m} \frac{(l+1)l}{r^2} \, , &&
    \nabla_r^2 V_t(r) = 4 \pi G \rho_t(r) \, .
\end{align}
Where $l$ is the angular momentum quantum number. The eigenvalue problem associated with this Hamiltonian can be solved, and the eigenvectors, $\phi_n^l$ found
\begin{align}
    H_l u_n^l = E_n^l u_n^l \, .
\end{align}
Here, the functions $u$ are related to the eigenvectors as $u = r\phi$. The eigenvectors are then given weights, $w_j$, such that their sum best approximates the target profile, i.e. 
\begin{align}
    \rho_t(r) &\approx \sum_l^{l_{max}} \sum_{m = -l}^{l} \sum_{n}^{e_{max}} w_n \, |\phi_n^l(r)|^2 \, .
\end{align}
Values of $E_i$ and $l$ included should be sufficiently large as to describe the dynamics of the halo. The final field is composed of the eigenfunctions summed using the weights and a random phase, i.e.
\begin{align}
    \psi(r, \theta, \varphi) = \sum_l^{l_{max}} \sum_{m = -l}^{l} \sum_{n}^{e_{max}} w_j \, Y^m_l(\theta, \varphi) \, \phi_n^l(r) \, e^{-i \, \omega_{lmj}} \, .
\end{align}
$\omega_{lmj}$ is chosen uniformly and randomly from $[0,2 \pi)$. An example of an ultralight dark matter halo generated using this method and the corresponding eigenspectrum is plotted in Figure \ref{fig:spectrum}.

\subsubsection{Soliton collisions}
Mergers of solitons have also been used to study halo properties \cite{Schwabe2016, Mocz2017, Zagorac_2022}. Generally, this works by populating a simulation box with multiple solitons, allowing them to merge, and then studying the final steady-state product. And while these confirm the conclusion that all halos should host a central core, they do not recover the core-halo mass scaling relation observed in cosmological simulations \cite{Schwabe2016, Mocz2017, Schive2014, Zagorac_2022}. A study of comparing the solitons resulting from different merger methods was carried out in Refs. \cite{Zagorac_2022, Blum_2025} which concluded that the collision method provided an upper bound on the soliton mass.

\subsection{Approximation schemes}
\subsubsection{Eigenvalue solvers} \label{sec:eigen_solvers}
It is always possible to write the field in terms of its energy eigenbasis, for eigenfunctions $\set{\phi_i}$ and values $\set{E_i}$ which solve
\begin{align}
    H \phi_i = E_i \phi_i \, .
\end{align}
Where $H$ is the Hamiltonian. If we assume that the ultralight dark matter is near equilibrium and that the energy distribution and eigenfunctions are not going to change much over the time period of interest, then we can use a spectral method to solve for the ``steady-state" evolution of the field. This is often useful as it can be less computationally intensive than a full spectral solver but still capture nonlinear dynamics like granular oscillation \cite{Dalal2021}. 

This works by populating the field with eigenvalues and then evolving their complex phase, i.e. 
\begin{align}
    \psi(x,t) = \sum_i \sqrt{f(E_i)} \, \phi_i(x) \, e^{-iE_it/\hbar + \theta_i} \, .
\end{align}
Where $f(E_i)$ is the energy distribution function, $\theta_i$ is an initial random phase. 

This method makes the assumption that the system is in a steady state, that is, that the energy distribution and eigenvectors do not change over the course of the evolution. This method has been employed to study collapsed and virialized halos where this approximation is thought to be true to a good approximation \cite{dalal2022,Dalal2021}. This method would, however, be inappropriate for studying systems where the eigenvalues/eigenvectors are not constant, for example a collapsing system.

\subsubsection{N-body schemes} \label{sec:num:N-body_approx}
Ultralight dark matter alters the initial matter density power spectrum. This effect is usually quantified using the transfer function discussed in Section \ref{subsec:transferFunction}. Simulating these modified initial conditions using traditional cold dark matter schemes allows a first-order approximation of the effect of these altered initial conditions. This idea is supported by the fact that on scales well above the de Broglie wavelength, the evolution of ultralight and cold dark matter agrees (see Section \ref{subsec:cdmCorrespondence}). The method misses the nonlinear effects captured by pseudo-spectral solvers; however, as long as our analysis is only concerned with relatively large-scale objects, such as the halo mass function, this method has generally been considered an accurate approximation. The largest comparison of N-body and Schr\"odinger-Poisson simulations is in \cite{May2022}, which generally corroborates the conclusions above.

\section{Observational constraints} \label{sec:constraints}

\begin{table*}
\begin{center}

 







\begin{tabular}{| c | c | c |}

\hline
Phenomena & Observations & Systematics \\

\hline \hline

\rule{0pt}{10pt} Transfer func. & (sub) halo-mass func., Lyman-alpha & emulator/(sub-)halo modeling \\[2ex]
\hline

\rule{0pt}{10pt} Solitons/cores & dwarf galaxy stellar dispersions & $M_c - M_h$ relation, core oscillation \\[2ex]
\hline

\rule{0pt}{10pt} Dyn. Heating & dwarf galaxy stellar dispersions & halo history, stellar potential \\[2ex]
\hline

\rule{0pt}{10pt} Rel. pressure & pulsar timing & pulsar variability \\[2ex]
\hline

\rule{0pt}{10pt} Granules & strong lensing & lens model \\[2ex]
\hline

\rule{0pt}{10pt} BHSR & blackhole spin & spin measurements \\[2ex]
\hline

\end{tabular}
\caption{\label{tab:pheno_constraint} Most common constraint types separated by related phenomena. Relevant observations and the most common systematic uncertainties are listed with prominent references. The first row refers to the impact on large-scale structure growth of the results from the way ultralight dark matter alters the transfer function of the linear density fluctuations. This is probed by observations attempting to measure the halo (or sub-halo) mass function \cite{Nadler_2025, Nadler2021, Garland2024, Sipple2024, Schutz2020,Ni2019,Corasaniti2017,winch2025, Schive2016} (UVLF stands for ultraviolet luminosity function, MW for Milky-Way) or the scales of fluctuations in the Lyman-alpha forest \cite{Rogers:2020ltq}. The main uncertainty with these methods is modeling the relationship between the numerical results and actual observations. The second row refers to constraints related to the solitonic cores predicted by ultralight dark matter \cite{Bar2018,Bar2019,Bar2022,zimmermann2025}. These can be compared with observations of galaxies' stellar dynamics. The halo mass-core mass relation is inferred from cosmological simulations at lower masses. The main uncertainty in these constraints is the accuracy of this relation (which has been demonstrated to be sensitive to simulation parameters \cite{Schwabe2016, Mocz2017, Schive2014, Zagorac_2022}) and the non-static behavior of the central core \cite{Schive2020}. The third row refers to constraints related to the dynamical heating of stellar distributions by ultralight dark matter granules \cite{Church2019, Marsh2019,dalal2022,teodori2025}. The predicted heating can be compared with the relevant structure (dwarf galaxies, star clusters, discs, etc) observed parameters. The main uncertainty is the history of the observed object, which may change the assumptions about past heating rates. The fourth row is the pulsar timing constraint \cite{Porayko2018} in which the relativistic pressure of the ultralight field creates timing delays. The largest uncertainty is the modeling of the pulsar intrinsic variation. The fifth row is the strong lensing constraint \cite{Powell2023} in which ultralight dark matter granules would alter the image of a lensed radio jet. The largest uncertainty is the lens model. The final row is black hole superradiance \cite{Davoudiasl2019, Stott2018} in which a blackhole loses its spin to existing ultralight fields with Compton wavelength of order the horizon scale. The largest uncertainties are those associated with accurately measuring the black hole spin.  }
\end{center}
\end{table*}

\begin{figure*}[!ht]
	\includegraphics[width = 1.0\textwidth]{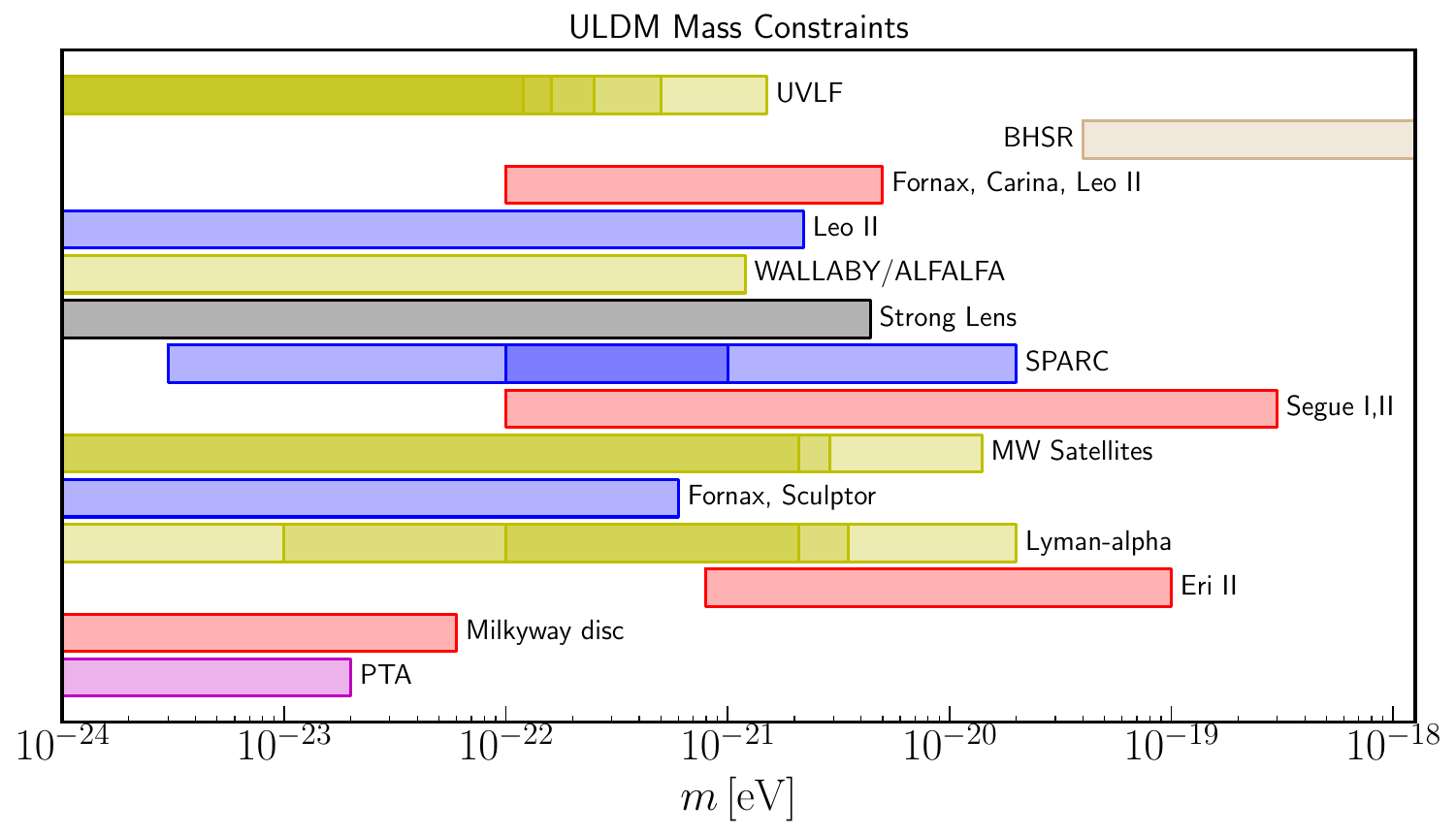}
	\caption{ Constraints on the mass of the ultralight dark matter particle assuming that all of dark matter is described by a single, classical, spin-0, ultralight field. Constraint types are grouped by color, red corresponds to dynamical heating constraints (Milkyway disc \cite{Church2019}; Eri II \cite{Marsh2019}; Segue I and II \cite{dalal2022}; Fornax, Carina, and Leo II \cite{teodori2025}), blue to constraints related to solitonic cores and stellar dynamics (Leo II \cite{zimmermann2025}, Fornax and Sculptor \cite{Safarzadeh2020}, SPARC catalog \cite{Bar2018,Bar2019,Bar2022}), yellow to constraint related to the halo mass or transfer functions (Ultraviolet luminosity function \cite{Sipple2024,Ni2019, Corasaniti2017, Schive2016, winch2025}, Milky-way satellites \cite{Nadler_2025, Nadler2021}, WALLABY/ALFALFA \cite{Garland2024}, Lyman-alpha \cite{Rogers:2020ltq,Kobayashi2017, Armengaud2017}), grey to the strong lens constraint \cite{Powell2023}, brown to black hole superradiance \cite{Davoudiasl2019,Stott2018}, and magenta to the pulsar timing array constraint \cite{Porayko2018}. }
	\label{fig:constraints_all}
\end{figure*}

\begin{figure*}[!ht]
	\includegraphics[width = 1.0\textwidth]{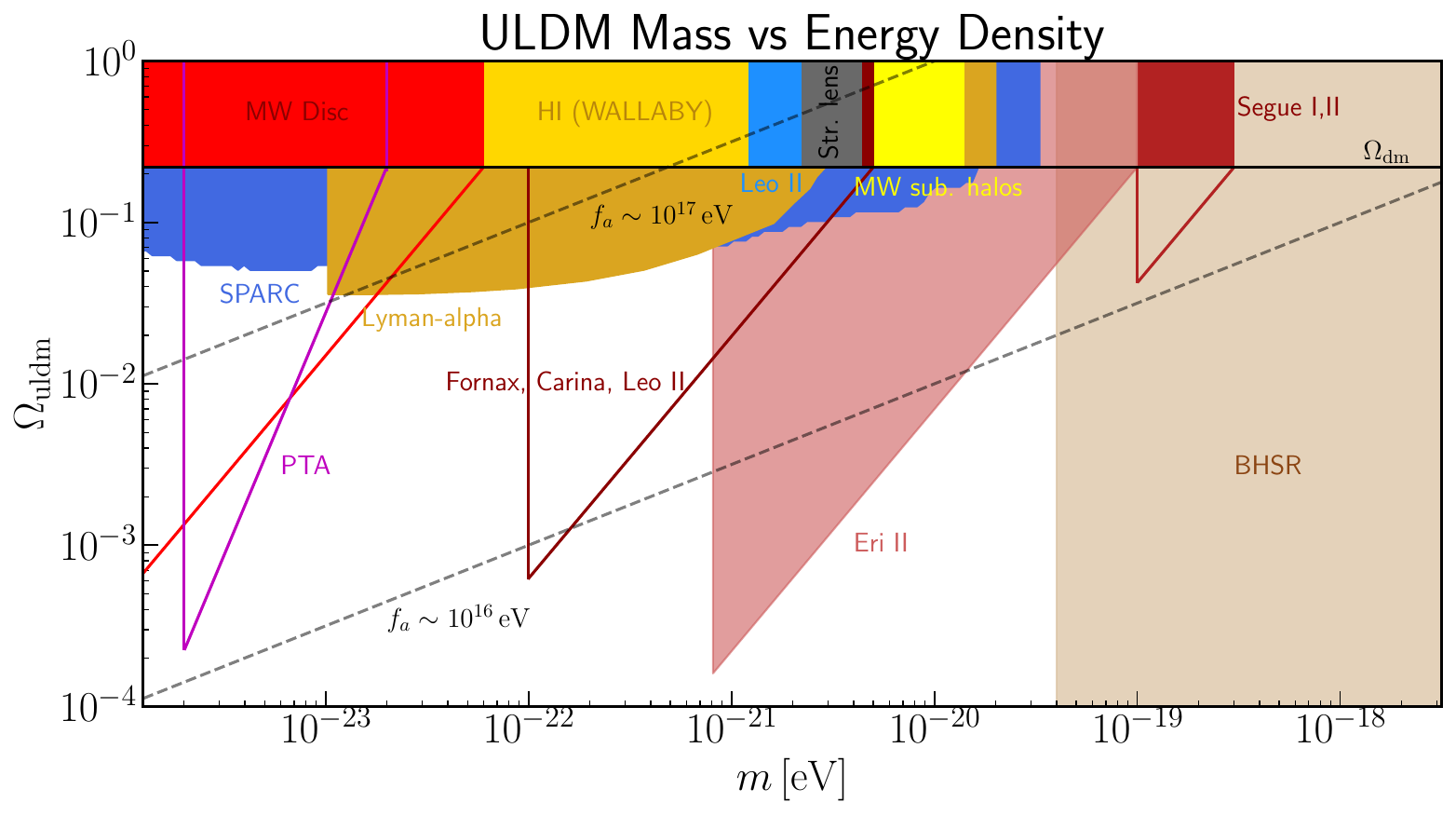}
	\caption{ Ultralight dark matter constraints as a function of the mass and density. Hollow lines indicate prospective work or work extended to include other regions of parameter space by scaling arguments without explicit testing. We note that not all constraints are equally rigorous. Many constraints are originally presented as lower bounds, and so identification of the left edge of each constraint is often ambiguous, see related discussions in Section \ref{sec:constraints}. Horizontal black line corresponds to current estimate of dark matter density, $\Omega_\mathrm{dm}$ \cite{Planck2020}. Dashed black lines correspond to the abundance for a given mass and decay constant $f_a$. Constraint types are grouped by color, red corresponds to dynamical heating constraints \cite{Church2019, Marsh2019, dalal2022, teodori2025}, blue to constraints related to solitonic cores \cite{zimmermann2025, Bar2018,Bar2019,Bar2022, Safarzadeh2020}, yellow to constraint related to the halo mass or transfer functions  \cite{Garland2024,Nadler2021, Nadler_2025, Rogers:2020ltq, Kobayashi2017, Irsic2017, Nori2019, Schutz2020, Sipple2024, winch2025,Ni2019, Corasaniti2017, Schive2016} (HI stands for Hydrogen intensity, MW for Milky-way, UVLF for ultraviolet luminosity function), gray to the strong lens constraint \cite{Powell2023}, brown to black hole superradiance \cite{Davoudiasl2019,Stott2018, Saha2022}, and magenta to the pulsar timing array constraint \cite{Porayko2018}. We present here the constraints as a function of the fraction of ultralight dark matter cosmic energy density, however, constraints related to individual galaxies (i.e., PTA, dynamical heating, and solitonic core) are technically only directly measuring the fraction of ultralight dark matter in the observed galaxy which some results indicate may differ from the total cosmic fraction \cite{luu2024}. Hollow dynamical heating curves are extended below $\Omega_\mathrm{dm}$ using the density-mass scaling relation for dynamical heating argued in \cite{Hui2017, Gosenca2023, teodori2025, dalal2022, Church2019}; in none of these constraints were mixed model simulations actually simulated. SPARC \cite{Bar2018,Bar2019,Bar2022} and Lyman Alpha \cite{Kobayashi2017} data provided by \href{https://keirkwame.github.io/DM_limits/}{Kier Rogers} via the respective cited collaborations. The code for making this plot is available at \href{https://github.com/andillio/FDM_Constraints}{FDM\_Constraints}. }
	\label{fig:constraints_m_vs_f}
\end{figure*}

In this section, we discuss current observational constraints on ultralight dark matter. Many of these have been originally phrased as lower mass bounds on the minimally coupled one classical ultralight field (vanilla) model. In each subsection, we will briefly describe the relevant phenomenology and the specific observation or dataset used to produce the constraint. We will then discuss the major systematic uncertainties and work done exploring how these might affect the constraint. Finally, for each, we will discuss current work extending these constraints to other nonstandard ultralight dark matter systems, e.g., multiple fields, higher spin fields, mixed dark matter models, etc. 

Table \ref{tab:pheno_constraint} contains a summary of the information in this section. Likewise, Figure \ref{fig:constraints_all} lists constraints organized by observable, phenomenology, and numerical method. 

Figure \ref{fig:constraints_m_vs_f} plots constraints as a function of both mass and cosmic energy density, $\Omega_\mathrm{uldm}$. The black dashed line indicates where ultralight dark matter comprises all the dark matter. Often, work was done assuming this to be the case, and the extension to other fractions is not obvious, and so many constraints live at and above this line, not because of a fundamental limitation of the method but because no work has yet been done to extend the constraint to other fractions. In this plot, we color-code constraints based on the related observation/phenomena. The main takeaway from this plot should be that the entire ultralight dark matter parameter space is ruled out as all the dark matter for masses $m \lesssim 10^{-18} \, \mathrm{eV}$ and for masses $m \lesssim 10^{-20} \, \mathrm{eV}$ by multiple complementary methods.  

Red constraints are related to dynamical heating of individual galaxies \cite{Church2019, teodori2025, Marsh2019, dalal2022}. The hollow curves relating to dynamical heating constraints that are extended below $\Omega_\mathrm{dm}$ are based on the relationship between mass and dark matter density/fraction argued in \cite{Hui2017, Church2019, dalal2022, Gosenca2023, teodori2025}, which is $\propto m^{3/2}$. In no case were mixed ultralight dark matter simulations actually run. The left edge of each extended line in these cases corresponds to the lowest mass simulation run in the corresponding work, as opposed to a fundamental limitation of the constraint. Often, some assumptions break down below this mass (e.g., the core radius approaches the half-light radius). We note that the mass-density scaling argued in the Eri II constraint \cite{Marsh2019} is $\propto m^3$, the left edge in this case identified analytically. We also note that in these constraints, technically, what is being measured is the ultralight dark matter density in the observed galaxy, not necessarily the cosmic energy density. Simulation work has pointed out that these two may differ \cite{luu2024}, though it is not immediately clear to what extent. 

Blue constraints are related to cores of galaxies and related rotation curves \cite{Bar2018,Bar2019,Bar2022,zimmermann2025}. Again, an important note here is that the constraint directly measures the ultralight dark matter present in specific galaxies, which may differ from the cosmic energy density in a scenario with multiple dark matter species \cite{luu2024, pozo2024}.

Yellow constraints are those relating to the transfer function and its impact on the halo mass function \cite{Nadler_2025, Nadler2021, Garland2024} or Lyman-Alpha \cite{Kobayashi2017,Rogers:2020ltq} observations. These constraints are sensitive to $\Omega_\mathrm{uldm}$ directly as opposed to the density in specific galaxies. Again, the left edges of constraints and those not extended into lower fractions represent existing work and not fundamental limitations of the probes.

The pulsar timing constraint \cite{Porayko2018}, plotted in magenta, is sensitive to the ultralight dark matter fraction in the Milky Way specifically. The left edge here corresponds to the mass with Compton time approximately equal to the longest pulsar observation time, i.e., $\tau_c \sim 30 \, \mathrm{yrs}$. The scaling of the right edge is $\propto m^{3}$, see arguments made in \cite{Porayko2018, Khmelnitsky2014}.

Hollow curves for the dynamical heating and pulsar timing array constraints do not extend past $\Omega_\mathrm{dm}$ because independent measurements bound the total amount of mass/dark matter in the observed galaxies, irrespective of the ultralight dark matter contribution to the cosmic energy density. 

It is important to note that many of the bounds discussed below depend on modeling assumptions and astrophysical uncertainties, including baryonic effects, halo structure, population-level diversity, and observational systematics. Measurement errors and assumptions in the comparison between theoretical predictions and data can also affect the inferred constraints. In addition, limitations in theoretical modeling and numerical implementations may impact the precision of the predictions. It has been shown in several cases that accounting for these effects can significantly broaden the allowed parameter space, shift the preferred mass ranges, or even weaken individual bounds derived under more restrictive assumptions. Therefore, the constraints should be interpreted with appropriate caution.

\subsection{Halo mass functions} \label{sec:halo_mass_funcs}

In this section we discuss constraints related to the halo mass function (and observational proxies). In the early universe, the ``quantum" pressure of the ultralight dark matter field resists the collapse of small scales. This effect leads to a transfer function that suppresses small-scale modes in the initial matter density power spectrum, which is discussed in Section \ref{subsec:transferFunction}. This means that ultralight dark matter then predicts a different halo mass function than its cold dark matter counterpart. The impact on the halo function mass function is usually studied using N-body simulations with initial conditions altered to be consistent with the ultralight dark matter transfer function, see Section \ref{sec:num:N-body_approx}. 

The halo mass functions of these altered simulations can be compared with observations of low mass halos. Generally, this will mean that the largest uncertainties in this method are using the simulations to make predictions which can accurately be compared with observational data. 

\subsubsection{Milky-way subhalos} \label{subsec:MW_subhalos}

Analysis of Milky Way-like halos and their satellites in modified N-body simulations was used to constrain the mass of the ultralight dark matter particle in \cite{Nadler2021, Nadler_2025}.

In \cite{Nadler2021}, estimates of the subhalo mass function associated with ultralight dark matter were constructed analytically using results of simulations in \cite{Du2018} and analytic transfer functions from \cite{Hu2000}. This was compared with observations of the number of Milky Way satellites predicted using models of the subhalo mass function and a census of MW satellites \cite{Drlica-Wagner2020} from the Dark Energy Survey \cite{Abbott2018} and Pan-STARRS1 \cite{chambers2019}. The resulting constraint was set at $2.9\times 19^{-21} \, \mathrm{eV}$ at $95\%$ confidence. 

The constraint is presented as a lower bound with the fiducial model used to produce the analytic model valid for $m \ge 10^{-21} \, \mathrm{eV}$ and halo masses $M \ge 10^{8} \, M_\odot$.

This method was updated in \cite{Nadler_2025} which used the COZMIC suite of simulations, which simulated 6 ultralight dark matter masses between $25.9 \, m_{22}$ and $490 \, m_{22}$ using modified transfer functions produced with axionCAMB \cite{Hlozek2015}. The resulting constraint was set at $m > 1.4 \times 10^{-20} \, \mathrm{eV}$ at $95\%$ confidence. 

The subhalo mass function was also studied in \cite{Schutz2020}, in which the perturbation of stellar streams and quasar lensing was used to infer the subhalo mass function. Originally, these observations were used to place constraints on warm dark matter \cite{Banik2021, Gilman2019}, which suppresses the halo mass function at low masses in a similar way to ultralight dark matter. The corresponding ultralight dark matter mass with the same suppression at low halo mass scales was computed and used to produce a constraint on the ultralight dark matter mass. The suppression of the transfer function was calculated using axionCAMB \cite{Hlozek2015}. It was found that masses around $m \lesssim 2.1 \times 10^{-21} \, \mathrm{eV}$ were inconsistent with data.  

There are several systematic uncertainties associated with this analysis. The modeling of the subhalo mass function relies on simulation modeling which may admit inaccuracies (see for example Ref. \cite{Mansfield2024}) and then is compared to observational surveys which have their own sources of uncertainties \cite{Nadler2024_dwarfs}. If halo finders fail to track certain subhalos this biases constraints towards higher mass. In these calculations ultralight dark matter dynamics only enter at the level of the transfer function. The nonlinear dynamic impact on small-scale structures is therefore not included. Large-scale studies comparing N-body and ultralight dark matter simulations, however, seem to corroborate the accuracy of this approximation\cite{May2021, Gough2024}. However, it has been demonstrated that tidal stripping can be significantly altered in the ultralight dark matter scenario \cite{Du2018} which observed a runaway effect. Baryons are also expected to reduce the subhalo mass function via enhanced tidal stripping by the galactic disk \cite{Wang_2025, Nadler_2025, Garrison-Kimmel2017}. This means the inclusion of nonlinear or non-DM physics could results in even fewer halos meaning excluding this physics biases constraints towards lower masses. 

\subsubsection{HI observations}

Similar work was done using simulation-informed halo mass functions and comparisons with those observationally inferred from HI surveys, ALFALFA, and WALLABY in \cite{Garland2024}. The $M_\mathrm{HI}-M_h$ relation is inferred from simulations \cite{Villaescusa-Navarro2018}. As a result, the largest source of systematic uncertainty was determined to be whether simulated data may predict an overabundance of HI content in low-mass galaxies. The inferred bounds were $m > 3.2 \times 10^{-22} \, \mathrm{eV}$ from ALFALFA \cite{Haynes2018} and $1.3 \times 10^{-21} \, \mathrm{eV}$ from WALLABY \cite{Koribalski2020}. This study included ultralight dark matter masses within the range $10^{-24} \, \mathrm{eV} \le m \le 10^{-20} \, \mathrm{eV}$.

\subsubsection{Ultraviolet luminosity function}

The ultraviolet luminosity function was studied \cite{Sipple2024, winch2025, Ni2019, Corasaniti2017,Schive2016}. Observations of high redshift galaxies are compared with the expected suppression of the halo mass function. The modeling of the halo mass function is generally done semi-analytically. Data is taken from the Hubble frontier fields, producing a constraint of $1.5 \times 10^{-21} \, \mathrm{eV}$. In addition to all dark matter studies, it was also determined that ultralight dark matter with masses between $10^{-26} \, \mathrm{eV} \lesssim m \lesssim 10^{-23}$ must account for less than $0.22$ fraction of the dark matter \cite{winch2025}. The main systematic uncertainty in these studies comes from the modeling of the halo luminosity function.

\subsection{Galaxy density profiles} \label{subsec:GalaxyDensityProfiles}

A highly general result of ultralight dark matter cosmological simulations is the presence of a core at the center of halos, see for example \cite{Schive2014, Schive2014_CoreHalo, Mocz2017}. Indeed, this is one of the original motivations for this model \cite{Hu2000}. Because the core simply represents the mass in the lowest energy modes of the system and its size is determined by the uncertainty principle \cite{Hui2021}, it is generally considered reasonable to assume that every ultralight dark matter halo should host some sort of core. However, it is also worth noting that it has been argued that the cores produced by ultralight fields are the wrong shape to describe data regardless of the field mass \cite{Deng2018}. We discuss the specifics of these cores in Section \ref{subsec:Pheno:solitons}. 

\subsubsection{SPARC catalog}

Comparisons of the rotation curves predicted for these cores and the observed rotation curves are used to place constraints on the ultralight dark matter mass in \cite{Bar2018, Bar2022}. The central argument made is that the specific kinetic energy in the soliton and the halo is the same \cite{Bar2018, Bar2019}, i.e. 
\begin{align}
    K/M_s \approx K/M_\mathrm{halo} \, .
\end{align}
The empirical scaling relations fitted to simulation results in \cite{Schive2014, Schive2014_CoreHalo} are used to make predictions for the expected rotation curves in galaxies assuming that the soliton is central and stationary in their respective halos. Comparison of these predictions for galaxy rotation curves in disc galaxies in the SPARC catalog \cite{Lelli2016}. The most recent analysis of the data disfavors ultralight dark matter from comprising all the dark matter in the mass range $3 \times 10^{-24} \, \mathrm{eV} \lesssim m \lesssim 2 \times 10^{-20} \, \mathrm{eV}$ \cite{Bar2022}.

This method admits a number of potential caveats. These caveats arise from the fact that the predicted cores are extrapolated from empirical fitting relations in dark matter only simulations \cite{Schive2014, Schive2014_CoreHalo} and then assumed to be central and stationary in their respective halos. However, it has been demonstrated that the presence of Baryons can impact the shape of the central core \cite{Chan2018, Veltmaat2019}, meaning that the fitting relation may not represent realistic dark matter cores. Additionally, the simulations used to derive the empirical halo-core mass scaling relations themselves only cover a specific region of mass and resolution parameter space. The constraints derived in \cite{Bar2018, Bar2022} go beyond the parameter space explored in these simulations and are therefore subject to potential extrapolation errors. Additionally, resolution constraints limit the range of halo masses, co-moving volumes, and redshifts studied in the original simulations, again producing a potential additional source of extrapolation error. Finally, the ``random walk" \cite{Schive2020} and oscillation \cite{Marsh2019} of the central core produce a time-varying potential which can be used to predict orbital parameters of stars from what would naively be predicted for a central stationary soliton. All of these potential sources of systematic uncertainty exist entirely within the more basic one classical field ultralight dark matter model. Multi-field studies \cite{pozo2024,luu2024} have demonstrated that a diversity of cores may be possible in scenarios where multiple species may be relevant. In such circumstances a population of satellites with small scale cores would not necessarily rule out the existence of an ultralight field as a component of the dark matter.

The shape of the solitonic core is sensitive to most extensions of the basic ultralight dark matter model. Additional fields of differing masses have been shown to produce cores that are not simple superpositions of the one field mass scaling relations \cite{Luu2023,luu2024}. Likewise, field self-interaction and interaction with other fields can affect the core profiles \cite{Mirasola2024}. 

\subsubsection{Milky-way dwarf satellites}

A study comparing the density profiles of Milky Way dwarf satellites was carried out in \cite{Safarzadeh2020}. Using the core mass relation derived in \cite{Schive2014} it was argued that the mass of the satellites (Fornax, Sculptor) necessary to produce cores consistent with observation would produce satellites too heavy to survive under dynamical friction (dynamical friction has been studied in the context of ultralight dark matter in \cite{Glennon2023b, Gorkavenko2024, Lancaster2020, Wang2022, Boey2024,Foote2023}). They concluded that $m < 6 \times 10^{-22} \, \mathrm{eV}$ was incompatible with data.

\subsubsection{Ultra-faint dwarfs}

Ultra-faint dwarf (UFD) galaxies are ideal laboratories for testing dark matter models, as they are among the most dark matter–dominated systems known in the Universe, thus less affected by baryonic feedback.

In~\cite{Hayashi2021}, a full Jeans analysis was performed using stellar kinematic data from 18 UFD galaxies in the Milky Way to constrain the mass of fuzzy dark matter (FDM). A Bayesian approach was adopted, fitting the FDM core profile given in (\ref{eq:full_FDM_profile}). For this work, the fitting function (\ref{eqn:solitonicCore}) for the core-halo relation was used.

This study did not derive exclusion bounds, but rather obtained posterior distributions for the FDM mass from individual systems. Most UFDs in the sample prefer relatively high FDM masses, indicating that their central cores are small. For instance, Segue 1 yields a posterior peak at $m = 1.1^{+8.3}_{-0.7} \times 10^{-19}\,\mathrm{eV}$ (1$\sigma$ credible interval). However, dwarf galaxies exhibit diverse inner density profiles, and interpreting any individual system as a precise measurement of the particle mass can be misleading~\cite{Chan:2021bja}.  In particular, allowing for diversity in the core--halo relation broadens the range of particle masses compatible with dwarf galaxy kinematics and can lead to multiple viable mass intervals \cite{Wardana:2026qag}. 

These results highlight the power of UFDs for probing dark matter: since their stellar mass and baryonic content are minimal, uncertainties from baryonic effects are significantly reduced. Future improvements in stellar velocity measurements, particularly at both small and large radii, are expected to sharpen these constraints considerably and test the viability of the FDM model at small scales.

\subsubsection{Leo II}

A comparison of observations of stellar kinematic data in Leo II and ultralight dark matter halos constructed via the eigenvalue decomposition method (see Section \ref{sec:eigen_halos}) was performed in \cite{zimmermann2025}. Data used to derive Leo II stellar kinematics is described in \cite{Spencer2017}. This data is compared with 5000 eigenvalue constructed halos with masses between $1.5 \times 10^{-21} \, \mathrm{eV} \le m \le 2.4 \times 10^{-21} \, \mathrm{eV}$. The resulting analysis concluded that the ultralight dark matter mass must be at least $m > 2.2 \times 10^{-21} \, \mathrm{eV}$ to be consistent with data. 

This result provides a more rigorous approximation of the bound associated with the central core of Leo II. They note the bound achieved by simply asserting that the core be less than the virial radius is $2 \times 10^{23} \, \mathrm{eV}$ \cite{zimmermann2025}.

\subsection{Lyman alpha forest} \label{sec:lymanAlpha}
As we discussed in Section~\ref{subsec:transferFunction}, ultralight dark matter suppresses the growth of structure on small scales. We need probes of the power spectrum on small scales so we can constrain this suppression. The Lyman-$\alpha$ forest arises from the absorption of quasar light by intervening clouds of neutral hydrogen along the line of sight. This makes it a sensitive probe of the matter power spectrum on small scales of $0.5 - 100, \, \mathrm{Mpc}/h$, where ultralight dark matter suppresses structure formation for the relevant mass range of current constraints\footnote{More precisely, the observable is the one-dimensional flux power spectrum, which is related to the three-dimensional matter power spectrum through the distribution and thermal state of the intergalactic medium (IGM), redshift-space distortions, and the nonlinear transformation between gas density and transmitted flux.}.

One of the most stringent constraints on fuzzy dark matter mass comes from Lyman-$\alpha$ forest observations. In~\cite{Rogers:2020ltq}, a reanalysis using a Bayesian emulator trained on simulations yields a lower bound of $m \gtrsim 2 \times 10^{-20}\,\mathrm{eV}$, strongly disfavoring the canonical $10^{-22}\,\mathrm{eV}$ mass typically invoked to address small-scale structure issues. This result confirms earlier findings~\cite{Irsic2017,Armengaud2017,Kobayashi2017}, while reducing interpolation biases and marginalizing over physically consistent intergalactic medium (IGM) models.

However, the constraints depend on IGM modeling. Fluctuations in temperature and ionization during reionization may alter the Lyman-$\alpha$ flux power spectrum, and while~\cite{Rogers:2020ltq} accounts for this, the full impact remains uncertain~\cite{Hui2017}. Upcoming surveys and independent probes, such as 21-cm cosmology, may help test these results and improve our understanding of the small-scale matter power spectrum.

\subsection{Strong lensing} \label{sec:strong_lensing}

Strong gravitational lensing is a powerful probe of the small-scale structure of matter and, in particular, of the nature of dark matter. Because lensing is purely gravitational, it is sensitive to the total mass distribution, independent of baryonic content, and can therefore be used to test dark matter models such as fuzzy dark matter.

Lensing is sensitive to a few characteristics of the ULDM model: the solitonic core at the halo center and granular density fluctuations on sub-kiloparsec scales due to interference. While the overall suppression of small-scale power in FDM leads to a reduced abundance of low-mass subhaloes (similar to warm dark matter), the granularity of FDM haloes introduces a unique form of structure, distinct from CDM. From a lensing perspective, this results in two competing effects: a reduction in the number of subhaloes that can perturb lensed images, and the emergence of granular fluctuations that can induce novel lensing signatures. 

In systems with compact sources (e.g., quadruply imaged quasars), FDM-induced interference patterns in the halo density can mimic or suppress the perturbations typically attributed to subhaloes. In~\cite{Laroche:2022pjm}, a sample of eleven such lenses was analyzed and showed that the granularity arising from FDM can significantly affect flux ratios. Their results disfavour FDM particle masses below a few $10^{-21.5}\,\mathrm{eV}$ at the $95\%$ confidence level, since lighter masses cannot reproduce the observed image anomalies.

Interestingly, in~\cite{Amruth2023} using the quadruply lensed system HS 0810+2554, it was shown that lens models based on ULDM with $m \sim 10^{-22}\,\mathrm{eV}$ successfully reproduce both the positions and fluxes of the observed images, including anomalies that CDM models fail to explain, showing a preference for ULDM over CDM.

In the case of extended sources, high-resolution observations are required to resolve subtle image distortions. Powell et al. (2023), using Very Long Baseline Interferometry of the lens system MG J0751+2716, derived a lower bound on the FDM mass of $m > 4.4 \times 10^{-21}\mathrm{eV}$ at $95\%$ confidence. This is currently the strongest lensing-based constraint on FDM, and more stringent than those derived from larger samples of lower-resolution data.

These results highlight the potential of strong lensing as a clean and complementary test of FDM, with future improvements in angular resolution and sample size expected to further tighten constraints.

In addition to lensing-based constraints on halo structure, strong lensing systems can also probe axion-like particles through their coupling to photons. In \cite{Basu:2020gsy}, a novel method was introduced to constrain ALPs using differential birefringence—a frequency-dependent rotation of the polarization angle between multiple lensed images. Using broadband polarization observations of the lens system B1152+199 from the Karl G. Jansky Very Large Array (VLA), they derived upper bounds on the ALP-photon coupling in the range $g_{a\gamma} \lesssim 9.2 \times 10^{-11}$–$7.7 \times 10^{-8},\mathrm{eV}^{-1}$ for ALP masses between $m_a = 3.6 \times 10^{-21},\mathrm{eV}$ and $4.6 \times 10^{-18},\mathrm{eV}$ at $95\%$ confidence. This demonstrates that polarization information from strong lensing systems can serve as an independent and complementary probe of ultralight axions and their interactions.

\subsection{Heating of stellar dispersions} \label{sec:heating}

It has been shown analytically \cite{Marsh2019, Church2019} and numerically \cite{dalal2022, Chowdhury2023, teodori2025} that the granular density pattern present in ultralight dark matter halos interacts with stellar test particles in such a way that energy is transferred to the stellar particles, heating their distribution. This effect is discussed in Section \ref{subsec:heating}. This results in an increase in the orbital radii of the stars, which then results in the predicted parameters of a stellar population being in conflict with observations. 

In general, this method relies on the observation of small-scale or compact objects (e.g.  ultra-faint dwarf galaxies, star clusters) which have compactness that would be disrupted by the presence of ultralight dark matter density oscillations. This method then allows constraints on the ultralight dark matter mass to come from observations of individual objects. However, the fact that the specific objects observed are typically small-scale objects at the edge of observational limits also represents the largest uncertainty in this method. And while simulations tend to corroborate the effect that ultralight dark matter halos with known parameters would have on compact stellar populations but the biggest uncertainty is in determining accurate parameters to describe those dark matter halos, particularly if they have an uncertain history. For example, it has been argued that tidal stripping may reduce the heating from the dark matter \cite{Schive2020} or stellar self-potential may change the heating over time \cite{teodori2025}. Full simulations may be needed to understand tidal stripping as ULDM halos may undergo different tidal disruption compared to their CDM counterparts \cite{Chan2025}.

In Figure \ref{fig:constraints_m_vs_f} the constraints are plotted with hollow lines indicating the extension to lower dark matter fractions arrived at analytically in \cite{teodori2025,Hui2021,Gosenca2023}, i.e., with a scaling on the right edge of $\propto m^{3/2}$(with the exception of the Eri II and Milky-way disc constraints, which are discussed below). An analytic estimate extending granular heating constraints to multiple ultralight field models is discussed in \cite{Gosenca2023}.

\subsubsection{Eri II} \label{subsec:heating:eriII}

This effect was studied in the context of a star cluster in the Milky Way satellite Eridanus II \cite{Marsh2019}. Analytic estimates of the heating rate of this star cluster concluded that the survival (i.e. the object was not completely disrupted by the gravitational kicks of granules) of this star cluster implied $m > 10^{-19} \, \mathrm{eV}$. Relevant observed parameters of Eri II were discussed in \cite{Li2017, Crnojevic2016}. The assumptions going into this calculation break down around $10^{-21}  \, \mathrm{eV}$, at which Eri II should not form at all, with the intervening region being affected by narrow ``resonate bands".

The original analysis contains an analysis of not only the ultralight dark matter mass but also the density fraction of dark matter that is in an ultralight field, see Figure \ref{fig:constraints_m_vs_f}. Interestingly, the authors report slope of the right edge of the bound is $\propto m^3$ \cite{Marsh2019}, this is in contrast to similar analyses of the relationship between heating at reduced ultralight dark matter density for granular heating which predict a slope of $\propto m^{3/2}$ \cite{teodori2025, Gosenca2023, Hui2021}. 

The effect of tidal stripping on this constraint was studied in Refs. \cite{Schive2020, Yang2025c}, for Eri II and Fornax respectively, which concluded that a tidally stripped satellite would have substantially less heating. 

\subsubsection{Segue I, II} \label{subsec:heating:segue}
The ultra-faint dwarf satellite galaxies, Segue I and II, were studied in the context of ultralight dark matter in \cite{dalal2022}. Ultralight dark matter halos were generated using the eigenvalue method described in Section \ref{sec:eigen_halos} and simulated using the method described in Section \ref{sec:eigen_solvers}. Test particles were initialized in a spherically symmetric distribution that decreased exponentially. The velocity distribution was chosen such that it was locally described by a Maxwellian. The test particles were coupled to the dark matter halo, and their half-light radius and stellar dispersion were tracked. It was found that the half-light radius of the distributions of test particles increased beyond what was consistent with observations (Segue I \cite{Simon2011} and II \cite{Kirby2013}). The reported constraint from both objects at $99\%$ confidence was $m > 3 \times 10^{-19} \, \mathrm{eV}$ with the majority of the constraining power coming from Segue II. 

Importantly, the soliton was explicitly excluded from the constructed halo and so its effect remains a systematic uncertainty. However, the authors argue that the inclusion of the soliton would only increase the amount of stellar heating and so their bound remains conservative \cite{dalal2022}. We note that for the halo parameters assumed in the paper that the solitonic core radius is of order the half-light radius for $m \lesssim 10^{-19} \, \mathrm{eV}$. The other systematic uncertainty addressed in the paper is the possibility that the halo has been tidally stripped, which could alter the eigenspectrum needed to model the granule dynamics.

The authors explicitly include simulations for masses $m \ge 10^{-19} \, \mathrm{eV}$ and argue that Segue I and II should provide sensitivity to masses $m \ge 10^{-22} \, \mathrm{eV}$. 

\subsubsection{Heating of Milky Way disc} \label{subsec:heating:MW}

Dynamical heating due to ultralight dark matter fluctuations was studied in the context of the Milky Way disc in \cite{Church2019}. Analytic estimates of the effect of dynamical heating on the Milky Way disc set a lower bound on the ultralight dark matter mass at $6 \times 10^{-23} \, \mathrm{eV}$. The authors argued that the heating of the disc by ultralight dark matter granules should not exceed estimations of the velocity dispersion at the disc's thickest point \cite{binney2010}, the uncertainty on this value drives the statistical uncertainty on the constraint. 

Unlike the dwarf galaxy analyzed in this Section, the Milky Way disc is dominated by baryons. Their effect is not considered in this work, and so it is the largest source of systematic uncertainty. The rate of heating found in this work was studied by simulations, which argued that the work overestimated the heating in their analytic approximation \cite{Yang2024, Chiang2022}. 

\subsubsection{Fornax, Carina, and Leo II}
The impact of ultralight dark matter dynamical heating on the stellar distributions of the Fornax, Carina, and Leo II dwarf galaxies was studied in \cite{teodori2025}. They ran simulations of ultralight dark matter halos matching the observed dwarf galaxy parameters (Fornax \cite{Walker_2009}, Carina and Leo II \cite{Koch2007}) coupled to massless point particles representing the dwarf galaxy stars, ultralight dark matter masses between $10^{-22} \, \mathrm{eV}$ and $5 \times 10^{-21} \mathrm{eV}$ were tested. Masses for which the stellar half-light radius grew beyond the observed values were ruled out. The authors find that their analysis rules out masses below $5 \times 10^{-21} \, \mathrm{eV}$. The stars are treated as massless point particles, and the authors identify stellar self-gravity as the largest uncertainty. For masses lower than tested, the author argues that the stars could be contained within the central core, which may change the heating history. The effect of tidal stripping on the Fornax constraint was studied in Ref. \cite{Yang2025c}.

\subsection{Pulsar timing arrays} \label{sec:PTA_constraints}

Ultralight dark matter is thought to produce a relativistic potential which oscillates at the Compton frequency; this effect is discussed in Section \ref{sec:relativistic_potential}. This oscillating potential can create a gravitational redshift, which can create time delays in pulsar signals \cite{Khmelnitsky2014}. 

This was searched for by the Parkes Pulsar timing array \cite{Porayko2018}. They place a constraint at $95\%$ that the local ultralight dark matter density must be $< 6 \, \mathrm{GeV \, cm^{-3}}$ for dark matter masses below $m \le 10^{-23} \, \mathrm{eV}$. While the current constraint does not reach the current measurement of the local dark matter density. The next generation of pulsar timing array experiments is thought to be sensitive to the local dark matter density for masses around $2 \times 10^{-23} \, \mathrm{eV}$. 

The left edge of the constraint is around $2 \times 10^{-24} \, \mathrm{eV}$, which corresponds to where the Compton timescale of the field would be longer than the total observation time for the pulsars. 

It is also interesting to point out that ultralight dark matter may make a gravitational wave signal similar to the one observed by pulsar timing arrays \cite{Blas2025, Brax2024}.

\subsection{Astrometry and Doppler shifts}

Ultralight dark matter granules may also affect the trajectories of stars in an observable way, see Section \ref{subsec:heating}. The effect of the granular density pattern on astrometry observations was considered in Refs. \cite{Kim2024_astro, dror2024, Furlanetto2024}, Doppler shift induced by gravitational kicks to the pulsar velocity in Ref. \cite{Kim2024}, and altering binary pulsar orbits in Ref. \cite{Kus2024}. 

\textbf{Astrometry} The authors of \cite{dror2024} studied the deflection in the apparent motion of quasars as viewed from Earth, and found a time-independent proper motion induced by the granules
\begin{align}
    k \phi \sim \frac{0.3 \, \mathrm{\mu as}}{\mathrm{year}} \left( \frac{10^{-29} \, \mathrm{eV}}{m} \right)^2 \left( \frac{k/m}{10^{-3}} \right) \frac{\rho}{0.3 \, \mathrm{GeV \, cm^{-3}}} \, ,
\end{align}
and an annually modulating signal 
\begin{align}
    \frac{v_0 k}{m} \phi \sim 0.1 \, \mathrm{\mu as} \left( \frac{10^{-28} \, \mathrm{eV}}{m} \right)^2 \left( \frac{k/m}{10^{-3}} \right) \frac{\rho}{0.3 \, \mathrm{GeV \, cm^{-3}}} \, .
\end{align}
The work here is most relevant for very low masses.

In Ref. \cite{Kim2024_astro} studied the deflection of the stellar trajectories due to the granular kicks. However, the data is currently not sensitive enough to provide a constraint given the measured local dark matter density. The peak sensitivity is around where the de Broglie time is of the order of the observation times, around $10^{-17} \, \mathrm{eV}$.

\textbf{Doppler effect} The random relative velocity between Earth and pulsars created by the gravitational kicks from ultralight dark matter granules was studied in Ref. \cite{Kim2024}. The time delay created by the relativistic Doppler shift these kicks induced was determined to go as
\begin{align}
    \delta t \sim  (\bar a\tau_\mathrm{db}^2) \left( \frac{T}{\tau_\mathrm{db}} \right)^{3/2} \sim 0.1 \left( \frac{10^{-17} \, \mathrm{eV}}{m} \right)^{3/2} \left( \frac{\rho}{0.4 \, \mathrm{GeV / cm^3}} \right) \left( \frac{T}{30 \, \mathrm{yrs}} \right)^{3/2} \, \mathrm{ns} \, .
\end{align}
Where $\bar a = Gm_\mathrm{eff} / \lambda_\mathrm{db}^2$ is the average acceleration over a de Broglie time. $T$ is the experimental integration time. 

\subsection{Superradiance}

Observations of rapidly spinning black holes constrain the possibility of ultralight boson clouds formed via superradiance. Thanks to theoretical modeling of the growth rates of superradiant instabilities, such observations can be used to search for black hole spin-down. Spin measurements, obtained from X-ray reflection spectroscopy and continuum fitting, as well as component spins from gravitational wave detections, constrain the presence of these clouds and set bounds on the parameters of ultralight bosonic fields.

Stellar-mass black holes ($M \sim 10\, M_\odot$) are sensitive to bosons with masses $m \sim 10^{-13}$--$10^{-12} \, \mathrm{eV}$. Spin measurements from x-ray reflection spectroscopy and continuum fitting exclude the existence of: (spin-0) $m \sim 10^{-13}$--$10^{-12} \, \mathrm{eV}$~\cite{Arvanitaki:2010sy}; (spin-1) $m \sim 10^{-14}$--$10^{-12} \, \mathrm{eV}$~\cite{Baryakhtar:2017ngi}; and (spin-2) $m \sim 10^{-14}$--$10^{-11} \, \mathrm{eV}$~\cite{Brito:2020lup}.

Supermassive black holes (SMBHs), such as those in galactic centers ($M \sim 10^6$--$10^9\, M_\odot$), probe much lighter bosons. Observations of high-spin SMBHs exclude: (spin-0) $m \sim 10^{-19}$--$10^{-16} \, \mathrm{eV}$~\cite{Cardoso:2016olt}; (spin-1) $m \sim 10^{-20}$--$10^{-17} \, \mathrm{eV}$~\cite{Baryakhtar:2017ngi}; and (spin-2) $m \sim 10^{-20}$--$10^{-15}\, \mathrm{eV}$~\cite{Brito:2020lup}.

In~\cite{Stott2018}, the authors study how the bounds change when accounting for different black hole mass distributions. In~\cite{Davoudiasl2019}, data from the Event Horizon Telescope (EHT) on M87* ($M \sim 6.5 \times 10^9\, M_\odot$) is used to constrain the mass of ultralight bosons to $m \in [2.9,\, 4.6] \times 10^{-21}\, \mathrm{eV}$ for spin-0 and $m \in [0.85,\, 4.6] \times 10^{-21}\, \mathrm{eV}$ for spin-1. There are also studies of superradiance around primordial black holes~\cite{Branco:2023frw,Calza:2023rjt}, constraining higher mass dark matter.

Finally, the presence of self-interactions can suppress the growth of the bosonic cloud, weakening the superradiant instability, and alter the gravitational wave signatures emitted by the cloud, affecting detectability~\cite{Yoshino:2012kn,Kawasaki:2015lpf,Baryakhtar:2020gao,Collaviti:2024mvh,Xie:2025npy}.

It is important to note that superradiance constraints are subject to systematic uncertainties~\cite{Brito:2025ojt,Hoof:2024quk,Ellis:2026gwt}. These include uncertainties in black hole mass and spin measurements~\cite{Witte:2024drg}, assumptions about the accretion history and environment~\cite{Ng:2020ruv}, and possible effects of additional interactions or multiple fields. Environmental effects such as accretion disks or companion stars can also modify the growth of the superradiant instability~\cite{Collaviti:2024mvh}. While these uncertainties do not eliminate the sensitivity of superradiance to ultralight fields, they can affect the precise exclusion regions and should be taken into account when interpreting the bounds.

\subsection{Extremely light axions - CMB and large-scale structure}
\label{sec:constrain_extremely_light}
As discussed in Section~\ref{subsec: DM_small_scales}, observations of the CMB and large-scale structure (LSS) provide the most stringent constraints on dark matter. For ULDM, these observations constrain the low-mass end of the parameter space, where ULDM is a subdominant component of the dark matter. This regime, $m < 10^{-24}\, \mathrm{eV}$, is often referred to as extremely light dark matter. These observations constrain the evolution and shape of the power spectrum. 

Using CMB and LSS data, Refs.~\cite{Hlozek2015,Hlozek2017,Hlozek:2017zzf,Lague:2020htq,Rogers:2023ezo} constrain the fraction of fuzzy dark matter for masses in the range $10^{-33}\,\mathrm{eV} < m < 10^{-24}\,\mathrm{eV}$, and find that ULDM cannot constitute all of the dark matter for $m < 10^{-24}\, \mathrm{eV}$. These constraints are shown in Figure~\ref{fig:constraints_m_vs_f}.

More recently, the Atacama Cosmology Telescope (ACT) Data Release 6 (DR6)~\cite{ACT:2025tim} has improved these bounds, further demonstrating the power of CMB observations to constrain ULDM in the extremely light mass range.


\section{Conclusions} \label{sec:conclusion}

In this review, we have discussed the models associated with ultralight dark matter as well as the motivations for these models. We have summarized the phenomena associated with an ultralight field and the primary numerical tools used to study these phenomena. We have also reviewed the current constraint space of the model and discussed the observational methods associated with each constraint. We conclude here with a discussion of the current state of and future directions for the field.

\textbf{Motivations.} Historically, ultralight dark matter was motivated from an observational astrophysics perspective largely by small-scale structure discrepancies between dark matter only simulations and observations \cite{Hu2000}. An initial estimate argued that a field of approximately $10^{-22} \, \mathrm{eV}$ would alleviate this discrepancy \cite{Hu2000}. However, the likelihood that ultralight dark matter of this mass addresses this discrepancy is increasingly implausible. Not only if this mass ruled out by a large number of independent observations \cite{Nadler2021, Nadler_2025, Garland2024, Bar2022, zimmermann2025, Rogers:2020ltq, Powell2023, teodori2025} (many of which are from the very observations of small-scale structure that ultralight dark matter was invented to address), but many of the small-scale structure problems themselves are now thought to admit observation or baryonic solutions \cite{Weinberg2015, Bullock2017}. Additionally, a single classical spin-0 ultralight field has been ruled out as all the dark matter for masses $\lesssim 10^{-18}  \, \mathrm{eV}$ by at least one method and by multiple complementary methods for masses $\lesssim 10^{-20}  \, \mathrm{eV}$. However, it is important to note that some work has found that the existence of ultralight fields of specific masses is consistent with or can help explain some observations \cite{pozo2024, Amruth2023, Lapi2022,Hayashi2021, Yang2024b, Yang2025b, Broadhurst_2025}, in particular, mixed ultralight and cold dark matter models remain of interest \cite{Rogers2023, pozo2024}. However, overall, small-scale structure problems are becoming a motivation that is increasingly difficult to reconcile with all observational constraints.

\textbf{Models.} However, identification of the dark matter particle continues to elude observational efforts \cite{Feng2010}. This is in spite of the fact that experiments are probing much of what was once considered the most promising parameter space \cite{aalbers2024}. From this perspective, any probe of dark matter microphysics remains interesting. Much of the work surrounding ultralight dark matter makes only weak assumptions about the underlying dark matter model (the most important assumption being the dark matter mass). This means that much of ultralight dark matter work can generally probe the lowest mass end of the dark matter parameter space. An increasing amount of work has focused on determining to what extent constraints can be generalized to extensions of the vanilla ultralight dark matter model. This work includes mixed dark matter models \cite{Tellez-Tovar2021, Street2022, Guo2021, Luu2020, Huang2023, Glennon2023, Vogt2022, Chen2023, Gosenca2023,Mirasola2024,Luu2023,luu2024, Dome2024, Schwabe2020}, higher spin models \cite{Amin2022, Amin2023}, and models with self-couplings \cite{Glennon2022,Glennon2023b,Mirasola2024}. Figure \ref{fig:constraints_m_vs_f} shows some of the initial progress on this front, where constraints are plotted not only as a function of the mass but also as a function of the fraction of the dark matter in an ultralight field. It is likely that future constraints will be examined in a mass-fraction parameter space.

\textbf{Phenomena.} More to this point, there has been great progress in understanding the phenomena associated with model extensions. Already interesting behavior associated with cores \cite{Street2022,Luu2020,Huang2023, Vogt2022, Mirasola2024, Luu2023}, granules \cite{Gosenca2023, Amin2022}, tidal disruption \cite{Glennon2022}, dynamical friction \cite{Glennon2023b}, and cosmological halos mass functions and structure growth \cite{Tellez-Tovar2021, Huang2023, Chen2023, luu2024, Amin2023} has been studied in extended models. However, there still exists much work to be done in understanding how sensitive different observational probes are to these extensions. 

\textbf{Numerical methods.} One of the largest hurdles to simulating ultralight dark matter continues to be the scaling of Schr\"odinger-Poisson solvers with increasing resolution. There is no solution to this problem that is appropriate in all settings. However, much progress has been made in identifying approximation techniques that circumvent these numerical requirements in specific circumstances. N-body simulations \cite{Nadler_2025}, adaptive mesh schemes \cite{Schwabe2020}, eigenvalue solvers \cite{dalal2022, Dalal2021}, and Gaussian random field approximations \cite{Powell2023} (to name a few) have all been applied to study masses beyond what has been possible in traditional Schr\"odinger-Poisson solvers. As the lower bound on the ultralight dark matter continues to creep upwards, clever approximation techniques like this will become increasingly relevant to compare observations with numerical results. The development and verification of these techniques remains an interesting future work. 

\textbf{Observational constraints.} The most important part of any dark matter model remains its ability to generate and test falsifiable predictions. Observational constraints on the lower bound of ultralight dark matter have steadily increased. Much of this improvement has hinged on increasingly powerful observations of small-scale structure \cite{dalal2022, teodori2025, zimmermann2025, Nadler_2025}. In addition to increasing the sensitivity of older methods, new constraints associated with studies of the full nonlinear granular structure of ultralight dark matter have become increasingly common \cite{dalal2022, teodori2025, Powell2023, Kim2024, Kim2024_astro, Eberhardt2024, dror2024, Marsh2019, Church2019}. Probes of ultralight dark matter continue to diversify, demonstrating the richness of this model's phenomenology. However, this has also introduced a new series of systematic uncertainties that need to be better understood. In particular, many new constraints are based on observations of individual dwarf galaxies \cite{Bar2022, dalal2022, teodori2025, Marsh2019}; a better understanding of how model extensions and satellite histories affect these constraints is needed. An additional confounding factor is the impact of baryons on many observations. Both baryonic feedback and self-potential are expected to alter the predictions of dark matter only studies, more work is needed to understand how this affects existing ultralight dark matter constraints. This, together with possible unknown systematics and modeling uncertainties, can significantly alter these bounds; therefore, current constraints should be interpreted with appropriate caution. As future observations and numerical techniques improve, the study of ultralight dark matter will likely continue to be able to probe smaller and smaller scales and higher and higher masses in the dark matter parameter space.

\textbf{Going forward.} Significant progress has been made understanding the impact of the dark matter mass on cosmological and astrophysical observations. And as a result remarkable progress has been made using these observations to determine the lower-bound on the dark matter mass. At this point, it is increasingly unlikely that the originally proposed ultralight dark matter mass of $10^{-22} \, \mathrm{eV}$ can compromise all the dark matter. Though the systematic uncertainties surrounding some constraints still prevents universal agreement in the literature. However, the need for this particular mass has been somewhat supplanted by alternative explanations to small scale structure problems. At the same time, interest in axiverse scenarios in which multiple fields of different mass may be relevant has increased. And while the original one field model appears ruled out, the mixed scenario remains both phenomenologically interesting and theoretically well motivated. As our observational data becomes more precise, numerical simulations more powerful, and theoretical modeling more complete, we can expect to significantly expand our understanding of these more complex ultralight dark matter parameter spaces.

\appendix

\section{Quantum to Classical} \label{app:ULDM:quantum}

\subsection{Quantum description}
In order to study the quantum field system, we first need to promote our classical field, $\psi(x,t)$, to a quantum field operator, $\hat \psi(x)$, where 
\begin{align}
    \hat \psi(x) = \sum_i \hat a_i u_i^\dagger(x) \,.
\end{align}
The spatial operator is a Fourier transformation of the normal creation and annihilation operators, $\hat a^\dagger$ and $\hat a$, respectively. $u_i^\dagger(x)$ is the plane wave with momentum $p_i$. The field evolves either by the Schr\"odinger equation, the equivalent Heisenberg equation, or the equivalent Von Neumann equation given
\begin{align}
    i \hbar \ket{\psi} &= \hat H \ket{\psi} \, , \\
    i \hbar \hat \psi &= [\hat \psi, \hat H] \, , \label{eqn:Heisenberg} \\
    i \hbar \partial_t \ket{\psi}\bra{\psi} &= \left[ \hat H, \ket{\psi}\bra{\psi} \right] \, .
\end{align}
which describe either the evolution of the quantum state $\ket{\psi}$, or the field operator itself, $\hat \psi$, or state operator $\ket{\psi} \bra{\psi}$. $\hat H$ is the operator-promoted Hamiltonian
\begin{align} \label{eqn_Ham}
    \hat H / \tilde \hbar &= \sum_j \omega_j \hat a_j^\dagger \hat a_j + \sum_{ijkl} \frac{\Lambda_{kl}^{ij}}{2} \hat a_k^\dagger \hat a_l^\dagger \hat a_i \hat a_j  \\
    &= \iint dx dy \, \hat \psi^\dagger(x) \frac{-\tilde \hbar \,  \nabla^2}{2} \hat \psi(y) +  \hat \psi^\dagger(x) \, \frac{\hat V(x)}{\tilde \hbar} \, \hat \psi(x) \, ,  \nonumber \\ 
    \nabla^2 \hat V(x) &= C m \, \hat \psi^\dagger(x) \hat \psi(x) \, .
\end{align}
Notice that the normalization of $\hat \psi$ is chosen such that $\hat \psi^\dagger \hat \psi$ gives the number density not the mass density. We will want to work in the quantum phase space so we define the Wigner function \cite{Wigner1932} of the quantum state which is given as the Weyl symbol of the state operator, $\hat \rho$, this is written
\begin{align} \label{WeylSymOmega}
    f_W[\psi, \psi^*] \equiv \frac{1}{\mathrm{Norm}} \int_{\mathbb{C}^{\mathbb{R}^3}} \int_{\mathbb{C}^{\mathbb{R}^3}} \mathcal{D} \eta \, \mathcal{D} \eta^* \braket{ \psi - \frac{\eta}{2} \, \ket{\psi} \bra{\psi}  \, \psi + \frac{\eta}{2} } e^{-|\psi|^2 - \frac{1}{4}|\eta|^2} e^{\frac{1}{2}(\eta^* \psi - \eta \psi^* )} \, .
\end{align}
The Wigner function is a real-valued functional of the field configuration $\psi(x), \, \psi^*(x)$ representing a pseudo-phase space for the field. We can solve the Von Neumann equation in this space. We use the fact that the Weyl symbol of the commutator is a Moyal bracket, i.e.
\begin{align}
    \set{\!\set{f,g}\!}_M &\equiv 2 f(\psi, \psi^*) \sinh \left( \frac{1}{2} ( \cev{\partial}_\psi \vec{\partial}_{\psi^*} - \cev{\partial}_{\psi^*} \vec{\partial}_{\psi} ) \right) g(\psi, \psi^*) \, , \nonumber \\
    &= 2 f(\psi, \psi^*) \left( \frac{1}{2} ( \cev{\partial}_\psi \vec{\partial}_{\psi^*} - \cev{\partial}_{\psi^*} \vec{\partial}_{\psi} ) + \frac{1}{4} ( \cev{\partial}_\psi \vec{\partial}_{\psi^*} - \cev{\partial}_{\psi^*} \vec{\partial}_{\psi} )^2 + \dots \right) g(\psi, \psi^*) \, ,
\end{align}
and find that 
\begin{align} \label{eqn:phaseSpaceSchr}
    \partial_t \hat \rho &= -\frac{i}{\hbar} \left[ \hat H, \hat \rho \right] \rightarrow \\
    \partial_t f_W[\psi, \psi^*;t] &= -\frac{i}{\hbar} \left\{ \left\{ H_W[\psi, \psi^*]\, , \, f_W [\psi, \psi^*;t] \right\} \right\}_M \, \\
    &= -\frac{i}{\hbar} \left\{ H_W[\psi, \psi^*]\, , \, f_W [\psi, \psi^*;t]  \right\}_c(1 + \mathcal{O}(1/n_\mathrm\mathrm{tot})) \, . \label{eqn:phaseSpaceShcrApprox} 
\end{align}
Where we have replaced $f$ and $g$ in equation \eqref{WeylSymOmega} with the Weyl symbol of the Hamiltonian and the Wigner function respectively. The expression in the last line is a Poisson bracket of the Wigner function and Hamiltonian (recall here, in anticipation of our final result, that the Poisson bracket of the phase space gives the classical equations of motion). Notice that the leading order correction term goes as $1/n_\mathrm\mathrm{tot}$ and does not explicitly depend on $\hbar$. Technically this approximation also requires that the phase space derivatives of the Wigner function are under sufficient control, which should be the case for systems with Wigner functions close to a ``coherent state" whose Wigner function is given
\begin{align} \label{eqn:wignerCoherent}
    f_W[\psi, \psi^*] = \frac{1}{\pi} e^{-\int dx |\psi(x) - z(x) |^2} \, .
\end{align}
This is the state that describes the initial conditions of ultralight dark matter produced via the misalignment mechanism.
When using Wigner functions, the expectation value of some quantum operator $\Omega(\set{\hat \psi, \hat \psi^\dagger})$ is calculated
\begin{align} \label{eqn:expectVal}
        \braket{\hat \Omega(\set{\hat \psi, \hat \psi^\dagger})} &= 
\int_{\mathbb{C}^{\mathbb{R}^3}} \int_{\mathbb{C}^{\mathbb{R}^3}} \mathcal{D} \psi \, \mathcal{D} \psi^* \,  f_W[\psi, \psi^*] \, \Omega_W[\psi, \psi^*] \, . 
\end{align}

\subsection{Taking the classical limit}
Equipped with our Wigner function pseudo phase space, the classical limit becomes straightforward as the result of two separate approximations. We first identify the classical field as the expectation value of the field operator (c.f. mean field theory) 
\begin{align}
    \psi^\mathrm{cl} \equiv \braket{\hat \psi} \, .
\end{align}
We solve for the evolution of this expectation value by taking an expectation value, equation \eqref{eqn:expectVal}, of Heisenberg's equation, equation \eqref{eqn:Heisenberg}, 
\begin{align}
    \partial_t \braket{\hat \psi(x,t)} &= \partial_t \psi^{cl}(x) = -\frac{i}{\hbar} \int_{\mathbb{C}^{\mathbb{R}^3}} D\psi \, \left\{ \set{H_W[\psi, \psi^*]\,,\,\psi(x)} \right\}_M\, f_W[\psi, \psi^*] \, .
\end{align}
If we are in the large occupation limit, we can approximate the Moyal bracket as a Poisson bracket, i.e,. equation \eqref{eqn:phaseSpaceShcrApprox},
\begin{align}
    \partial_t \psi^{cl}(x) \approx -\frac{i}{\hbar} \int_{ \mathbb{C}^{\mathbb{R}^3} } D\psi \, \left\{ H_W[\psi, \psi^*]\,,\,\psi(x) \right\}_c\, f_W[\psi, \psi^*] \, .
\end{align}
Next we assert that the distribution is tightly peaked around the mean field value, i.e. $|\braket{\hat \psi}|^2 \gg \braket{\delta \hat \psi^\dagger \, \delta \hat \psi}$; the distribution can therefore be approximated as a delta function at the classical field value, $f_W = \delta[\psi(x) - \psi^{cl}(x)]$
\begin{align}
    \partial_t \psi^{cl}(x) &\approx -\frac{i}{\hbar} \int_{\mathbb{C}^{\mathbb{R}^3}} D\psi \, \left\{ H_W[\psi, \psi^*]\,,\,\psi(x) \right\}_c\, \delta[\psi - \psi^{cl}] \, \\
    &= -\frac{i}{\hbar} \left\{ H_W[\psi^{cl}, \psi^{cl*}]\,,\,\psi^{cl}(x) \right\}_c \, , \\
    &= -\frac{i}{\hbar} \left( \frac{\nabla^2}{2m} + m\,V \right) \psi^{cl}(x) \, .
\end{align}
The tightly peaked approximation will generally be true for coherent states with large occupation number, see equation \eqref{eqn:wignerCoherent}, which is generally what we assume for the initial conditions of ultralight dark matter produced via the misalignment mechanism. The last line is the classical Schr\"odinger-Poisson equations when $\nabla^2 V(x) = C m | \psi^{cl}(x)|^2$. We see that the derivation of the classical field equations of motion relies on two assumptions. The first is that the Moyal bracket was well approximated by a Poisson bracket, which is true in the large occupation limit, i.e. $n_\mathrm{tot} \gg 1$. The second is that the quantum distribution is tightly localized around the classical field value, i.e. $|\braket{\hat \psi}|^2 \gg \braket{\delta \hat \psi^\dagger \, \delta \hat \psi}$, see discussions in \cite{Eberhardt2022Q, Eberhardt2021, Eberhardt2023, sikivie2017, Guth2015}.

Of course we have taken this limit using the initial conditions of the quantum state. If we simple evolve the quantum equations of motion here, as was done in \cite{Eberhardt2023}, we will see that the quantum corrections grow quickly in systems which exhibit classical chaos and slowly otherwise. 

\subsection{Decoherence}

Decoherence is the process by which an open quantum system, $\ket{\psi(t)}$, is ``measured" by its environment, $\ket{\mathcal{E}(t)}$, and projected onto its basis of ``pointer states", see \cite{Zurek2003} for an indepth discussion. We will generally assume that at $t=0$ the system and environment can be written as a product state and thet joint system evolves as
\begin{align} \label{eqn:ICs_decohere}
    \ket{A(0)} &= \ket{\psi(0)}\ket{\mathcal{E}(0)} = \sum_i c_i(t=0) \ket{\phi}_i \otimes \sum_j b_j(t=0) \ket{\epsilon}_j \, .
\end{align}
The state evolves via a Hamiltonian which describes the evolution of each independent system, $\hat H_\psi$ and $\hat H_\mathcal{E}$, and also the interaction between them, $\hat H_{\mathrm{int}}$, i.e. 
\begin{align}
    \hat{H}_A = \hat H_\psi + \hat H_\mathcal{E} + \hat H_{\mathrm{int}} \, .
\end{align}
The evolution at time $T$ is 
\begin{align}
    \ket{A(T)} = e^{-i \, \hat H_A \, T} \ket{A(0)} \, .
\end{align}
At this time, because of the influence of the interaction term in our Hamiltonian, there is no guarantee that the state can be written as a simple tensor product as in equation \eqref{eqn:ICs_decohere}. Instead the system may be entangled with the environment. The state can more generally be written as 
\begin{align}
    \ket{A} = \sum_{ij} c_{ij} \ket{\phi}_i \ket{\epsilon}_j \, . 
\end{align}
This allows for the presence of entanglement. We also consider the state operator 
\begin{align}
    \hat \rho_A = \sum_{ijkl} c_{ij} c^*_{kl} \ket{\phi}_i \ket{\epsilon}_j \bra{\phi}_k \bra{\epsilon}_l \, .
\end{align}
Now, if we assert that an observer measures the environment to be in the eigenbasis $\ket{\tilde \epsilon}$. We then have a reduced density matrix tracing over the environment eigenbasis
\begin{align}
    \hat \rho^R_A = \mathrm{Tr}_\epsilon[\rho_A] = \sum_i \bra{\tilde \epsilon} \hat \rho_A \ket{\tilde \epsilon}_i \, .
\end{align}
When $\braket{\tilde \epsilon| \tilde \epsilon}_{ij} = \delta_{ij}$, the reduced density matrix is now a classical ensemble of pointer states of the system. Pointer states are the states that develop the least entanglement over time with the preferred environmental basis states. The rate at which quantum corrections to the classical field equations grow (the quantum breaktime) has been estimated \cite{Eberhardt2021,Eberhardt2022,Eberhardt2022Q,Eberhardt2023,sikivie2017,Dvali2017,Dvali2018,Kopp2021}. So too has the rate at which environmental interactions decohere the quantum state back into the pointer basis (the decoherence time) \cite{Eberhardt2023, Allali_2020,Allali2021,Allali2021b}. It is generally argued that the decoherence timescales are short compared to the quantum breaktime timescales \cite{Eberhardt2023, Allali_2020,Allali2021,Allali2021b}. However, a detailed study identifying qualities of the pointer basis does not currently exist.

\section{Metrics of classicality} \label{app:matricsClassicality}

There are many methods by which the ``quantumness" or ``classicality" of a system may be measured. There is no one that is universally best, it is instead important that we understand what each is measuring and how they relate to the quantities we are interested in. We will list a few here that have been studied in terms of the quantum nature of ultralight dark matter. This section closely follows \cite{Eberhardt2022Q}. A number of metrics have been used in the context of ultralight dark matter, including the Penrose-Onsager criterion \cite{Eberhardt2022Q, Ferreira2021}, squeezing \cite{Kopp2021}, ``$Q$" factor \cite{Eberhardt2022Q, Eberhardt2021, Eberhardt2023, Eberhardt_testing, Eberhardt2022}, and negativity in the Wigner function. 

\subsection{Penrose-Onsager criterion} The Penrose-Onsager (PO) criterion is the statement that there exists a complex vector $z_i$ such that \cite{Penrose1956}

\begin{equation} \label{POcriterion}
    \braket{\hat a^\dagger_i \hat a_j} = \vec{z}_i^\dagger \vec z_j \, .
\end{equation}

Where $\hat a^\dagger_i$ and $\hat a_i$ are the creation and annihilation operators, respectively, for the $i$th momentum modes. This means the second moment matrix $M_{ij} \equiv \braket{\hat a^\dagger_i \hat a_j}$ can be written as an outer product of a single field $\vec z$ and as a result has a single non-zero eigenvalue. 

One can test the degree to which this criterion is satisfied by looking at the eigenvalues of $M_{ij}$. When the PO criterion is satisfied, there will be a single nonzero ``principal" eigenvalue, $\lambda^p$ equal to the squared norm of $\vec z$. Where $\vec z^* / \sqrt{\sum_i|z_i|^2}$ is the corresponding principal eigenvector, $\vec \xi^p$.  When the system is well described by the classical theory, we expect that the principal eigenvalue is very close to $n_\mathrm{tot}$ \cite{Leggett2001}. If the principal eigenvalue deviates too far from $n_\mathrm{tot}$, a single classical field description is insufficient.

It should be noted that satisfying the PO criterion does not imply that the conjugate of the principal eigenvector obeys the classical field equations of motion, though this is often, at least approximately, the case \cite{Eberhardt2022Q}.

Often, when simulating a physical system, the mode occupation or spatial densities are of specific interest. The PO criterion is a useful measure of classicality because when it is satisfied, this implies that there exists a single field that captures these occupation numbers. Computationally, this requires solving matrix eigenvalues and therefore has cubic scaling with the mode number, $M$.

\subsection{Q factor}

The equations of motion for the expectation value of the field operator (i.e., the classical field) can be written in terms of the classical equations of motion plus a series of terms proportional the higher order moments of the field (see for example \cite{Eberhardt2021}), in momentum space this looks like 
\begin{align} \label{series}
    \partial_t \braket{\hat a_p} &= \braket{f(\hat a_p)} = f( \braket{\hat a_p} ) + \sum_{ij} \braket{\delta \hat a_i^\dagger \delta \hat a_j} \frac{\partial^2}{\partial \braket{\hat a_i^\dagger} \partial \braket{\hat a_j}} f( \braket{\hat a_p} ) + \dots  \, .
\end{align}
Where the first term in the series on the right-hand side of the equation is the classical equations of motion and the next term is the lowest order quantum correction. Note $\delta \hat a_i = \hat a_i - \braket{\hat a_i}$. The leading order correction term is proportional to the second moments of the field operators and the second derivative of the time evolution function with respect to the field expectation. We can define a parameter $Q$ which approximates the ratio of the first and second order terms in the expansion above as
\begin{align} \label{Qparam}
    Q \equiv \sum_j \frac{\braket{\delta \hat a_j^\dagger \delta \hat a_j}}{n_\mathrm{tot}} \, .
\end{align}
When this parameter is small, we can say that the underlying quantum distribution is well localized around the classical field value \cite{Eberhardt2021,Eberhardt2022,Eberhardt2022Q,Eberhardt2023}. Overtime, nonlinearities in the Hamiltonian will take this quantity away from $0$. 

This parameter may be a preferable metric of classicality to others (such as the Penrose-Onsager criterion) because in cases where the initial quantum distribution is known to be well approximated by the classical field theory, for example when the initial quantum state is a coherent state as we expect for ultralight dark matter created by the misalignment mechanism. In this case, the $Q$ parameter gives more direct information about the deviation from this initial state and the corrections to the classical field equations of motion. Likewise, in lower order quantum correction methods, this parameter is often more straightforward and inexpensive to calculate. This has been used in the context of ultralight dark matter to estimate the quantum breaktime \cite{Eberhardt2021,Eberhardt2022,Eberhardt2022Q,Eberhardt2023}.

\subsection{Squeezing}

Quantum squeezing is present when the uncertainty of some operator becomes smaller than the vacuum state. This metric of classicality is common in quantum optics and has also been studied in the case of ultralight dark matter \cite{Kopp2021}. For some operator $\hat O$ with $[\hat O, \hat O^\dagger] = 1$. Then, a Hermitian operator (a quadrature) can be defined
\begin{equation}
     \hat X_\theta = \hat O e^{-i \theta}+\hat O^\dagger e^{i \theta} \,, 
\end{equation} 
for parameter $\theta$. The variance of $\hat X_\theta$ is 
\begin{align} 
    \mathrm{Var}(\hat  X_\theta) & =  1+ 2 \langle \delta \hat O^\dagger \delta \hat O\rangle   \notag\\
    & \qquad+ e^{-2i \theta} \mathrm{Var}(\hat O) + e^{2i \theta} \mathrm{Var}(\hat O^\dagger)
\end{align}
For angle $e^{2i\theta_-} = \sqrt{\frac{\mathrm{Var}(\hat O)}{\mathrm{Var}(\hat O^\dagger)}}$ the variance $V^{\hat O}_- \equiv \mathrm{Var}(\hat  X_{\theta_-})$ is minimized and given by
\begin{equation}\label{generalVminus}
    V^{\hat O}_-(t) = 1+ 2 \langle \delta \hat O^\dagger \delta \hat O\rangle  - 2 |\mathrm{Var}(\hat O)|\,.
\end{equation}
If the quantum state is the vacuum or a coherent state, then $V^{\hat O}_- =1$. If $V^{\hat O}_- < 1$ the state is defined to be squeezed \cite{drummond2013}.

\section{Cold dark matter correspondence expanded} \label{app:cdm_correspond}

\subsection{Phase space}

\begin{figure*}[!ht]
	\includegraphics[width = .97\textwidth]{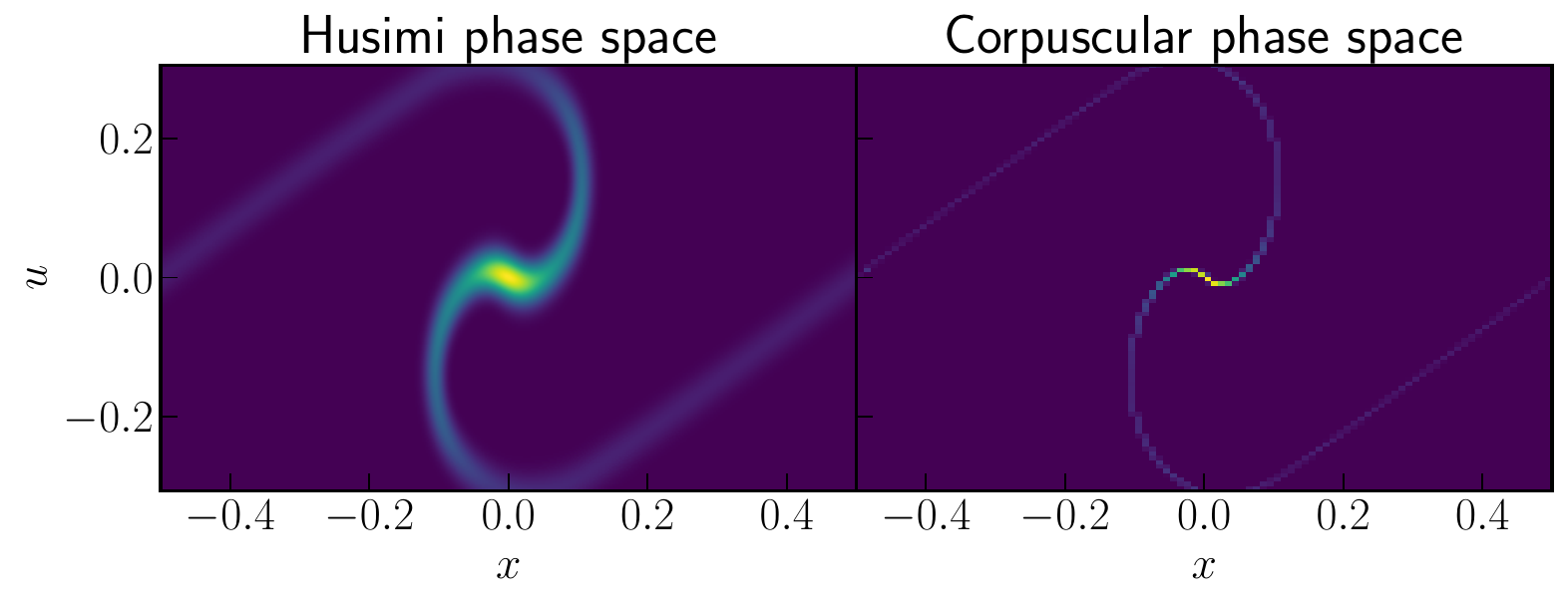}
	\caption{ Phase space for two simulations of the gravitational collapse of an initial overdensity in a single spatial dimension. \textbf{Left:} Husimi transform (equation \eqref{eqn:Husimi}) of ultralight dark matter simulation. \textbf{Right:} phase space of corpuscular particle simulation. Notice that each phase space has similar features, each showing the characteristic spiral associated with this system. For phase space structures large compared to $\hbar$ we expect that these two phase spaces should roughly agree.}
	\label{fig:phaseSpace}
\end{figure*}
When discussing the evolution of an ensemble of corpuscular particles, we often think of their phase space, i.e., their distribution function as a function of joint position and velocity, $f(q,v)$. The equations of motion for a collisionless phase space evolving under gravity are the Vlasov-Poisson equations, given
\begin{align}
    \partial_t f(q,u) &= - u \cdot \nabla f(q,u) + \nabla V(q) \cdot \nabla_u f(q,u) \, , \\
    \nabla^2 V(q) &= 4 \pi G \rho(q) \, .
\end{align}
which describes cold dark matter.

Unlike an ensemble of corpuscular particles, a field does not have a traditional phase space. However, there exist descriptions of fields that retain many of the useful qualities of phase space. Perhaps the two most common of these are the Wigner \cite{Wigner1932} and Husimi \cite{Husimi1940} distributions. Formally, the Wigner function is the Weyl symbol of the state (or density) operator of the field. In bracket notation,n our field is 
\begin{align}
    \psi(x) =  \braket{x | \psi} \, .
\end{align}
And the state operator is then 
\begin{align}
    \hat \rho = \ket{\psi}\bra{\psi} \, .
\end{align}
The state operator is a more general representation of the state vector, which can be used to describe the evolution of mixed states using the Von Neumann equations. Ultralight dark matter is generally described by a pure state, therefore, further discussion of mixed states is beyond the scope of this review, and we will proceed assuming that we are in a pure state. The Wigner function can then be written
\begin{align}
    W(q,u) = (2 \pi \hbar)^{-1} \int_{-\infty}^{\infty} \braket{q - \frac{1}{2}y | \, \hat \rho \, | q + \frac{1}{2} y} e^{im \, u \, y/\hbar} dy \, .
\end{align}
Where $q,u$ are phase space position and velocity coordinates, respectively. Like a traditional phase space, the Wigner function satisfies
\begin{align}
    \int_{-\infty}^{\infty} W(q,u) \, dq = \braket{u| \hat \rho | u} \, , \\
    \int_{-\infty}^{\infty} W(q,u) \, du = \braket{q| \hat \rho | q} \, .
\end{align}
The equation of motion of the Wigner function is written
\begin{align}
    \partial_t W(q,u) &= -u \cdot \nabla W(q,u) + \sum_{n \, \mathrm{ odd}} \frac{1}{n!} (-i \hbar / 2)^{n-1} \frac{\partial^n V(q)}{\partial q^n} \frac{\partial^n_u W(q,u)}{\partial u^n} \, , \\
    &= - u \cdot \nabla W(q,u) + \nabla V(q) \cdot \nabla_u W(q,u) + \mathcal{O}(\hbar^2) \, . \nonumber
\end{align}
We see that the equation of motion is simply the Vlasov equation with correction terms. Interesting, just like the ``quantum" pressure, the first order correction term is proportional to $\hbar^2$ and additional derivatives of the spatial density. The powers of $\hbar$ give us the intuition that the corrections again become small as the dynamical scales in our system become large compared to $\hbar$, however, like the quantum pressure, the additional derivatives make the exact limit somewhat unclear.

However, unlike a traditional phase space, the Wigner function is not generally everywhere positive. In fact, the Wigner function is only everywhere positive for a Gaussian field. Additionally, the Wigner function may have oscillatory features that make it difficult to use directly.  The drawbacks of the Wigner function often motivate the other pseudo-phase space descriptions of fields, the Husimi distribution. The Husimi distribution can be represented as a product of two Wigner functions. It is often useful to write the Husimi distribution in terms of the Husimi phase space basis states
\begin{align}
    \braket{x|q,u} = (2 \pi s^2)^{-1/4} e^{-(x-q)^2/4s^2} e^{im \, u q/\hbar} \, .
\end{align}
Where $s$ is the spatial smoothing scale, a non-dynamical parameter chosen arbitrarily. We note that these basis states are not orthogonal and are over-complete. The Wigner function can then be written in terms of these basis states as 
\begin{align} \label{eqn:Husimi}
    H(q,u) = (2 \pi \hbar)^{-1} |\braket{q,u | \psi}|^2 \, .
\end{align}
Unlike the Wigner function the Husimi function is everywhere positive. This makes it more useful in some applications, especially those that require a visual representation of the phase space. However, the Husimi function does not have an equation of motion with a straightforward relation to the Vlasov equation, nor does it recover the correct integrated distributions.

The evolution of both phase spaces will closely approximate the evolution of a traditional phase space obeying the Vlasov-Poisson equations for scales in phase space large compared to $\hbar$. In Figure \ref{fig:phaseSpace}, we compare the phase space for corpuscular dark matter and ultralight dark matter simulations of the gravitational collapse of an initial overdensity in a single spatial dimension. We can see that the two phase spaces are remarkably similar on large scales. 

\subsection{Hamilton's equations} \label{subsec:hamEqns}

In this subsection, we will show the relationship between Hamilton's equations and Schr\"odinger's equation. This is also worked out in \cite{Eberhardt2020}. It is also somewhat instructive to show how one can recover Hamilton's equations for corpuscular particles. We usually write these in terms of the position of the particle, $r$, and velocity $u$ as
\begin{align} \label{eqn:Ham}
    &\partial_t r = u \, , \\ 
    &\partial_t u = -\partial_x V(r) \, . \nonumber
\end{align}
Let us again consider the Schr\"odinger equation and its solution given
\begin{align}
    \partial_t \psi(x,t) = \frac{-i}{\hbar} \left( \frac{\hat p^2}{2m} + m V(x,t) \right) \psi(x,t) \, \\
    \psi(x, t + T) = e^{-i \int_t^{T+t} dt \left( \frac{\hat p^2}{2m} + m V(x,t) \right) / \hbar } \psi(x,t) \, . \label{eqn:SP_soln}
\end{align}
and the representation of the field in terms of phase and amplitude
\begin{align}
    \psi(x,t) = A(x,t) \, e^{iS(x,t) / \hbar} \, , \\
    \tilde \psi(u,t) =\tilde A(u,t) \, e^{i \tilde S(u,t) / \hbar} \, .
\end{align}
represented in the position and velocity basis respectively. We can split the exponential in equation \eqref{eqn:SP_soln} into one function of momentum and one function of position. Let us also consider the evolution of the solution over a small moment in time, $\delta t$, then we can say
\begin{align}
    e^{-\frac{i \delta t}{\hbar} \left( \frac{\hat p^2}{2m} + m V(x,t) \right)} \approx e^{-i \delta t \frac{\hat p^2}{2m} / \hbar} e^{-i \delta t \, m V(x,t) / \hbar} \left( 1 + \mathcal{O}(\delta t^2 [\nabla^2, V(x)]) \right) \, .
\end{align}
We can write the evolution now as an update of the field in the velocity basis, followed by an update of the field inthe  position basis. We write these two updates as updates of the phases in each basis, i.e.
\begin{align}
    \delta S(x,t) &= \delta t \,  m V(x,t) \, , \\
    \delta \tilde S(u,t) &= \delta t \, \frac{p^2}{2m} \, .
\end{align}
Recall, the argument we made in previous sections that the derivative of the complex phase is related to the representation in the other basis. By this we mean that $u(x,t) = \frac{1}{m} \nabla S(x,t)$ and $x(u,t) = \nabla_v \tilde S(u,t)$, recall also that $p = mu$. Taking the appropriate derivatives of the equations above and restoring $\delta t$ to a derivative, we find
\begin{align}
    &\partial_t \partial_x S(x,t) = \partial_t u(x,t) = m \, \partial_x V(x,t) \, , \\
    &\partial_t \partial_u \tilde S(u,t) = \partial_t x(u,t) = p/m = u \, , .
\end{align}
Which are, of course, equations \eqref{eqn:Ham}. Though the arguments above only really make sense for the case where the phase space is single valued.

\section{Length, and time scales}

The scale at which wavelike effects manifest in astrophysical and cosmological systems is generally a function of the mass of the ultralight dark matter field (especially in the minimally coupled case in which only gravity is relevant). It is useful, therefore, to track a few relevant quantities which can help us understand ultralight dark matter dynamics. The Compton time, $\tau_\mathrm{C} = \hbar / mc^2$, is the timescale associated with the mass energy of the field. This timescale is usually short compared to dynamical times of astrophysical systems at masses which could plausibly account for the universe's dark matter content and is therefore often ignored (albeit with some notable exceptions, e.g., \cite{Khmelnitsky2014}). 

\begin{align}
    \tau_\mathrm{C} &= 2 \pi \hbar / m c^2 \sim 10^{-5} \left( \frac{10^{-17} \, \mathrm{eV}}{m} \right) \, \mathrm{yrs} = \left( \frac{10^{-22} \, \mathrm{eV}}{m}\right) \, \mathrm{yrs} \, , \\
    \tau_\mathrm{db} &= 2 \pi \hbar / m \sigma^2 \sim 30 \left( \frac{10^{-17} \, \mathrm{eV}}{m} \right) \left( \frac{200 \, \mathrm{km/s}}{\sigma} \right)^{2} \, \mathrm{yrs} = 3 \left( \frac{10^{-22} \, \mathrm{eV}}{m} \right) \left( \frac{200 \, \mathrm{km/s}}{\sigma} \right)^{2} \, \mathrm{Myr} \, ,  \\ 
    \lambda_\mathrm{db} &= 2 \pi \hbar / m \sigma \sim 6 \times 10^{-6} \left( \frac{10^{-17} \, \mathrm{eV}}{m} \right) \, \mathrm{kpc} = 0.6 \left( \frac{10^{-22} \, \mathrm{eV}}{m} \right) \left( \frac{200 \, \mathrm{km/s}}{\sigma} \right) \, \mathrm{kpc} \, .
\end{align}

\vspace{0.5cm}

 \begin{figure*}[!ht]
	\includegraphics[width = .97\textwidth]{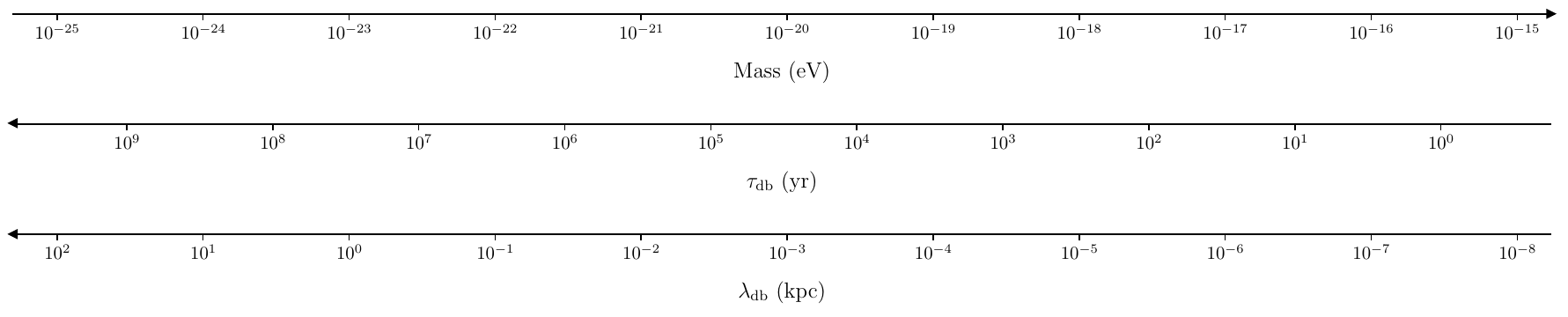}
	\caption{Comparison between mass of ULDM, de Broglie time, and de Broglie wavelength for a Milky Way-like galaxy.}
	\label{fig:mass_time_lambda}
\end{figure*}

\bibliography{BIB.bib}
\end{document}